\documentclass[12pt,a4paper]{article}
\usepackage[utf8x]{inputenc}
\usepackage[L7x]{fontenc}
\usepackage[lithuanian]{babel}
\usepackage{amsmath}
\usepackage{feynarts}
\usepackage{float}
\usepackage{slashed}
\usepackage[dvips]{graphicx}
\usepackage{caption}
\usepackage{subcaption}
\usepackage{chngcntr}
\usepackage{rotating}
\usepackage{pifont}

\counterwithin{figure}{section}
\numberwithin{equation}{subsection}

\usepackage[top=2cm, bottom=3cm, left=3cm, right=2cm]{geometry}

\begin{document}

\thispagestyle{empty}
\begin{center}
\section*{Predicting $ \tau $ lepton polarization at the LHC}
\textsc{Adomas Jelinskas}\\
\end{center}

\subsection*{Abstract}

This is a master's thesis and it is written in Lithuanian language.

The goal of this thesis is to predict the polarization of a $ \tau $ particle produced at the LHC and coming from a real $ W $ boson. This is achieved by calculating the projection of the expectation value of the polarization vector of the $ \tau $ particle. Calculations are done in the frame of the Standard Model. In this model only left-chiral currents couple to $ W $ bosons. The mass of the $\tau$ particle is left non-zero, therefore one can see the influence of helicity mixing on polarization. 

The initial colliding particles at the LHC are protons and the most important channels for $ W $ production are quark fusion: $ q\overline{q}\rightarrow W^+ \rightarrow \tau^+ \nu_{\tau} $. Calculation for $\tau $ production from $Z$ and $\gamma$ bosons, $ q\overline{q}\rightarrow Z^0,A^\gamma \rightarrow \tau^+ \tau^- $, are also included for investigation of possible back\-ground events. Because quarks are confined in a proton, this thesis presents a thorough treatment on discribing proton's inner structure with Parton distribution functions.

The results section shows plots of differential cross sections for $ \tau^+ $ production from different quark flavours and different energy intervals of the $ \tau^+ $ particle. A projection of the expectation value of the polarization vector depends on the chosen energy intervals. The vector on which it is projected is chosed to be the direction of the $ \tau^+ $ particle's motion. The influence of the mass is best seen for low energy particles, because then the helicity mixing is higher.

\vspace{0.5cm}
\noindent
PACS numbers: 13.88.+e, 14.60.Fg

\newpage
\thispagestyle{empty}
\begin{center}
{\large \textbf{VILNIAUS UNIVERSITETAS\\ FIZIKOS FAKULTETAS\\ TEORINĖS FIZIKOS KATEDRA} }\\
\vspace{144pt}
\large Adomas Jelinskas\\
\vspace{54pt}
{\Large $ \tau $ leptono poliarizacija LHC}\\
\vspace{108pt}
Magistro studijų baigiamasis darbas\\ \vspace{12pt}
(Studijų programa – Teorinė fizika ir astrofizika)\\
\vspace{80pt}
\begin{flushleft}
\begin{minipage}{0.35\textwidth}\begin{flushleft}
Studentas\\
Darbo vadovas\\
Recenzentas\\
Katedros vedėjas\\
\end{flushleft}\end{minipage}
\begin{minipage}[c]{0.46\textwidth}\begin{flushright}
Adomas Jelinskas \hspace{1cm}\\
Doc. Dr. Thomas Gajdosik\hspace{1cm}\\
Dr. Andrius Juodagalvis\hspace{1cm}\\
Prof. Habil. Dr. Leonas Valkūnas\hspace{1cm}\\
\end{flushright}\end{minipage}
\end{flushleft}
\vfill
Vilnius 2013
\end{center}

\newpage
\setcounter{page}{1}
\pagenumbering{arabic}
\tableofcontents

\newpage
\section*{Įvadas}
\addcontentsline{toc}{section}{Įvadas}

Šiame darbe nagrinėjami protono-protono susidūrimai Didžiajame priešpriešinių srautų greitintuve (angl. \textit{Large Hadron Collider}, LHC), kurių metu susidaro $ \tau $ leptonas ir prognozuojama jo poliarizacija. Labiausiai domina silpnosios sąveikos procesai, kurių metu susidaro $ \tau $ leptono ir atitinkamo $ \nu_\tau $ neutrino pora. Standartiniame modelyje laikoma, kad silnosios sąveikos nešėjas, $ W $ bozonas, sąveikauja tik su kairiojo chirališkumo srovė. Chirališkumas yra susijęs su poliarizacija. Norint aptikti dešiniojo chirališkumo srovės sąveiką, reikia nuspėti, kokiems matuojamiems dydžiams ši srovė turės įtakos. Vienas tokių dydžių yra vidutinės poliarizacijos projekcija į tam tikrą pasirinktą ašį. Taigi, šio darbo tikslas yra prognozuoti, Standartinio modelio ribose, kokią vidutinės poliarizacijos vektoriaus projekcijos vertę turės $ \tau^+ $ leptonai, susidarę Didžiajame hadronų greitintuve.

Didžiajame priešpriešinių srautų greitintuve pradinės dalelės yra protonai, bet protonų energijos yra tokios didelės, kad tiesiogiai sąveikauja protonus sudarančios dalelės (vadinamos partonais), todėl šiame darbe didelis dėmesys skiriamas protonų struktūros aprašymui. Nagrinėjama, kokia tikimybė protone aptikti tam tikrą partoną, kaip tos tikimybės priklauso nuo sąveikos energijos. Struktūra aprašoma partono pasiskirstymo funkcijomis, kurios yra naudojamos skaičiuoti protono-protono susidūrimų sklaidos skerspjūvius.

\newpage
\section{Protono aprašymas naudojant partono pasiskirstymo funkcijas}

Protonas yra sudėtinė dalelė, sudaryta iš kvarkų, antikvarkų ir gliuonų, bendrai vadinamų partonais. Šių dalelių dinamika yra aprašoma stipriosios sąveikos teorija, vadinama kvantine chromodinamika. Bet dabartiniai skaičiavimo metodai nėra pajėgūs išspręsti tokio sudėtingo uždavinio, kaip partonų energijos tikrinių verčių nustatymas. Todėl praktiškai ši problema sprendžiama pasitelkus eksperimentinius rezultatus ir protonas aprašomas struktūros funkcijomis. Šiame skyriuje yra aprašyta, kaip teoriškai yra įvedamos partono pasiskirstymo funkcijos ir kaip jos gali būti suskaičiuojamos pasitelkus eksperimentą.

\subsection{Protono struktūros funkcijos}

\begin{figure}[h]
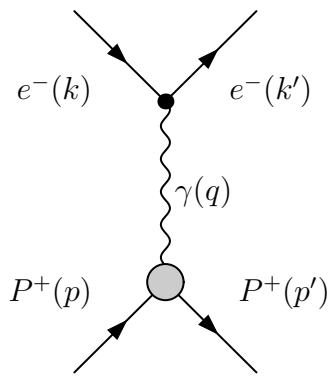

\centering
\begin{feynartspicture}(150,150)(1,1)\FADiagram{}
\FAProp(5.,20.)(10.,15.)(0.,){/Straight}{1}
\FALabel(5.98,16.48)[tr]{$e^-(k)$}
\FAProp(10.,15.)(15.,20.)(0.,){/Straight}{1}
\FALabel(13.52,16.48)[tl]{$e^-(k')$}
\FAProp(10.,15.)(10.,5.)(0.,){/Sine}{0}
\FALabel(10.52,10.)[l]{$\gamma(q)$}
\FAProp(10.,5.)(5.,0.)(0.,){/Straight}{-1}
\FALabel(5.98,3.52)[br]{$P^+(p)$}
\FAProp(10.,5.)(15.,-0.)(0.,){/Straight}{1}
\FALabel(14.02,3.5199)[bl]{$P^+(p')$}
\FAVert(10.,15.){0}
\FAVert(10.,5.){-1}
\end{feynartspicture}
\caption{Elektrono-protono sąveikos Feinmano diagrama}\label{1pav}
\end{figure}

Tarkime, kad elementarioji dalelė (elektronas) sąveikauja su protonu elektromagnetine sąveika. Feinmano diagrama tokiam procesui pavaizduota \ref{1pav} paveiksle,
kur fotono-protono viršūnė pavaizduota dideliu apskritimu norint parodyti, kad protonas nėra taškinė dalelė. Matricinis elementas tokiam procesui yra išreikštas dviejų tenzorių sandauga, iš kurių vienas yra elektrono-fotono viršūnės tenzorius ir nepriklauso nuo fotono-protono sąveikos viršūnės \cite{hey}. Metodai, kaip apskaičiuoti matricinius elementus, pateikti \cite{hey,peskin,griffiths}. Fotono-protono viršūnės tenzorių galima parametrizuoti remiantis tam tikromis prielaidomis. Laikoma, kad protono galinė būsena susumuota ir suvidurkinta pagal protono sukinius. Šis tenzorius bus žymimas $ W_{\mu\nu} $. Elektrono-fotono viršūnės tenzorius yra \cite{hey}
\begin{equation}
L^{\mu\nu}=2[k^\mu k'^\nu+k'^\mu k^\nu+(q^2/2)g^{\mu\nu}]
\end{equation}  
ir matricinis elementas yra
\begin{equation}
\langle|\mathcal{M}^2|\rangle=\left(\frac{e^2}{q^2}\right)^2 L^{\mu\nu}W_{\mu\nu}.
\end{equation} 
Čia $ q^\mu=k^\mu-k'^\mu $ yra fotono momento vektorius ir $ q^2=2(k\cdot k') $, kai nepaisoma elektrono rimties masės.  Pagrindinis kinematinis sąryšis 
\begin{equation}
p^\mu+k^\mu=p'^\mu+k'^\mu
\end{equation}
gali būti perrašomas
\begin{equation}
p'^\mu=p^\mu+q^\mu.
\end{equation}
Pakėlus kvadratu gaunama
\begin{equation}\label{sarysis}
\frac{-q^2}{2(p\cdot q)}=1.
\end{equation}
Šis sąryšis rodo, kad elastinės sklaidos metu vektoriai $ p^\mu $ ir $ q^\mu $ nėra nepriklausomi ir sistema aprašoma tik vienu nepriklausomu kintamuoju, dažniausiai $ q^2 $.

Tenzoriaus $ W_{\mu\nu} $ Lorenco indeksai gali būti sudaryti iš turimų sistemos vektorių ir tenzorių, o jo funkcinė priklausomybė - iš skaliarų. Rašant tenzorių bendru pavidalu reikia atsižvelgti į tam tikrus reikalavimus, kuriuos jis turi tenkinti: Lorenco kovariantiškumas (turi būti tenzorius visose atskaitos sistemose), turi tenkinti srovės tvermės dėsnį. 

Dydžiai, kurie gali sudaryti šį tenzorių, yra nepriklausomi vektoriai $p^\mu,\, q^\mu $ ir tenzorius $g^{\mu\nu}$. Srovės tvermės dėsnis yra $q^{\mu}W_{\mu\nu}=q^{\nu}W_{\mu\nu}=0$. Iš turimų dydžių galima sukonstruoti du tenzorius, kurie tenkina šį tvermės dėsnį:
\begin{equation}
\left(p^{\mu}-\frac{(q\cdot p)}{q^2}q^{\mu}\right)\left(p^{\nu}-\frac{(q\cdot p)}{q^2}q^{\nu}\right)
\end{equation}
ir
\begin{equation}
\left(-g^{\mu\nu}+\frac{q^{\mu}q^{\nu}}{q^2}\right).
\end{equation}
Šie du tenzoriai padauginami iš nežinomų funkcijų, kurios priklauso nuo turimų skaliarinių dydžių. Elastinės sąveikos metu yra vienas nepriklausomas kintamasis, dažniausiai imamas $q^2$.
Taigi, protono tenzorius, tenkinantis srovės tvermės dėsnį, parametrizuojamas taip:
\begin{eqnarray}
W^{\mu\nu}=4A(q^2)\left(p^{\mu}-\frac{(q\cdot p)}{q^2}q^{\mu}\right)\left(p^{\nu}-\frac{(q\cdot p)}{q^2}q^{\nu}\right)+2M^2B(q^2)\left(-g^{\mu\nu}+\frac{q^{\mu}q^{\nu}}{q^2}\right).
\end{eqnarray}
Daugikliai prie struktūros funkcijų $ A(q^2) $ ir $ B(q^2) $ yra tokie dėl patogumo. $ M $ yra protono masė. Reikia atkreipti dėmesį, kad tenzorius yra visiškai bendras ir nėra jokių prielaidų dėl protono struktūros.

Sklaidos skerspjūvis laboratorinėje atskaitos sistemoje gaunamas toks:
\begin{eqnarray}
\frac{d\sigma}{d\Omega}=\frac{\alpha^2}{4k^2\sin^4\!(\theta/2)}\frac{k'}{k}\left(A(q^2)\cos^2\!(\theta/2)+B(q^2)\sin^2\!(\theta/2)\right).
\end{eqnarray}
$ k $ ir $ k' $ yra elektrono energijos prieš ir po sąveikos. Šioje lygtyje yra dvi struktūros funkcijos, kurios atitinka elektrono sklaidą nuo protono elektrinio ir magnetinio momentų. Kintamieji $ k' $ ir $ \theta $ elastinės sklaidos metu yra susieti paprastu sąryšiu, kuris yra ekvivalentus (\ref{sarysis}):
\begin{equation}
\frac{k}{k'}=1+\frac{2k}{M}\sin^2\!(\theta/2).
\end{equation}
Eksperimente suskaičiavus išlekiančius elektronus kampais $ \theta(q^2) $ galima suskaičiuoti struktūros funkcijas $ A(q^2) $ ir $ B(q^2) $.

\subsection{Neelastinė sklaida}

\begin{figure}[h]
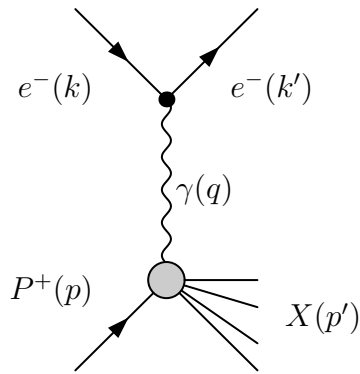

\centering
\begin{feynartspicture}(150,150)(1,1)\FADiagram{}
\FAProp(5.,20.)(10.,15.)(0.,){/Straight}{1}
\FALabel(5.98,16.48)[tr]{$e^-(k)$}
\FAProp(10.,15.)(15.,20.)(0.,){/Straight}{1}
\FALabel(13.52,16.48)[tl]{$e^-(k')$}
\FAProp(10.,15.)(10.,5.)(0.,){/Sine}{0}
\FALabel(10.52,10.)[l]{$\gamma(q)$}
\FAProp(10.,5.)(5.,0.)(0.,){/Straight}{-1}
\FALabel(5.98,3.52)[br]{$P^+(p)$}
\FAProp(10.,5.)(15.,0.)(0.,){/Straight}{0}
\FALabel(16.52,2.0)[bl]{$X(p')$}
\FAProp(10.,5.)(15.,1.5)(0.,){/Straight}{0}
\FAProp(10.,5.)(15.,3.5)(0.,){/Straight}{0}
\FAProp(10.,5.)(15.,5.)(0.,){/Straight}{0}
\FAVert(10.,15.){0}
\FAVert(10.,5.){-1}
\end{feynartspicture}
\caption{Elektrono-protono neelastinė sklaida}\label{2pav}
\end{figure}

Kai elektrono energija pakankamai didelė, elastinės sklaidos tikimybė yra maža palyginus su neelastinės sklaidos tikimybe. Būtent šio tipo sklaida ir parodo, remiantis taškinių susidūrimų teorija, kad protonas sudarytas iš taškinių dalelių partonų \cite{hey}. Neelastinės sklaidos metu galinė būsena gali būti viena iš daugybės galimų, todėl, esant didelėms energijoms, neelastinė sklaida yra daug tikėtinesnė. Lengviausia yra skaičiuoti sudėtinį sklaidos skerspjūvį (angl. \textit{inclusive cross section}), kuris gaunamas susumuojant visas galimas galines būsenas ir matuojant tik išlekiančio elektrono energiją. Schematiškai tai galima pavaizduoti diagrama \ref{2pav} paveiksle,
kur $ X(p') $ reiškia visas galimas galines būsenas, turinčias momentą $ p' $. Čia, kaip ir elastinės sklaidos atveju, laikoma, kad matricinis elementas yra suvidurkintas pagal poliarizacijas ir papildomai dar susumuotas pagal visas galimas galines būsenas $ X $ ne tik sukinio, bet ir kitų kvantinių skaičių atžvilgiu.
Kadangi susidūrimas neelastinis ir neišskiriama jokia galinė būsena, tai tos būsenos momentas yra nepriklausomas kintamasis. Taigi, elektrono momentas po smūgio $ k' $ ir kampas $ \theta $ yra nepriklausomi dydžiai, priešingai nei elastinio susidūrimo metu, kur tie dydžiai yra susieti paprasta formule. Struktūros funkcijos priklausys nuo dviejų kintamųjų.

Pagrindinis kinematinis sąryšis yra 
\begin{eqnarray}
p'^\mu=p^\mu+q^\mu,
\end{eqnarray}
kur $p'^\mu$ yra galinės būsenos momentas, $p^\mu$ - protono momentas, o $q^\mu$ yra perduodamas (virtualaus fotono) momentas. Pakėlus kvadratu
\begin{eqnarray}
p'^2=M^2+2(q\cdot p)+q^2.
\end{eqnarray}
$ p'^2 $ yra galinės būsenos invariantinė masė pakelta kvadratu. Ji nelaikoma lygi 0, nes paprastai galinė būsena būna sudaryta iš daugybės hadronų, todėl masė nebūtinai bus maža. Šiuo atveju galinės būsenos invariantinė masė $p'^2=W^2$ yra kintamasis, o ne konstanta $M^2$, kaip buvo elastinėje sklaidoje.\\
Pažymėjus
\begin{eqnarray}
q\cdot p=M\nu,\\
-q^2=Q^2,
\end{eqnarray}
kintamieji susieti taip:
\begin{equation}
2M\nu=Q^2+W^2-M^2.
\end{equation}

Tenzorius, apibūdinantis fotono-protono sąveiką, parametrizuojamas taip pat, kaip elastinės sklaidos atveju, bet struktūros funkcijos priklauso nuo dviejų kintamųjų, dažniausiai nuo $Q^2$ ir $\nu$:
\begin{eqnarray}
W^{\mu\nu}=W_1(Q^2,\nu)\left(-g^{\mu\nu}+\frac{q^{\mu}q^{\nu}}{q^2}\right)+\frac{W_2(Q^2,\nu)}{M^2} \left(p^{\mu}-\frac{(q\cdot p)}{q^2}q^{\mu}\right)\left(p^{\nu}-\frac{(q\cdot p)}{q^2}q^{\nu}\right).
\end{eqnarray}

Matricinis elementas turi papildomą daugiklį $ 4\pi M $, nes protono tenzorius taip apibrėžtas:
\begin{equation}
\langle|\mathcal{M}^2|\rangle=\left(\frac{e^2}{q^2}\right)^2 4\pi M L^{\mu\nu}W_{\mu\nu}.
\end{equation} 

Skaičiuojant diferencialinį sklaidos skerspjūvį, galinių būsenų tankis rašomas tik išlekiančiam elektronui, nes kitų galinės būsenos dalelių impulsai jau susumuoti protono tenzoriuje.
Laboratorinėje atskaitos sistemoje 
\begin{equation}
\frac{d^2\sigma}{d\Omega dk'}=\frac{\alpha^2}{4k^2\sin^4(\theta/2)}\left(W_2\cos^2(\theta/2)+2W_1\sin^2(\theta/2)\right),
\end{equation}
arba pakeitus kintamuosius 
\begin{eqnarray}
Q^2=2kk'(1-\cos \theta),\\
\nu=k-k',
\end{eqnarray}
\begin{equation}
d\Omega dk'=\frac{\pi}{kk'}dQ^2 d\nu
\end{equation} 
gaunama sklaidos skerspjūvio formulė
\begin{equation}
\label{d}
\frac{d^2\sigma}{dQ^2 d\nu}=\frac{\pi\alpha^2}{4k^2\sin^4(\theta/2)}\frac{1}{kk'}\left(W_2\cos^2(\theta/2)+2W_1\sin^2(\theta/2)\right).
\end{equation}

Nagrinėjant partonų modelį yra patogu pereiti prie naujų bedimensinių kintamųjų:
\begin{eqnarray}
x=\frac{Q^2}{2M\nu},\\
y=\frac{\nu}{k},
\end{eqnarray}
\begin{equation}
dQ^2 d\nu=2Mk^2ydxdy.
\end{equation}

Šie sklaidos skerspjūviai yra labai bendri, tačiau priklauso nuo nežinomų funkcijų $ W_1 $ ir $ W_2 $, kurios priklauso nuo turimų skaliarinių dydžių $ Q^2 $ ir $ \nu $. Esant mažai energijai, sklaidos skerspjūvis turi susivesti į elastinės sklaidos, o dvi funkcijos $ W_1 $ ir $ W_2 $ atitinkamai į gautąsias elastinei sklaidai. Šios funkcijos bendru atveju yra labai sudėtingos priklausomybės nuo kintamųjų, tačiau esant didelei energijai, atsiranda priklausomybė tarp $ Q^2 $ ir $ \nu $ kintamųjų ir struktūros funkcijos priklauso jau ne nuo  $ Q^2 $ ir $ \nu $ atskirai, bet nuo jų santykio (šis efektas angliškai vadinamas \textit{Bjorken scaling}). Tai labai supaprastina teorinę analizę ir veda prie partonų modelio.

\subsection{Partonų modelis}

Partonų modelį, aprašantį protono vidinę struktūrą, pasiūlė Bjorken ir Feynman \cite{hey, peskin}. Šiame modelyje laikoma, kad protono sudedamosios dalys yra kvarkai, antikvarkai ir elektrinio krūvio neturinčios kitos dalelės, vėliau identifikuotos kaip gliuonai. Modelis remiasi tuo, kad eksperimentiškai nustatyta, jog protono sudėtinės dalelės elgiasi lyg laisvos taškinės dalelės. Tačiau kyla klausimas, kodėl galima laikyti partonus nesąveikaujančiais trumpais laiko intervalais?\cite{peskin} Problema kyla dėl to, kad kvantinio lauko teorijoje laikoma, kad dalelės sąveikauja keisdamosis virtualiais bozonais, kurie gali turėti neribotai didelius momentus, o dėl to gali gyvuoti labai trumpai. Tai reiškia, kad sąveikos trukmė gali būti be galo trumpa - sąveika negali išsijungti mažiems laiko tarpams ir dalelės negali tapti laisvomis. Šią problemą išsprendžia efektas, angliškai vadinamas \textit{asymptotic freedom}, kuris reiškia, kad sąveikos stipris priklauso nuo energijos (arba atstumo) skalės. Stipriosios sąveikos teorijoje QCD parodoma, kad didėjant energijai (arba mažėjant atstumui) sąveikos stipris mažėja. Todėl galima laikyti, kad stiprioji sąveika išsijungia esant labai trumpiems sąveikos laikams.

Praeitame skyriuje nagrinėtos protono struktūros funkcijos, kaip spėjo Bjorken, tampa 
\begin{align*}
&MW_1(Q^2,\nu)\rightarrow F_1(x),\\
&\nu W_2(Q^2,\nu)\rightarrow F_2(x),  
\end{align*}
kai energija labai didelė (angl. \textit{deep inelastic scatering}). Kintamieji $Q^2$ ir $\nu$ tampa susieti: kintamasis $x=Q^2/2M\nu$ tampa fiksuotas (jei $Q^2$ didelis, tai ir $\nu$ didelis).
Eksperimentai tokius spėjimus patvirtino \cite{hey,perkins}. Toks funkcijų ir kinematinių kintamųjų elgesys paaiškinamas tuo, jog elektronas elastiškai sąveikauja su protone esančiais kvarkais. Pažiūrėjus į elastinės sklaidos surištus kintamuosius $Q^2=2M\nu$ matomas panašumas su neelastinės sklaidos kintamaisiais $Q^2=2M\nu x$. Tai duoda užuominą apie sąveikų panašumą. Taigi, galima teigti, kad sąveika vėl tampa elastine, tik ne su protonu, o su partonu, jei elastinės sąveikos formulėje protono masę $ M $ pakeisime kvarko mase $ m_i=xM $ ir dydį $ x $ laikysime protono rimties masės dalimi, kurią sudaro kvarko masė.
Teoriškai tai galima paaiškinti taip: didelio momento $ Q^2 $ perdavimo metu fotonas mato labai smulkius atstumo ir laiko intervalus, todėl fotonas gali sąveikauti tiesiogiai su labai mažomis sudedamosiomis protono dalimis, kvarkais, kurie turi tik dalį protono momento.

Atskaitos sistemoje, kurioje protono momentas yra didelis, masės galima nepaisyti ir momento vektorius yra \textit{lightlike} vektorius, nukreiptas susidūrimo ašimi. Partonai taip pat turi \textit{lightlike} momentus, beveik lygiagrečius susidūrimo ašiai. Jų statmenos momentų komponentės yra labai mažos, nes tam reikėtų sugerti didelės energijos gliuoną, bet tai yra negalima dėl \textit{asymptotic freedom}: esan didelės energijos gliuonui, sąveikos stipris $ \alpha_s $ yra labai mažas. Todėl galima laikyti, kad partonai neturi statmenos momento komponentės ir kvarko momentas yra
\begin{equation}
p^\mu=fP^\mu,
\end{equation}
kur $ f $ yra kvarko turima protono momento dalis, $ P^\mu $ - protono momentas. Žemiausios eilės artutinumu nepaisoma procesų su gliuonais, t.y. kvarkai laikomi laisvais pradinėje būsenoje ir po sąveikos tie patys kvarkai laikomi esą galinės būsenos. Tuomet elektrono-protono sąveikos skerspjūvis bus skaičiuojamas procesui, kuriame elektrono emituotas fotonas sąveikauja su vienu laisvu partonu. Sąveikos skerspjūvio formulė priklausys nuo partono momento dalies $ f $ ir tikimybės $ f_i(f) $, kad $ i $ skonio kvarkas turi $ f $ dalį momento.

Taigi, laikoma, kad sklaida yra elastinė, o sąveikauja elektronas su laisvu kvarku, turinčiu dalį $ f $ protono momento. Momento tvermės dėsnis fotono-kvarko viršūnėje yra
\[ q^\mu+fp^\mu=p'^\mu, \]
\[ q^2+f^2M^2+2f(p\cdot q)=m^2\approx 0,\]		
kur $p'^\mu$ yra kvarko momentas po susidūrimo. Tarus, kad partonas nėra virtualus (kaip anksčiau aptarta, jis yra laisvas), momento kvadratas lygus masės kvadratui. Taip pat galime nepaisyti nario $ f^2M^2 $. Tuomet  
\begin{equation}
f=\frac{Q^2}{2M\nu}\equiv x.
\end{equation}
Todėl šiame modelyje kintamasis $ x $ yra kvarko turima protono momento ir rimties masės dalis.

Jei sąveika su kvarku yra elastinė, tai galima naudoti paprastas formules elastinei taškinei sklaidai, kuri priklausys nuo kvarko savybių (masės, protono momento dalies, jo krūvio). Elastinės taškinės elektrono-kvarko sklaidos diferencialinis sklaidos skerspjūvis $ i $-ojo skonio kvarkui užrašomas
\begin{equation}\label{onequark}
\frac{d^2\sigma^i}{dQ^2d\nu}=\frac{\pi \alpha^2}{4k^2\sin^4(\theta/2)}\,\frac{1}{kk'}\,\left(Q_i^2\cos^2(\theta/2)+Q_i^2\frac{Q^2}{4m_i^2}2\sin^2(\theta/2)\right)\delta(\nu-Q^2/2m_i),
\end{equation}
kur $ m_i $ ir $ Q_i $ yra $ i $ skonio kvarko masė ir krūvis $ e $ vienetais. Delta funkcija atsiranda, nes sklaida elastinė ir turi galioti atitinkamas sąryšis tarp kintamųjų.

Palyginus vienos rūšies kvarko sąveikos formulę (\ref{onequark}) su sudėtine formule (\ref{d}), kur įtrauktos sąveikos su visais kvarkais, matoma, kad vieno kvarko indėlis į struktūros funkcijas yra
\begin{align}
W^i_1&=Q_i^2\frac{Q^2}{4M^2x^2}\delta(\nu-Q^2/2Mx),\\
W^i_2&=Q_i^2\delta(\nu-Q^2/2Mx).
\end{align}
Norint gauti pilnas funkcijas $W$, reikia sudėti visų kvarkų indėlius ir suintegruoti pagal visas galimas momento dalies $x$ vertes. Po integralu reikia įrašyti tikimybės funkciją $f_i(x)$, kuri reiškia, kad $i$ skonio kvarkas turi $x$ dalį protono momento. Šios tikimybės $f_i(x)$ ir yra vadinamos partono pasiskirstymo funkcijomis (angl. \textit{Parton Distribution Functions, PDF}) ir jos nėra gaunamos iš modelio, bet įeina į modelį kaip parametrai, apibūdinantys protoną.
\begin{align}
W_1(\nu,Q^2)&=\sum_i\int_0^1f_i(x)Q_i^2\frac{Q^2}{4M^2x^2}\delta(\nu-Q^2/2Mx)dx,\\
W_2(\nu,Q^2)&=\sum_i\int^1_0f_i(x)Q_i^2\delta(\nu-Q^2/2Mx)dx.
\end{align}
Integralai lengvai suskaičiuojami dėl delta funkcijos:
\begin{align}
MW_1(\nu,Q^2)&=\sum_i\frac{1}{2}Q_i^2f_i(x)\equiv F_1(x),\\
\label{f2}
\nu W_2(\nu,Q^2)&=\sum_iQ_i^2xf_i(x)\equiv F_2(x)
\end{align}
ir gaunamas sąryšis
\begin{equation}
2xF_1(x)=F_2(x).
\end{equation}

Pasinaudojus gautais rezultatais dėl struktūros funkcijų ir pakeitus kintamuosius $x=Q^2/2M\nu$ ir $ y=\nu/k $, sąveikos skerspjūvį galima užrašyti
\begin{equation}
\frac{d^2\sigma}{dxdy}=\frac{2\pi\alpha^2}{Q^4}sF_2(x)\left(1+(1-y)^2\right)
\end{equation}

Iš pradžių galima manyti, kad funkcijos $F_1$ ir $F_2$ bus didelės tik taško $x=1/3$ aplinkoje, nes yra trys kvarkai, t.y. kiekvienas jų turės trečdalį protono momento. Tačiau taip nėra \cite{hey, peskin}: struktūros funkcijų priklausomybė gana sudėtinga nuo $x$ ir jos gali turėti dideles vertes taškuose toli nuo $x=1/3$ taško.

\subsection{Partono pasiskirstymo funkcijos}

Laikant, kad elektronas sąveikauja su krūvį turinčiais kvarkais, tikimybės funkcijas galima užrašyti išreikštai kiekvieno skonio kvarkams ir antikvarkams. Tokios pačios funkcijos kiekvienam kvarko skoniui bus ne tik protono atveju, bet ir neutrono, nes stiprioji sąveika nepriklauso nuo kvarkų skonių. Taip pat tos pačios funkcijos naudojamos aprašyti protono sąveiką su neutrinais. Taigi, yra keli eksperimentiniai šaltiniai šioms funkcijoms gauti. 

Gautoje funkcijos $ F_2(x) $ išraiškoje (\ref{f2}) tikimybės funkcijos užrašomos išreikštai kiekvienam kvarko skoniui, o daugikliai $ Q_i^2 $ yra atitinkami krūviai \cite{hey}:
\begin{equation}\label{eproton}
F_2^{ep}(x)=x\lbrace \frac{4}{9}\left[u(x)+\overline{u}(x)\right]+\frac{1}{9}\left[d(x)+\overline{d}(x)+s(x)+\overline{s}(x)\right]\ldots \rbrace.
\end{equation}
Čia $ u(x)\equiv f_u(x) $ yra $ u $ skonio kvarko pasiskirstymo funkcija, $ \overline{u}(x) $ - $ u $ skonio antikvarko ir t.t., o taškai reiškia kitus galimus kvarkų skonius. Viršutinis indeksas $ ep $ rodo, kad funkcija skirta elektrono protono sklaidai.

Partono pasiskirstymo funkcijos turi tenkinti tam tikras sąlygas, kad atspindėtų tinkamus protono kvantinius skaičius, kaip antai: visa tikimybė rasti $ s $ skonio kvarką ir antikvarką yra lygi 0:
\begin{equation}
\label{ssuma}
\int_0^1[s(x)-\overline{s}(x)]dx=0.
\end{equation}
Taip pat protone turi būti du $ u $ kvarkai ir vienas $ d $ kvarkas:
\begin{eqnarray}
\label{usuma}
\int_0^1[u(x)-\overline{u}(x)]dx=2,\\
\label{dsuma}
\int_0^1[d(x)-\overline{d}(x)]dx=1.
\end{eqnarray}
Visas partonų momentas turi būti lygus protono momentui
\begin{equation}
\int_0^1x[u(x)+\overline{u}(x)+d(x)+\overline{d}(x)+g(x)+\ldots]dx=1,
\end{equation}
kur funkcija $ g(x) $ yra gliuono pasiskirstymo funkcija. Yra nustatyta, kad maždaug pusę protono momento turi gliuonai.

Kiekvienas hadronas turi savo partonų pasiskirstymo funkcijas, tenkinančias sąlygas, kurios atspindi tinkamus hadrono kvantinius skaičius. Dėl hadronų simetrijos savybių funkcijos turi atspindėti tas simetrijas, pavyzdžiui, protono ir neutrono partonų pasiskirstymo funkcijos apytiksliai tenkina tokias lygybes:
\begin{align}
&u^p(x)=d^n(x)\equiv u(x),\\
&d^p(x)=u^n(x)\equiv d(x),
\end{align}
kur $ p $ ir $ n $ indeksai nurodo, kad funkcijos yra atitinkamai protono ir neutrono. Dėl to struktūros funkciją elektrono-neutrono sklaidai galima užrašyti
\begin{equation}
F_2^{en}(x)=x\lbrace \frac{4}{9}[d(x)+\overline{d}(x)]+\frac{1}{9}[u(x)+\overline{u}(x)+s(x)+\overline{s}(x)]\ldots \rbrace.
\end{equation}
Čia reikia atkreipti dėmesį, kad naudojamos tos pačios funkcijos, kaip ir elektrono-protono sklaidai, ir pirmosios dvi partono pasiskirstyymo funkcijos reiškia, kad neutrone tikimybės rasti $ u $ kvarką ir antikvarką yra tokios pačios, kaip protone rasti $ d $ skonio kvarką ir antikvarką.

Struktūros funkcijas galima dar labiau supaprastinti tarus, kad tikimybė susidaryti $ u, d, s $ kvarkų ir antikvarkų poroms dėl kvantinių fliuktuacijų yra vienoda. Tuomet jų pasiskirstymo funkcijos bendrai pavadinamos $ q_s $. Indeksas $ s $ čia reiškia \textit{sea} kvarkus. $ u $ ir $ d $ kvarkų funkcijos bus valentinių kvarkų (tai kvarkai, kurie egzistuoja protone ne dėl kvantinių fliuktuacijų) $ u_v $ ir $ d_v $ ir \textit{sea} kvarkų funkcijų suma, o \textit{sea} kvarkai bus $ u $ ir $ d $ antikvarkai bei visi $ s $ kvarkai ir antikvarkai:
\begin{eqnarray}
u=u_v+q_s,\\
d=d_v+q_s,\\
\overline{d}=\overline{u}=s=\overline{s}=q_s.
\end{eqnarray}
Toks modelio supaprastinimas pakeičia šešias funkcijas (\ref{eproton}) lygtyje trimis funkcijomis.

\subsection{Neutrinų sklaida protonais}

\begin{figure}[h]
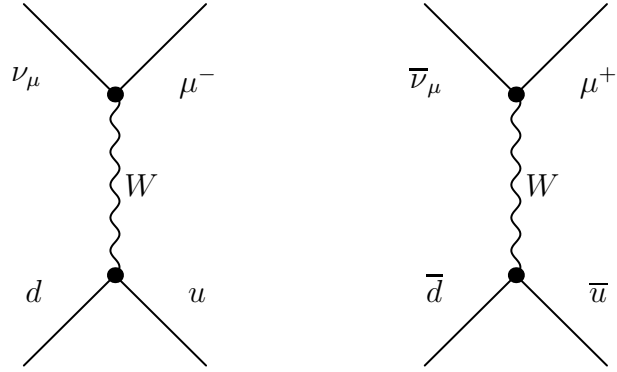

\centering
\begin{feynartspicture}(300,150)(2,1)\FADiagram{}
\FAProp(5.,20.)(10.,15.)(0.,){/Straight}{0}
\FALabel(5.98,16.48)[tr]{$\nu_{\mu}$}
\FAProp(10.,15.)(15.,20.)(0.,){/Straight}{0}
\FALabel(13.52,16.48)[tl]{$\mu^-$}
\FAProp(10.,15.)(10.,5.)(0.,){/Sine}{0}
\FALabel(10.52,10.)[l]{$W$}
\FAProp(10.,5.)(5.,0.)(0.,){/Straight}{0}
\FALabel(5.98,3.52)[br]{$d$}
\FAProp(10.,5.)(15.,-0.)(0.,){/Straight}{0}
\FALabel(14.02,3.5199)[bl]{$u$}
\FAVert(10.,15.){0}
\FAVert(10.,5.){0}
\FADiagram{}
\FAProp(5.,20.)(10.,15.)(0.,){/Straight}{0}
\FALabel(5.98,16.48)[tr]{$\overline{\nu}_{\mu}$}
\FAProp(10.,15.)(15.,20.)(0.,){/Straight}{0}
\FALabel(13.52,16.48)[tl]{$\mu^+$}
\FAProp(10.,15.)(10.,5.)(0.,){/Sine}{0}
\FALabel(10.52,10.)[l]{$W$}
\FAProp(10.,5.)(5.,0.)(0.,){/Straight}{0}
\FALabel(5.98,3.52)[br]{$\overline{d}$}
\FAProp(10.,5.)(15.,-0.)(0.,){/Straight}{0}
\FALabel(14.02,3.5199)[bl]{$\overline{u}$}
\FAVert(10.,15.){0}
\FAVert(10.,5.){0}
\end{feynartspicture}
\caption{Neutrino-kvarko sklaida}\label{3pav}
\end{figure}

\begin{figure}[h]
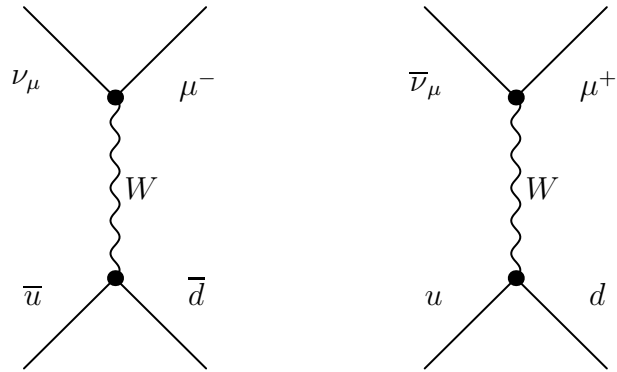

\centering
\begin{feynartspicture}(300,150)(2,1) \FADiagram{}
\FAProp(5.,20.)(10.,15.)(0.,){/Straight}{0}
\FALabel(5.98,16.48)[tr]{$\nu_{\mu}$}
\FAProp(10.,15.)(15.,20.)(0.,){/Straight}{0}
\FALabel(13.52,16.48)[tl]{$\mu^-$}
\FAProp(10.,15.)(10.,5.)(0.,){/Sine}{0}
\FALabel(10.52,10.)[l]{$W$}
\FAProp(10.,5.)(5.,0.)(0.,){/Straight}{0}
\FALabel(5.98,3.52)[br]{$\overline{u}$}
\FAProp(10.,5.)(15.,-0.)(0.,){/Straight}{0}
\FALabel(14.02,3.5199)[bl]{$\overline{d}$}
\FAVert(10.,15.){0}
\FAVert(10.,5.){0}
 \FADiagram{}
 \FAProp(5.,20.)(10.,15.)(0.,){/Straight}{0}
\FALabel(5.98,16.48)[tr]{$\overline{\nu}_{\mu}$}
\FAProp(10.,15.)(15.,20.)(0.,){/Straight}{0}
\FALabel(13.52,16.48)[tl]{$\mu^+$}
\FAProp(10.,15.)(10.,5.)(0.,){/Sine}{0}
\FALabel(10.52,10.)[l]{$W$}
\FAProp(10.,5.)(5.,0.)(0.,){/Straight}{0}
\FALabel(5.98,3.52)[br]{$u$}
\FAProp(10.,5.)(15.,-0.)(0.,){/Straight}{0}
\FALabel(14.02,3.5199)[bl]{$d$}
\FAVert(10.,15.){0}
\FAVert(10.,5.){0}
\end{feynartspicture}
\caption{Neutrino-kvarko sklaida}\label{4pav}
\end{figure}

Analogiškai praeituose poskyriuose aptartai elektrono-protono sklaidai galima užrašyti sklaidos skerspjūvius ir struktūros funkcijas neutrino-protono sklaidai. Esant pakankamai didelei energijai, neutrinas sąveikauja su viena taškine dalele, todėl reikia žinoti formules neutrinų (antineutrinų) ir kvarkų sklaidai. Diferencialiniai sklaidos skerspjūviai skirtingoms silpnosios sąveikos reakcijoms laboratorinėje atskaitos sistemoje išreiškiami lygtimi per kintamąjį $ y $ \cite{perkins}:
\begin{equation}
\frac{d\sigma^{\nu q}}{dy}=\frac{G^2 s}{\pi},
\end{equation}
kuri tinka dviem procesams
\begin{eqnarray}
\nu_{\mu}+d\longrightarrow \mu^- + u\nonumber,\\
\overline{\nu}_{\mu}+\overline{d}\longrightarrow \mu^+ + \overline{u},
\end{eqnarray}
vaizduojamiems Feinmano diagramomis \ref{3pav} paveiksle
ir lygtimi
\begin{equation}
\frac{d\sigma^{\overline{\nu}q}}{dy}=\frac{G^2 s}{\pi}(1-y)^2,
\end{equation}
kuri tinka dviem procesams
\begin{eqnarray}
\nu_{\mu}+\overline{u}\longrightarrow \mu^- + \overline{d}\nonumber,\\
\overline{\nu}_{\mu}+u\longrightarrow \mu^+ +d,
\end{eqnarray}
vaizduojamiems Feinmano diagramomis \ref{4pav} paveiksle.
Kaip ir praeitame skyriuje, pilnutinį sąveikos skerspjūvį neutrino (antineutrino)-protono arba neutrino (antineutrino)-neutrono sklaidai galima užrašyti su partonų pasiskirstymo funkcijomis:
\begin{equation}
\frac{d\sigma^{\nu p}}{dydx}=\frac{G^2 s}{\pi}x\left[d(x)+\overline{u}(x)(1-y)^2\right],
\end{equation}
\begin{equation}
\frac{d\sigma^{\nu n}}{dydx}=\frac{G^2 s}{\pi}x\left[u(x)+\overline{d}(x)(1-y)^2\right],
\end{equation}
kur indeksai $ \nu p $ reiškia neutrino sklaidą nuo protono, o $ \nu n $ - nuo neutrono.

Sklaidai nuo izoskaliarinių taikinių (kurie turi po lygiai protonų ir neutronų)
\begin{equation}
\label{neutrin}
\frac{d\sigma^{\nu N}}{dydx}=\frac{G^2 s}{2\pi}x\left[[u(x)+d(x)]+[\overline{u}(x)+\overline{d}(x)](1-y)^2\right].
\end{equation}
Atitinkamai antineutrinams sklaidos formulė gaunama sukeitus vietomis daugiklius 1 ir $ (1-y)^2 $:
\begin{equation}
\label{antineutrin}
\frac{d\sigma^{\overline{\nu} N}}{dydx}=\frac{G^2 s}{2\pi}x\left[[u(x)+d(x)](1-y)^2+[\overline{u}(x)+\overline{d}(x)]\right].
\end{equation}
Įvedus struktūros funkcijas
\begin{equation}
\frac{F_2^{\nu N}(x)}{x}=u(x)+d(x)+\overline{u}(x)+\overline{d}(x),
\end{equation}
\begin{equation}
F_3^{\nu N}(x)=u(x)+d(x)-\overline{u}(x)-\overline{d}(x),
\end{equation}
formulės (\ref{neutrin}) ir (\ref{antineutrin}) tampa
\begin{equation}
\frac{d\sigma^{\nu N,\overline{\nu} N}}{dydx}=\frac{G^2 s}{2\pi}\left\lbrace\left[\frac{F_2(x)\pm xF_3(x)}{2}\right]+\left[\frac{F_2(x)\mp xF_3(x)}{2}\right](1-y)^2\right\rbrace.
\end{equation}

Kol kas šiame skyriuje pateiktas formalizmas yra pirmos eilės aproksimacija skaičiuojant leptono-protono sąveikos skerspjūvį. Aukštesnės eilės pataisos turės įskaityti modifikacijas partonų pasiskirstymo funkcijoms. Čia nagrinėtos partonų pasiskirstymo funkcijos esą priklauso tik nuo kintamojo $ x $, ir ši savybė vadinama \textit{Bjorken scaling}, tačiau dėl aukštesnės eilės pataisų funkcijos priklausys ir nuo dalelių energijos. Tai vadinama \textit{scaling violation}: funkcijos lėtai kinta nuo $ Q^2 $. Tai yra pataisa dėl QCD efektų. 

\subsection{Protono-protono sklaida}

Remiantis ankstesniuose skyriuose aprašytu partonų modelių ir partono pasiskirstymo funkcijomis galima prognozuoti protono-protono susidūrimų rezultatus. Kadangi partono  pasiskirstymo funkcijos aprašo protoną iki sąveikos su vektoriniu bozonu, tai turi būti įmanoma, naudojantis tomis pačiomis PDF, aprašyti protonų susidūrimus. Būtent šis procesas ir vyksta Didžiajame hadronų greitintuve ir jį reikia aprašyti, norint suskaičiuoti greitintuve susidariusios $ \tau $ dalelės sklaidos skerspjūvį.
Nagrinėjamas procesas yra 
\begin{equation}
P+P \rightarrow \tau^+ +\nu_\tau +X.
\end{equation}
Susidūrus dviem protonams susidaro $ \tau $ ir $ \nu_\tau $ leptonai ir daugybė hadronų $ X $, kurie nėra analizuojami.

Šiame procese du leptonai susidaro silpnosios sąveikos metu po dviejų protone esančių kvarkų susiliejimo. Visi kiti kvarkai po sąveikos išlekia hadronais $ X $. 

Atskaitos sistemoje, kurioje protonų trimačių momentų suma lygi 0 (masės centro sistema, angl. \textit{center-of-mass frame, CM}), protonų momentų vektoriai gali būti parametrizuojami taip:
\begin{equation}
P_1^\mu=\left(\begin{array}{c} P\\0\\0\\P \end{array}\right),\quad P_2^\mu=\left(\begin{array}{c} P\\0\\0\\-P \end{array}\right).
\end{equation}
Nepaisant kvarkų rimties masių ir statmenų momento komponenčių, kvarkų momentų vektoriai yra:
\begin{equation}
p_1^\mu=x_1\left(\begin{array}{c} P\\0\\0\\P \end{array}\right),\quad p_2^\mu=x_2\left(\begin{array}{c} P\\0\\0\\-P \end{array}\right).
\end{equation}
Pagrindinio vykstančio proceso $ q+\overline{q}\rightarrow \tau^+ +\nu_\tau $ sklaidos skerspjūvis pažymimas $ \sigma(q\overline{q}\rightarrow \tau^+ \nu_\tau) $. Padauginus šį sklaidos skerspjūvį iš partono pasiskirstymo funkcijų kvarkui su momento dalimi $ x_1 $ ir antikvarkui su momento dalimi $ x_2 $ ir suintegravus pagal visas momentų dalis $ x_1 $ ir $ x_2 $, gaunamas pilnutinis protonų sklaidos skerspjūvis
\begin{align}\label{kvark}
\nonumber \sigma(PP\rightarrow \tau \nu_\tau X)=\sum_{AB}\int\int dx_1 dx_2\Big(& \sigma\left(q_A(x_1)\overline{q}_B(x_2)\rightarrow \tau^+ \nu_\tau\right) q_A(x_1)\overline{q}_B(x_2)+\\
&+\sigma\left(\overline{q}_B(x_1)q_A(x_2)\rightarrow \tau^+ \nu_\tau\right)\overline{q}_B(x_1)q_A(x_2)\Big)
\end{align} 
Antrasis narys sumoje atsiranda, nes kvarkas gali turėti momento dalį $ x_2 $, o antikvarkas - $ x_1 $.

\newpage
\section{Partonų pasiskirstymo funkcijų diferencialinės\\ lygtys}

Partono pasiskirstymo funkcijos priklauso nuo būdingo sąveikos momento $ Q^2 $ \cite{hey, peskin}. Ši išvada gaunama įskaičius aukštesnės eilės pataisas partonų modeliui. Atsižvelgus į stipriąją sąveiką pirmuoju artutinumu $ \alpha_s $ atžvilgiu, t.y. į procesus su gliuonais, partono pasiskirstymo funkcijos nebeturi \textit{Bjorken scaling} savybės. Žinoma, pačios funkcijos vistiek bus nustatomos eksperimentiškai, bet atsiranda galimybė nustatyti funkcijų priklausomybę nuo momento $ Q^2 $ ir išvesti diferencialines lygtis, kurias turi tenkinti šios funkcijos. Eksperimente suskaičiavus funkcijas viename taške $ Q^2 $, galima, naudojantis tomis diferencialinėmis lygtimis, suskaičiuoti funkcijas visuose kituose taškuose. Eksperimentiškai nustatyti funkcijas visoms $ Q^2  $ vertėms yra neįmanoma.

\begin{figure}[h]
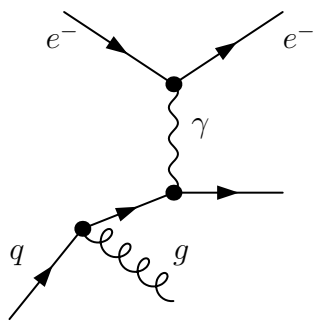

\begin{feynartspicture}(150,150)(1,1)
\FADiagram{}
\FAProp(10.,15.)(10.,9.)(0.,){/Sine}{0}
\FAProp(10.,15.)(4.,19.)(0.,){/Straight}{-1}
\FAProp(10.,15.)(16.,19.)(0.,){/Straight}{1}
\FAProp(10.,9.)(5.,7.)(0.,){/Straight}{-1}
\FAProp(5.,7.)(1.,2.)(0.,){/Straight}{-1}
\FAProp(5.,7.)(10.,3.)(0.,){/Cycles}{0}
\FAProp(10.,9.)(16.,9.)(0.,){/Straight}{1}
\FAVert(10.,15.){0}
\FAVert(10.,9.){0}
\FAVert(5.,7.){0}
\FALabel(3.,17.)[lb]{$ e^- $}
\FALabel(16.,17.)[lb]{$ e^- $}
\FALabel(11.,12.)[lb]{$ \gamma $}
\FALabel(1.,5.)[lb]{$ q $}
\FALabel(10.,5.)[lb]{$ g $}
\end{feynartspicture}
\centering
\caption{Kvarko ir elektrono sąveikos Feinmano diagrama}\label{gliuon}
\end{figure}

Taigi, šiame skyriuje bus nagrinėjami protone esančių kvarkų sąveika su gliuonais (\ref{gliuon} pav.).
Kvarkas, anksčiau laikytas realiu, dabar dėl išspinduliuoto gliuono taps virtualus, pasikeis jo momentas, todėl pasikeis ir reakcijos amplitudė, kuri priklauso nuo kvarko momento. Jei kažkuris kvarkas turėjo protono momento dalį $ x $ ir pasiskirstymo funkciją $ f(x) $, tai, išspinduliavęs gliuoną, jis turės momentą $ y<x $, todėl prie partono pasiskirstymo funkcijos $ f(y) $ reikia pridėti tikimybę, kad prieš tai buvęs kvarkas išspinduliuos gliuoną, ir šita tikimybė bus proporcinga $ f(x) $.

Kvarkui yra didelė tikimybė išspinduliuoti kolinearų gliuoną, nes, skaičiuojant tokį matricinį elementą, vardiklyje yra daugiklis $ \sin^2 \theta $ ir integravimas pagal kampą $ \theta $ yra nuo $ 0 $ iki $ 1 $, taigi, tikimybė labai išauga esant mažies emisijos kampams, o mažais kampais spinduliuojami kolinearūs mažos energijos gliuonai. Taigi, skaičiuojant tokių procesų amplitudes, galima tiesiogiai jas skaičiuoti, arba pataisas dėl gliuonų radiacijos įtraukti į partonų pasiskirstymo funkcijas. Būtent taip ir yra daroma - partono pasiskirstymo funkcijose yra susumuotos visos aukštesnės eilės amplitudės, kurios yra didelės tik dėl kolinearaus gliuono emisijos.

Reikia nagrinėti pradinį kvarką, aukščiau esančioje Feinmano diagramoje prieš sąveiką su fotonu. Šis kvarkas gali išspinduliuoti gliuoną ir Feinmano diagrama tokiam paprastam procesui yra pavaizduota \ref{2diag} paveiksle.

\begin{figure}[H]
\begin{feynartspicture}(100,100)(1,1)\FADiagram{}
\FAProp(0.,10.)(10.,10.)(0.,){/Straight}{1}
\FAProp(10.,10.)(20.,15.)(0.,){/Straight}{1}
\FAProp(10.,10.)(20.,5.)(0.,){/Cycles}{0}
\end{feynartspicture}
\centering
\caption{Kvarko skilimas}\label{2diag}
\end{figure}

Tarkime, kad pradinio kvarko momentas yra
\begin{equation}
p^\mu=\left(\begin{array}{c} p\\0\\0\\p \end{array}\right)
\end{equation}

Susidariusių dalelių momentai užrašomi taip, kad dalelės būtų beveik realios, tai yra, su mažomis statmenomis komponentėmis $ p_\perp $. Priklausomai nuo to, kuri iš susidariusių dalelių sąveikauja toliau, ta dalelė laikoma virtualia ir jos propagatorius turi būti užrašomas $ p_\perp^2 $ tikslumu, o kitos dalelės momentas gali būti užrašomas $ p_\perp $ tikslumu. Taigi, skaičiuojant spinorus ir poliarizacijos vektorius, momentai kvarkui ir gliuonui atitinkamai yra

\begin{equation}
k^\mu=\left(\begin{array}{c} (1-z)p\\-p_\perp \\0\\(1-z)p \end{array}\right),\qquad q^\mu=\left(\begin{array}{c} zp\\p_\perp \\ 0 \\zp \end{array}\right)
\end{equation}
Šie momentai tenkina realumo sąlygas $ q^2=k^2=0+\mathcal{O}(p_\perp^2) $.

Čia pateikiamas tik kvarko pasiskirstymo funkcijos skaičiavimas, todėl kvarkas laikomas virtualiu. Gliuono pasiskirstymo funkcija apskaičiuojama analogiškai. Skaičiuojant kvarko propagatorių, reikia naudoti tokius momentus

\begin{align}\label{prop}
&q^\mu=\left(\begin{array}{c} zp \\ p_\perp \\ 0 \\ zp-\frac{p_\perp^2}{2zp} \end{array}\right), &k^\mu=\left(\begin{array}{c} (1-z)p\\-p_\perp \\0\\(1-z)p+\frac{p_\perp^2}{2zp} \end{array}\right)
\end{align}
Kadangi gliuonas laikomas realiu, tai $ q^2=0+\mathcal{O}(p_\perp^4) $, tuo tarpu kvarkui $ k^2=-\frac{p_\perp^2}{z}+\mathcal{O}(p_\perp^4) $.

Matricinis elementas (\ref{2diag} pav.) diagramai yra
\begin{eqnarray}
i\mathcal{M}=\overline{u}(k)(-ig_s\gamma^\mu)u(p)\epsilon^\ast_\mu(q)
\end{eqnarray}
Spalvos daugiklis yra
\begin{equation}
\chi^\dagger_\alpha(k)t^a_{\alpha\beta}\chi_\beta(p)a^a(q)
\end{equation}
Pakeltas kvadratu ir suvidurkintas
\begin{align*}
&\frac{1}{3}\sum_{p,k,q} \left[\chi^\dagger_\alpha (k)t^a_{\alpha\beta}\chi_\beta(p)a^a(q) \right] \left[(a^b)^\dagger(q) \chi^\dagger_\gamma(p)t^b_{\gamma\delta}\chi_\delta(k) \right]
\\&=\frac{1}{3}\mathrm{Tr}\left[t^a t^a\right]=\frac{4}{3}
\end{align*}
Matricinį elementą paprasta apskaičiuoti naudojant \textit{helicity} tikrines funkcijas, nes kvarko masė laikoma lygia 0.
Bendra matricinio elemento išraiška, panaudojus spinorus
\begin{equation}
u(p)=\left(\begin{array}{c} \sqrt{p\cdot\sigma}\,\xi(p) \\ \sqrt{p\cdot\overline{\sigma}}\,\xi(p) \end{array}\right)
\end{equation}
yra (kol kas nerašant spalvos daugiklio)
\begin{align}\label{a}
\nonumber i\mathcal{M}&=-ig_s\left(\begin{array}{cc} \xi^\dagger(k)\sqrt{k\cdot\sigma} & \xi^\dagger(k)\sqrt{k\cdot\overline{\sigma}}\end{array}\right)\left(\begin{array}{cc} 0&1\\1&0 \end{array}\right)\left(\begin{array}{cc} 0&\sigma^\mu\\\overline{\sigma}^\mu &0 \end{array}\right)\left(\begin{array}{c} \sqrt{p\cdot\sigma}\,\xi(p)\\\sqrt{p\cdot\overline{\sigma}}\,\xi(p) \end{array}\right)\epsilon^\ast_\mu(q)\\
&=-ig_s\left( \xi^\dagger(k)\sqrt{k\cdot\sigma}\,\overline{\sigma}^\mu \sqrt{p\cdot\sigma}\,\xi(p) + \xi^\dagger(k)\sqrt{k\cdot\overline{\sigma}} \,\sigma^\mu\sqrt{p\cdot\overline{\sigma}}\,\xi(p)\right)\epsilon^\ast_\mu(q)
\end{align}

Norint gauti \textit{left-handed} Dirako spinorus ir atsižvelgus į pasirinktą elektrono momentą, spinorai yra tokie:
\begin{equation}
\xi_L(p)=\left( \begin{array}{c} 0\\1 \end{array}\right),\qquad \xi_L(k)=\left( \begin{array}{c} \frac{p_\perp}{2(1-z)p}\\1 \end{array}\right)
\end{equation}
Tuomet (\ref{a}) lygties antras dėmuo lygus 0 ir, paėmus konkrečius gliuono poliarizacijos vektorius
\begin{equation}
\epsilon_L^{\ast\mu}(q)=\frac{1}{\sqrt{2}}\left(\begin{array}{c} 0\\1\\i\\-\frac{p_\perp}{zp}\end{array}\right),\qquad \epsilon_R^{\ast\mu}(q)=\frac{1}{\sqrt{2}}\left(\begin{array}{c} 0\\1\\-i\\-\frac{p_\perp}{zp}\end{array}\right),
\end{equation}
matriciniai elementai skirtingoms poliarizacijoms yra
\begin{eqnarray}
i\mathcal{M}(q_L\rightarrow q_L g_L)=ig_s\frac{\sqrt{2(1-z)}}{z(1-z)}p_\perp\\
i\mathcal{M}(q_L\rightarrow q_L g_R)=ig_s\frac{\sqrt{2(1-z)}}{z}p_\perp
\end{eqnarray}
Dėl lyginumo (angl. \textit{parity}) invariantiškumo matriciniai elementai nesikeičia, jei pakeičiamos visos poliarizacijos, todėl iškart galima parašyti susumuotą ir suvidurkintą amplitudę
\begin{equation}\label{mat}
\frac{1}{2}\sum_{\mathrm{pol.}}|\mathcal{M}|^2=\frac{2g_s^2p_\perp^2}{z(1-z)}\left(\frac{1+(1-z)^2}{z}\right) \equiv \langle|\mathcal{M}|^2\rangle
\end{equation}
Tarkim, kad kvarkas, išspinduliavęs kolinearų gliuoną ir turintis momentą $ k^\mu $, toliau sąveikauja su kažkokiu taikiniu $ X $. Pažymėkime tokius dydžius:\\
$\mathcal{M}(q X\rightarrow Y)$ - suvidurkintas ir susumuotas pagal poliarizacijas matricinis elementas kvarko sklaidai nuo kažkokio taikinio $ X $, kai susidaro galinė būsena $ Y $;\\
$ \int d\Pi_Y $ - galinės būsenos $ Y $ fazinės erdvės integralas.\\
Sklaidos skerspjūvio formulė, kai kvarkas su momentu $ p^\mu $, išspinduliavęs gliuoną, sąveikauja su taikiniu $ X $, yra (naudojant atitinkamą momentą kvarko propagatoriui (\ref{prop}))
\begin{equation}
d\sigma=\frac{1}{2p2E_X(1+v_X)}\frac{d^3q}{(2\pi)^3 2q^0} \langle|\mathcal{M}|^2\rangle \left(\frac{1}{k^2}\right)^2\mathcal{M}(q X\rightarrow Y)d\Pi_Y
\end{equation}
Integralą pagal gliuono momentą galima perrašyti taip
\begin{eqnarray}
d^3q=dq^3\cdot d^2q_\perp =pdz\cdot p_\perp 2\pi dp_\perp=pdz\cdot\pi dp_\perp^2
\end{eqnarray}
Tuomet sklaidos skerspjūvį galima užrašyti
\begin{align}
\nonumber \sigma&=\int\frac{pdz\pi dp_\perp^2}{8\pi^3 2zp}\langle|\mathcal{M}|^2\rangle \left(\frac{z^2}{p_\perp^4}\right)\frac{(1-z)}{2(1-z)p2E_X(1+v_X)}\int d\Pi_Y\mathcal{M}(q X\rightarrow Y)\\
&=\int \frac{dzdp_\perp^2}{16\pi^2z}\langle|\mathcal{M}|^2\rangle \frac{(1-z)z^2}{p_\perp^4} \cdot \sigma (q X \rightarrow Y)
\end{align}
Įrašius matricinio elemento išraišką (\ref{mat})
\begin{align}
\nonumber \sigma&=\int \frac{dzdp_\perp^2}{16\pi^2}\frac{z(1-z)}{p_\perp^4}\frac{2g_s^2p_\perp^2}{z(1-z)}\left[\frac{1+(1-z)^2}{z}\right]\cdot \sigma (q X \rightarrow Y)\\
&=\int_0^1 dz\int \frac{dp_\perp^2}{p_\perp^2}\frac{\alpha_s}{2\pi}\left[\frac{1+(1-z)^2}{z}\right] \cdot \sigma (q X \rightarrow Y)
\end{align}
Integralas pagal statmeną komponentę turėtų sumuoti tik mažas vertes, nes buvo laikoma, kad ji maža, tačiau galima integruoti nuo $ m^2 $ iki $ s $, nes kitaip neaišku, kokia turėtų būti viršutinė riba, ir vistiek didelių statmenų komponenčių indėlis yra mažas dėl to, kad vienas daugiklis yra vardiklyje.
\begin{equation}\label{c}
\sigma(qX\rightarrow gY)=\int_0^1 dz\frac{\alpha_s}{2\pi}\,\mathrm{ln}\!\left(\frac{s}{m^2}\right)\left[\frac{1+(1-z)^2}{z}\right] \cdot \sigma (q X \rightarrow Y)
\end{equation}
Šią formulę galima interpretuoti taip: prieš sklaidos skerspjūvį dešinėje lygybės pusėje esantis daugiklis reiškia tikimybę, kad pradiniame kvarke galima rasti kvarką su momentu $ (1-z)p $. Tos tikimybės išraiška:
\begin{equation}
f_q^{(1)} (z)=\frac{\alpha_s}{2\pi}\,\mathrm{ln}\!\left(\frac{s}{m^2}\right)\left[\frac{1+(1-z)^2}{z}\right] 	
\end{equation}

\begin{figure}[H]
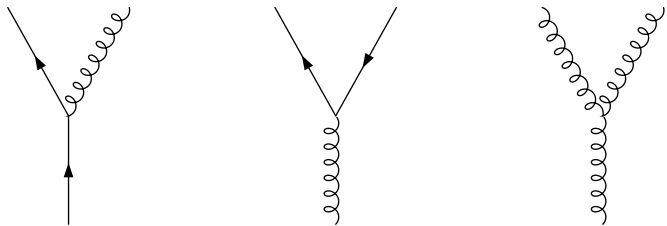

\begin{feynartspicture}(400,100)(3,1)
\FADiagram{}
\FAProp(10.,1.)(10.,10.)(0.,){/Straight}{1}
\FAProp(10.,10.)(5.,19.)(0.,){/Straight}{1}
\FAProp(10.,10.)(15.,19.)(0.,){/Cycles}{0}
\FADiagram{}
\FAProp(10.,1.)(10.,10.)(0.,){/Cycles}{0}
\FAProp(10.,10.)(5.,19.)(0.,){/Straight}{1}
\FAProp(10.,10.)(15.,19.)(0.,){/Straight}{-1}
\FADiagram{}
\FAProp(10.,1.)(10.,10.)(0.,){/Cycles}{0}
\FAProp(10.,10.)(5.,19.)(0.,){/Cycles}{0}
\FAProp(10.,10.)(15.,19.)(0.,){/Cycles}{0}
\end{feynartspicture}
\caption{Partonų skilimo diagramos}\label{3diag}
\end{figure}

Formulėje (\ref{c}) neįskaityta tikimybė, kad kvarkas neišspinduliuoja nieko ir sąveikauja pilnos energijos kvarkas. Jeigu toje formulėje pakeisime kintamąjį $ x=1-z $ ir $ x $ reikš kvarko momento dalį, tai formulė atrodys taip
\begin{equation}
\sigma(qX\rightarrow gY)=\int_0^1 dx\frac{\alpha_s}{2\pi}\,\mathrm{ln}\!\left(\frac{s}{m^2}\right)\left[\frac{1+x^2}{1-x}\right] \cdot \sigma(qX \rightarrow Y)
\end{equation}
Jei norėtume suskaičiuoti, kaip sąveikauja pilnos energijos kvarkas su $ x=1 $,  integralas diverguotų. Reikia kažkaip įskaityti tą pilnos energijos kvarką ir izoliuoti divergenciją. Paprasčiausia būtų užrašyti tikimybės funkciją su Dirako delta funkcija
\begin{equation}
f^{(0)}_q(x)=\delta(1-x)
\end{equation}
Tuomet tikimybės funkcija yra
\begin{equation}
f_q(x)=f^{(0)}_q(x)+f^{(1)}_q(x)=\delta(1-x)+\frac{\alpha_s}{2\pi}\, \mathrm{ln}\!\left(\frac{s}{m^2}\right)\left[\frac{1+x^2}{1-x}\right] 
\end{equation}
Tačiau ir ji dar netinka pilnam kvarko aprašymui, nes, pirma, antroje funkcijoje yra singuliarumas, kai $ x=1 $, ir, antra, funkcija nėra tinkamai normuota, nes $ \delta $ funkcija normuota į 1 ir įskaito pilnos energijos kvarką su momento dalimi $ x=1 $, bet reikia atsižvelgti į tai, kad kvarkas, turėdamas momentą $ x=1 $ gali išspinduliuoti gluoną ir tuomet jis nebepriklausys šiai funkcijai - reikia iš $ \delta $ funkcijos atimti tą dalį kvarkų. Divergencija, kai $ x \rightarrow 1 $, atsiranda dėl kolinearių gliuonų emisijos, tačiau yra žinoma, kad tokio tipo divergenciją panaikina virtualių gliuonų diagramos, todėl tikimybės funkcija gali būti užrašyta
\begin{equation}
f_q(x)=\delta(1-x)+\frac{\alpha_s}{2\pi}\, \mathrm{ln}\!\left(\frac{s} {m^2}\right)\left[\frac{1+x^2}{1-x}-A\delta(1-x)\right] 
\end{equation}
ir reikia surasti konstantą A, kad funkcija būtų tinkamai normuota. Konstanta A iš tikrųjų ir atitinka virtualių gliuonų diagramas, kurios čia neskaičiuojamos. Pilna tikimybės funkcija turi tenkinti ($ \alpha_s $ eilės tikslumu) normavimo sąlygą
\begin{equation}
\int_0^1 dx f_q(x)=1
\end{equation}
Kadangi integralas vistiek diverguoja, tai yra įprasta apibrėžti diverguojančią pasiskirstymo dalį taip:
\begin{equation}
\frac{1}{(1-x)_+}
\end{equation}
yra lygi funkcijai $ 1/(1-x) $ visuose taškuose, išskyrus $ x=1 $, ir tolydžios funkcijos $ g(x) $ integralas su šiuo pasiskirstymu yra 
\begin{equation}
\int_0^1 dx \frac{g(x)}{(1-x)_+}=\int_0^1 dx \frac{g(x)-g(1)}{(1-x)}
\end{equation}
Pasinaudojus šiuo apibrėžimu, galima surasti konstantą A
\begin{equation}
\int_0^1 dx \frac{1+x^2}{(1-x)_+}=\int_0^1 dx \frac{x^2-1}{(1-x)}=-\frac{3}{2}
\end{equation}
Galiausiai kvarko tikimybės pasiskirstymas yra
\begin{equation}
f_q(x)=\delta(1-x)+\frac{\alpha}{2\pi}\, \mathrm{ln}\!\left(\frac{s} {m^2}\right)\left[\frac{1+x^2}{(1-x)_+}+\frac{3}{2}\delta(1-x)\right] 
\end{equation}
Ši funkcija yra tinkamai normalizuota, bet vistiek singuliari šalia taško $ x=1 $, todėl galima tikėtis, kad aukštesnės eilės pataisos bus svarbios šioje srityje - reikia įskaityti kelių kolinearių gliuonų emisiją. Tarkime, kad kvarkas išspinduliuoja du gliuonus\\
\begin{center}
\begin{feynartspicture}(100,100)(1,1)\FADiagram{}
\FALabel(8.,17.5)[br]{1}
\FALabel(1.5,16.)[br]{2}
\FAProp(11.5,11.5)(19.,12.)(0.,){/Straight}{1}
\FAProp(5.,7.)(3.,16.5)(0.,){/Cycles}{0}
\FAProp(11.5,11.5)(10.,19.)(0.,){/Cycles}{0}
\FAProp(1.,2.)(5.,7.)(0.,){/Straight}{1}
\FAProp(5.,7.)(11.5,11.5)(0.,){/Straight}{1}
\end{feynartspicture}
\end{center}
kur gliuonas 1  yra išspinduliuojamas su statmena momento dedamąja $ p_{1\perp} $, o gliuonas 2 - su momento dedamąja $ p_{2\perp} $. Gliuono 2 emisija gali būti apskaičiuojama lygiai, kaip ir aukščiau, o tuomet, jei $ p_{2\perp}\ll p_{1\perp} $, tai pirmasis virtualus kvarkas gali būti laikomas realiu ir gliuono 1 emisijos matricinis elementas bus toks pat, kaip ir gliuono 2, todėl dviejų gliuonų emisija duos 
\begin{equation}
\left(\frac{\alpha}{2\pi}\right)^2 \int_{m^2}^s \frac{dp_{1\perp}^2}{p_{1\perp}^2} \int_{m^2}^{p_{1\perp}^2} \frac{dp_{2\perp}^2}{p_{2\perp}^2}=\frac{1}{2}\left(\frac{\alpha}{2\pi}\right)^2\, \mathrm{ln}^2\!\left(\frac{s}{m^2}\right)
\end{equation}
Visi kolinearūs gliuonai gali duoti svarų indėlį, jei tenkina sąlygą $ p_{1\perp} \gg p_{2\perp} \gg p_{3\perp}\gg\ldots$ ir tai duos daugiklį
\begin{equation}
\frac{1}{n!}\left(\frac{\alpha}{2\pi}\right)^n\, \mathrm{ln}^n\!\left(\frac{s}{m^2}\right)
\end{equation}
Kadangi tarpiniai kvarkai yra vis labiau virtualūs, galima juos interpretuoti kaip originalaus kvarko sudėtines dalis, kai jo struktūra yra analizuojama vis mažesnėje atstumo skalėje. Kvarkas su momentu $ k^2 \sim p^2_\perp $ gali būti laikomas kvarko sudėtine dalimi ir būti matomas, kai kvarkas yra stebimas $ \Delta r \sim p_\perp^{-1} $ skyra. Tai reiškia, kad, priklausomai nuo skalės, kvarkas gali būti matomas kaip virtualių kvarkų ir gliuonų visuma. Galima įsivaizduoti, kad kvarko ir gliuono skilimas yra nuolatinis procesas, priklausantis nuo statmenos kvarko momento komponentės. Norint tai aprašyti matematiškai, reikia į pasiskirstymo funkcijas įtraukti priklausomybę nuo statmenos komponentės, todėl įvedame naujas funkcijas $ f_g(x,Q) $ ir $ f_q(x,Q) $, kurios reiškia tikimybę rasti gliuoną ar kvarką su momento dalimi $ x $, atsižvelgiant į kolinearių gliuonų emisiją su statmena momento komponente $ p_\perp < Q $. Jei $ Q $ truputį padidinsime, reiks taip pat atsižvelgti į galimą gliuonų emisiją su statmenu momentu $ Q < p_\perp < Q+\Delta Q$. Diferencialinė tikimybė, kad kvarkas išspinduliuos gliuoną (dabar jau įskaitant spalvos daugiklį), yra
\begin{equation}
\frac{4}{3}\frac{\alpha_s}{2\pi}\frac{dp_\perp^2}{p_\perp^2}\frac{1+(1-z)^2}{z}
\end{equation}
Tai įskaičius, nauja gliuono tikimybės funkcija išreiškiama taip
\begin{align}
\nonumber f_g(x,Q+\Delta& Q)\\
\nonumber =&f_g(x,Q)+\int_0^1 dx'\int_0^1 dz\left[\frac{4}{3}\frac{\alpha_s}{2\pi}\frac{\Delta Q^2}{Q^2}\frac{1+(1-z)^2}{z}\right]f_q(x',p_\perp)\delta(x-zx')\\
=&f_g(x,Q)+\frac{\Delta Q}{Q}\int_x^1 \frac{dz}{z} \left[\frac{4}{3}\frac{\alpha_s}{\pi}\frac{1+(1-z)^2}{z}\right]f_q(\frac{x}{z},p_\perp)
\end{align}
Šioje lygtyje galima pereiti prie ribos $ \Delta Q \rightarrow 0 $ ir gauti diferencialinę lygtį, kurią tenkina pasiskirstymo funkcija:
\begin{equation}
\frac{d}{d\, \mathrm{ln}Q}f_g(x,Q)=\frac{\alpha_s}{\pi}\int_x^1 \frac{dz}{z} \left[\frac{4}{3}\frac{1+(1-z)^2}{z}\right]f_q(\frac{x}{z},p_\perp)
\end{equation}
Ši lygtis vadinama partono evoliucijos lygtimi, o funkcija 
\begin{equation}
\frac{4}{3}\frac{1+(1-z)^2}{z}\equiv P_{g \leftarrow q}(z)
\end{equation}
vadinama skilimo funkcija (angl. \textit{spliting function}). Ji parodo tikimybę kvarkui išspinduliuoti gliuoną.\\
Atsižvelgus į tai, kad, turint gliuoną arba kvarką, jie gali skilti, reikia įskaityti visus galimus skilimus, užrašant evoliucijos lygtis. Feinmano diagramos šiems skilimams yra pavaizduotos \ref{3diag} paveiksle.\\
Tuomet pilnos evoliucijos funkcijos atrodo taip:
\begin{align}
\nonumber &\frac{d}{d\, \mathrm{ln}Q}f_g(x,Q)=\frac{\alpha_s(Q^2)}{\pi}\int_x^1 \frac{dz}{z}\left\lbrace P_{g\leftarrow q}(z)\sum_f \left[ f_f(\frac{x}{z},Q)+f_{\overline{f}}(\frac{x}{z},Q)\right] +P_{g\leftarrow g}(z)f_g (\frac{x}{z},Q)\right\rbrace,\\
\nonumber &\frac{d}{d\, \mathrm{ln}Q}f_f(x,Q)=\frac{\alpha_s(Q^2)}{\pi}\int_x^1 \frac{dz}{z}\left\lbrace P_{q\leftarrow q}(z)f_f(\frac{x}{z},Q)+P_{q\leftarrow g}(z)f_g (\frac{x}{z},Q)\right\rbrace,\\
&\frac{d}{d\, \mathrm{ln}Q}f_{\overline{f}}(x,Q)=\frac{\alpha_s(Q^2)}{\pi}\int_x^1 \frac{dz}{z}\left\lbrace P_{q\leftarrow q}(z)f_{\overline{f}}(\frac{x}{z},Q)+P_{q\leftarrow g}(z)f_g (\frac{x}{z},Q)\right\rbrace
\end{align}
Jos dar vadinamos \textit{Altarelli-Parisi} lygtimis. 

Skilimo funkcijos yra
\begin{align}
\nonumber &P_{q \leftarrow q}(z)=\frac{4}{3}\left[\frac{1+z^2}{(1-z)_+}+\frac{3}{2}\delta(1-z)\right],\\
\nonumber &P_{g \leftarrow q}(z)=\frac{4}{3}\left[\frac{1+(1-z)^2}{z}\right],\\
\nonumber &P_{q \leftarrow g}(z)=\frac{1}{2}\left[z^2+(1-z)^2\right],\\
&P_{g \leftarrow g}(z)=6\left[\frac{(1-z)}{z}+\frac{z}{(1-z)_+}+z(1-z)+\left(\frac{11}{12}-\frac{n_f}{18}\right)\delta(1-z)\right]
\end{align}

Partonų pasiskirstymo funkcijos įtraukia visus galimus singuliarumus vietoj to, kad jie būtų skaičiuojami atskirai skaičiuojant matricinius elementus. 

\newpage
\section{MSTW grupės partono pasiskirstymo funkcijos}\label{pdfskyrius}

Šiame darbe skačiavimams naudojamos MSTW grupės partono pasiskirstymo funkcijos \cite{mstw}. Jos pasirinktos dėl laisvo ir paprasto priėjimo ir $C\!+\!+$ kodo pateikimo. 
Pasižiūrėjus į konkrečias funkcijų vertes įvairiems kvarkų skoniams galima nuspręsti, kuriuos kvarkus reikia įtraukti į skaičiavimus, o kurie yra nesvarbūs.  Esant labai didelės energijos protonams (Laboratorinėje atskaitos sistemoje, kuri yra ir protonų masės centro sistema, vieno protono energija siekia $ P=7$ TeV), svarbūs gali būti ir sunkesni kvarkai, kurie protone atsiranda dėl fliuktuacijų. 

\begin{figure}[H]
		\begin{subfigure}[b]{0.32\textwidth}
			\centering
			\includegraphics[width=\textwidth]{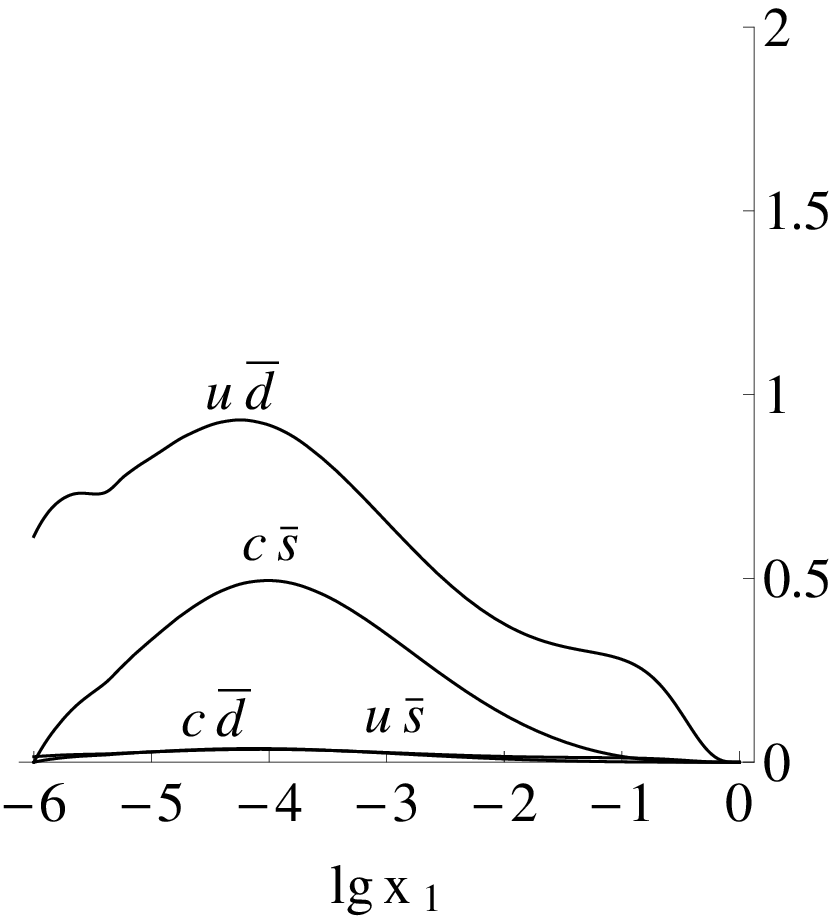}
		\end{subfigure}
		\begin{subfigure}[b]{0.32\textwidth}
			\centering
			\includegraphics[width=\textwidth]{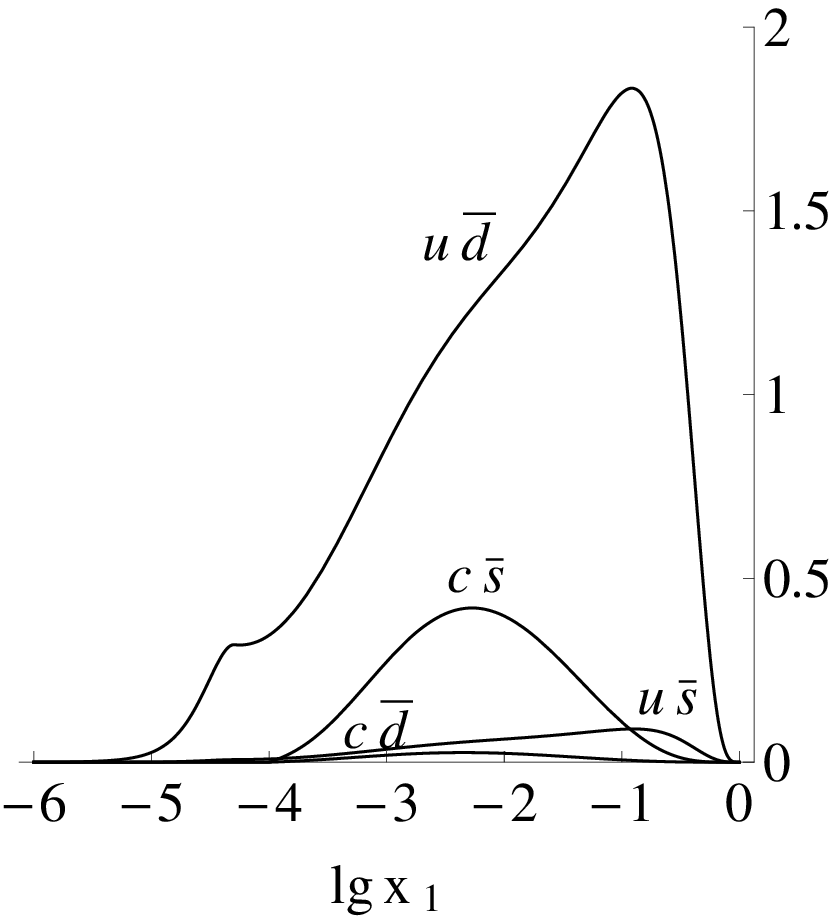}
		\end{subfigure}
		\begin{subfigure}[b]{0.32\textwidth}
			\centering
			\includegraphics[width=\textwidth]{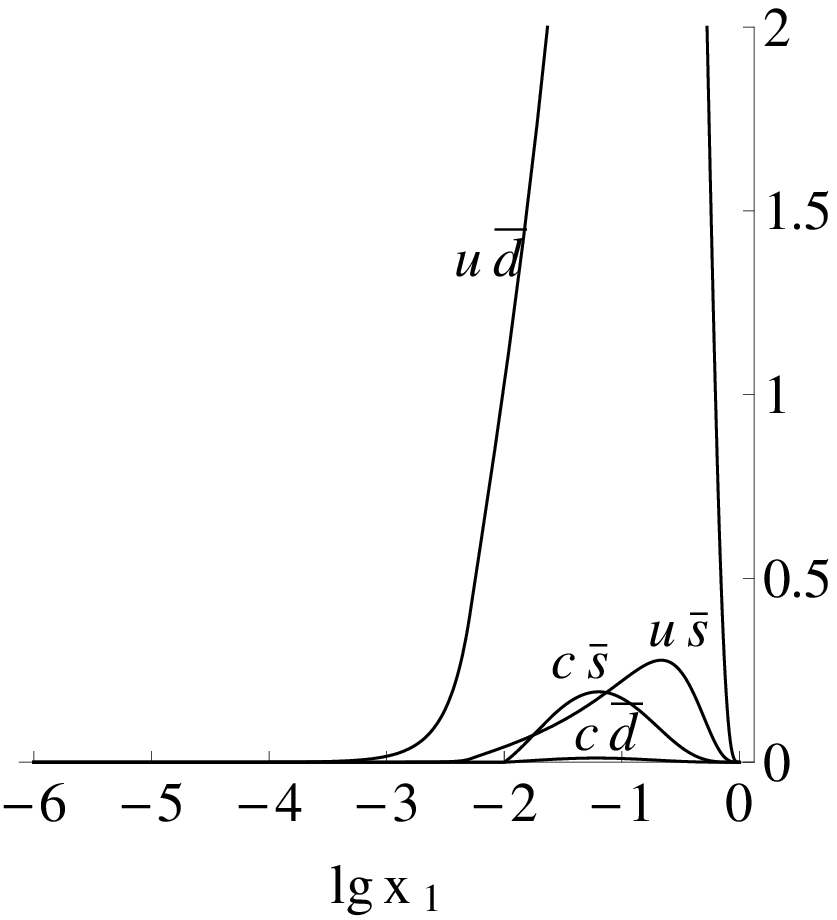}
		\end{subfigure}
\caption{Partono pasiskirstymo funkcijos (\ref{pdfW}), padaugintos iš $ x_1x_2 $. $ \mathrm{lg} x_2=-2 $ (kairėje), $ \mathrm{lg} x_2=-4 $ (viduryje), $ \mathrm{lg} x_2=-6 $ (dešinėje).}\label{pdfWpav}
\end{figure}

Skaičiuojant $ W^+ $ kanalą, galimos kvarkų poros yra $ u\overline{d},\,u\overline{s},\,c\overline{s},\,c\overline{d} $. CKM matricos komponenčių vertės yra \cite{data}
\begin{equation}
|V|=\left(\begin{array}{ccc} |V^{ud}|&|V^{us}|&|V^{ub}| \\ |V^{cd}|&|V^{cs}|&|V^{cb}| \\|V^{td}|&|V^{ts}|&|V^{tb}| \end{array}\right)=\left(\begin{array}{ccc} 0.974 &0.225 &0.003 \\ 0.225 &0.973 &0.041 \\0.009& 0.040 & 0.999 \end{array}\right)
\end{equation}
Kvarkų lygio diferencialiniai sklaidos skerspjūviai, kaip ir (\ref{kvark}) lygtyje, bus dauginami iš partono pasiskirstymo funkcijos ir atitinkamo CKM matricinio elemento, pakelto kvadratu. PDF kvarkui, priklausanti nuo protono momento dalies $ x_i $, kur $ i=1,2 $, ir energijos skalės $ Q $, žymima pagal kvarko skonio pavadinimą, pavyzdžiui, $ u $ kvarkui $ u(x_i,Q) $. Taigi, bus tokie daugikliai
\begin{align}\label{pdfW}
\nonumber 	&u\left(x_1,Q\right)\,\overline{d}\left(x_2,Q\right)|V^{ud}|^2\\
\nonumber 	&u\left(x_1,Q\right)\,\overline{s}\left(x_2,Q\right)|V^{us}|^2\\
\nonumber	&c\left(x_1,Q\right)\,\overline{d}\left(x_2,Q\right)|V^{cd}|^2\\
			&c\left(x_1,Q\right)\,\overline{s}\left(x_2,Q\right)|V^{cs}|^2
\end{align} 
ir čia $ Q=2P\sqrt{x_1 x_2} $. \ref{pdfWpav} paveiksluose parodytos šios funkcijos. Kadangi yra du kintamieji $ x_1 $ ir $ x_2 $, tai abscisių ašyje atidėtas kintamasis $ x_1 $, o kintamasis $ x_2 $ parinktas trijų skirtingų verčių.
Visuomet dominuoja $ u\overline{d} $ kvarkų pora - ji yra svarbiausia. Kvarkų $ u\overline{s} $ pora, palyginus su $ c\overline{s} $ kvarkais, duoda mažesnį indėlį į $ \tau $ dalelės susidarymą, tačiau šiame darbe vistiek ši pora yra įskaitoma, o $ c\overline{s} $ pora - ne, dėl to, kad $ c $ kvarko masė yra tos pačios eilės, kaip ir $ \tau $ masė ir ją reikėtų įskaityti, bet čia analizuojama situacija, kai kvarkai neturi masės.

\begin{figure}[H]
		\begin{subfigure}[b]{0.32\textwidth}
			\centering
			\includegraphics[width=\textwidth]{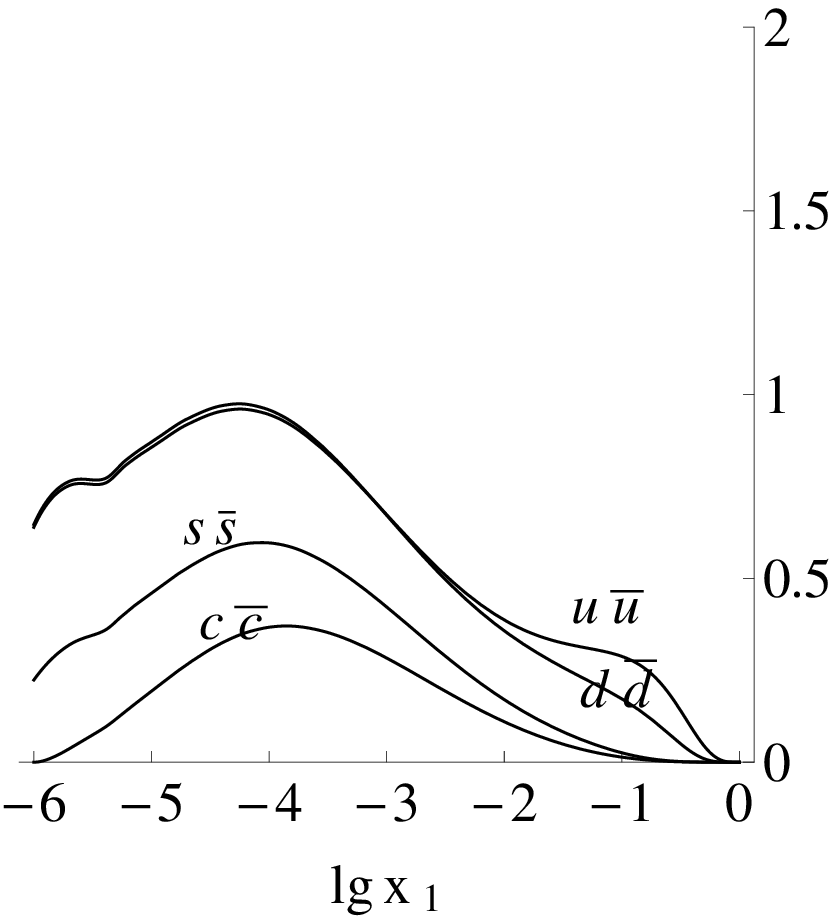}
		\end{subfigure}
		\begin{subfigure}[b]{0.32\textwidth}
			\centering
			\includegraphics[width=\textwidth]{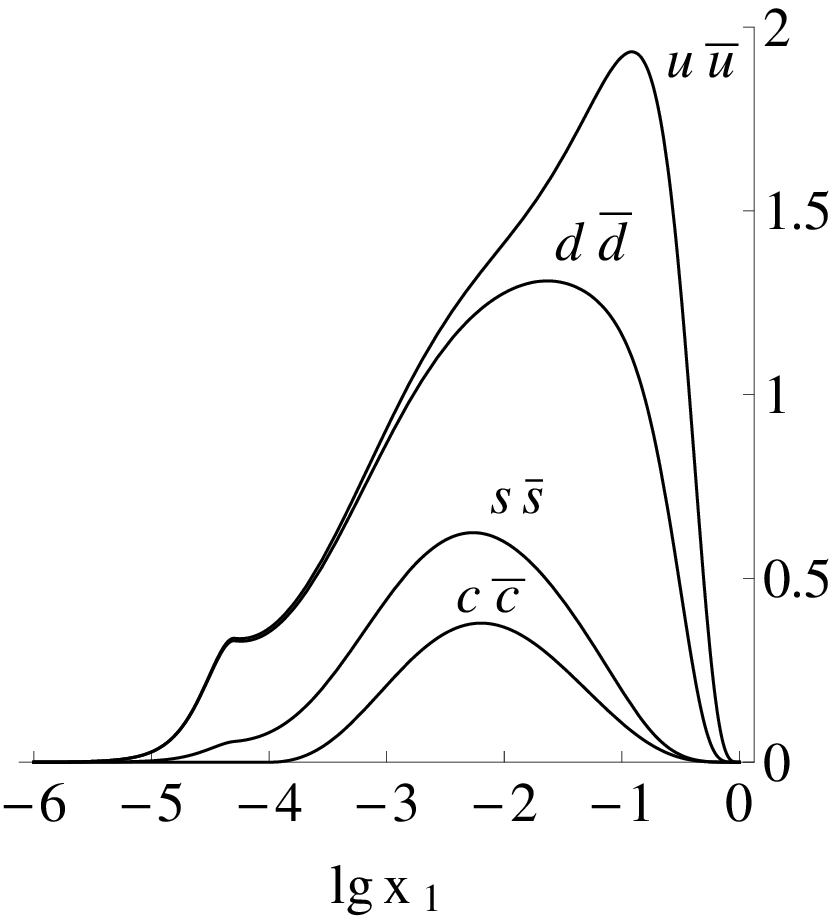}
		\end{subfigure}
		\begin{subfigure}[b]{0.32\textwidth}
			\centering
			\includegraphics[width=\textwidth]{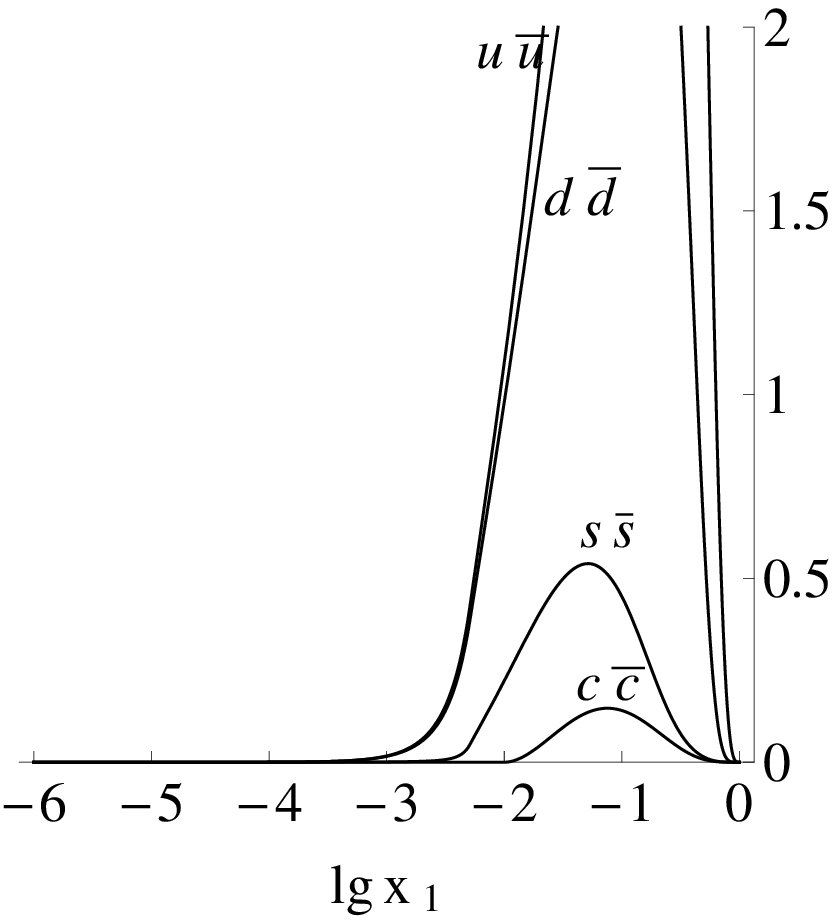}
		\end{subfigure}
\caption{Partono pasiskirstymo funkcijos (\ref{pdfZ}), padaugintos iš $ x_1x_2 $. $ \mathrm{lg} x_2=-2 $ (kairėje), $ \mathrm{lg} x_2=-4 $ (viduryje), $ \mathrm{lg} x_2=-6 $ (dešinėje). }\label{pdfZpav}
\end{figure}

Analogiškai, skaičiuojant $ \tau^+ \tau^- $ leptonų poros susidarymą dėl $ Z^0 $ ir $ A^\gamma $ bozonų, galimos kvarkų poros yra $  u\overline{u},\,d\overline{d},\,s\overline{s},\,c\overline{c} $. Matriciniai elementai bus dauginami iš tokių funkcijų: 
\begin{align}\label{pdfZ}
\nonumber 	&u\left(x_1,Q\right)\,\overline{u}\left(x_2,Q\right)\\
\nonumber 	&d\left(x_1,Q\right)\,\overline{d}\left(x_2,Q\right)\\
\nonumber	&s\left(x_1,Q\right)\,\overline{s}\left(x_2,Q\right)\\
			&c\left(x_1,Q\right)\,\overline{c}\left(x_2,Q\right)
\end{align} 
Šios funkcijos pavaizduotos \ref{pdfZpav} paveiksluose. Matoma, kad dominuoja $ u $ ir $ d $ kvarkai, o $ s $ kvarko indėlis šiuo atveju didesnis už $ c $ kvarko.

\newpage
\section{Leptono poliarizacijos vektorius}

Matriciniame elemente prieš nagrinėjamos dalelės spinorą įrašius projekcijos operatorių $ P_n=(1-\slashed{n}\gamma^5)/2 $ ir susumavus pagal dalelės poliarizacijas, gaunamas dydis, kuris reiškia vidutinės leptono poliarizacijos vektoriaus projekciją į $ n^\mu $ ašį \cite{vector}. Dydis $ P_n v^s(k) $, kur $ v^s(k) $ yra Dirako spinoras, yra spinoro projekcija į $ n^\mu $ ašį. $ v^s(k) $ išraiška nėra svarbi, nes šios funkcijos sudaro pilną funkcijų sistemą, ir susumavus pagal spinoro poliarizacijas $ s $ gaunamas pilnumo sąryšis, kuris nepriklauso nuo spinoro $ v^s(k) $ išraiškos. Gautą matricinį elementą padalinus iš įprasto elemento, kuris skaičiuotas be projekcijos operatoriaus, gaunama normuota vidutinės leptono poliarizacijos vektoriaus projekcija į $ n^\mu $ ašį. 

Projekcijos operatorius yra Lorenco invariantas, todėl vektorius $ n^\mu $ turi būti apibrėžtas vienareikšmiškai ir kovariantiškai visose atskaitos sistemose t.y., turi tenkinti tam tikras sąlygas. Visų pirma, leptono rimties atskaitos sistemoje (angl. \textit{rest-frame}, RF) vektorius $ n^\mu $ gali būti parenkamas turįs tik erdvines koordinates. Leptono, kurio masė $ m $, momento vektorius $ k^\mu $ turi tik laikinę koordinate, taigi, jų dviejų Lorenco sandauga lygi 0 šioje ir visose kitose atskaitos sistemose. Taip pat vektorius parenkamas vienetinio ilgio:
\begin{align}
(n\cdot k)=&0\\
n^2=&-1\\
k^2=&m^2
\end{align}
Rimties atskaitos sistemoje vektoriaus kryptis neapibrėžta, bet laboratorinėje atskaitos sistemoje (angl. \textit{lab-frame}, LAB) galima reikalauti, kad vektoriaus $ n^\mu $ erdvinės koordinatės sutaptų su $ \tau $ dalelės momento vektoriaus $ k^\mu $ erdvinėmis koordinatėmis - trimačiai vektoriai lygiagretūs. Laikinė vektoriaus $ n^\mu $ komponentė apibrėžiama per kitus sistemos vektorius: kvarkų arba protonų momentų vektorius. Tuomet LAB sistemoje kampu $ \theta $ susidariusios dalelės kryptis taip pat bus ir poliarizacijos vektoriaus projekcijos kryptis.

Kadangi reikalaujama, kad vektoriaus $ n^\mu $ erdvinės koordinatės būtų lygiagrečios vektoriaus $ k^\mu $ erdvinėms koordinatėms, tai jam vienareikšmiškai apibrėžti reikalingas dar vienas vektorius, turintis tik laikinę komponentę. Jis gali būti toks: Protonų momentai LAB sistemoje yra
\begin{equation}
P_1^\mu=\left(\begin{array}{c} P\\0\\0\\P \end{array}\right),\qquad P_2^\mu=\left(\begin{array}{c} P\\0\\0\\-P \end{array}\right)
\end{equation}
ir galima apibrėžti vektorių
\begin{equation}
T^\mu=\frac{1}{2P}(P_1^\mu + P_2^\mu)=\left(\begin{array}{c} 1\\0\\0\\0 \end{array}\right),
\end{equation}
kuris turi tik laikinę komponentę ir gali būti panaudotas sistemos laiko matavimui. Tuomet vektorius $ n^\mu $ išreiškiamas dviem vektoriais
\begin{equation}
n^\mu=\alpha T^\mu + \beta k^\mu
\end{equation}
Iš aukščiau užrašytų sąlygų randami koeficientai
\begin{equation}
n^\mu=\frac{1}{\sqrt{(k\cdot T)^2-m^2}}\left(-m T^\mu + \frac{(k\cdot T)}{m}k^\mu\right)
\end{equation}
arba
\begin{equation}\label{vektorius}
n^\mu=\frac{1}{|\vec{k}|}\left(-m T^\mu + \frac{E}{m}k^\mu\right)
\end{equation}
kur $ E $ ir $ |\vec{k}| $ yra nagrinėjamos dalelės energija ir impulsas LAB sistemoje. Kai dalelė nejuda, $ |\vec{k}|=0 $, vektorius yra neapibrėžtas.

\newpage
\section{$ \tau^+ $ leptono susidarymas}

\begin{figure}[H]
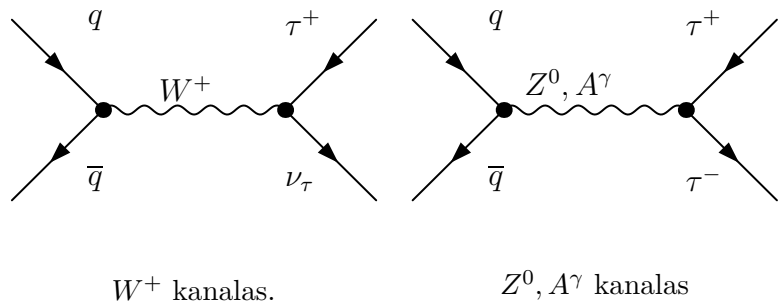

\centering
\begin{feynartspicture}(300,150)(2,1)
\FADiagram{$ W^+ $ kanalas.}
\FAProp(0.,15.)(5.,10.)(0.,){/Straight}{1}
\FALabel(5.,15.5)[tr]{$q$}
\FAProp(0.,5.)(5.,10.)(0.,){/Straight}{-1}
\FALabel(15.,15.5)[tl]{$\tau^+ $}
\FAProp(5.,10.)(15.,10.)(0.,){/Sine}{0}
\FALabel(11.,10.5)[br]{$W^+$}
\FAProp(15.,10.)(20.,15.)(0.,){/Straight}{-1}
\FALabel(5.,5.5)[br]{$\overline{q}$}
\FAProp(15.,10.)(20.,5.)(0.,){/Straight}{1}
\FALabel(15.,5.5)[bl]{$\nu_\tau $}
\FAVert(5.,10.){0}
\FAVert(15.,10.){0}
\FADiagram{$ Z^0, A^\gamma $ kanalas}
\FAProp(0.,15.)(5.,10.)(0.,){/Straight}{1}
\FALabel(5.,15.5)[tr]{$q$}
\FAProp(0.,5.)(5.,10.)(0.,){/Straight}{-1}
\FALabel(15.,15.5)[tl]{$\tau^+$}
\FAProp(5.,10.)(15.,10.)(0.,){/Sine}{0}
\FALabel(11.,10.5)[br]{$Z^0, A^\gamma $}
\FAProp(15.,10.)(20.,15.)(0.,){/Straight}{-1}
\FALabel(5.,5.5)[br]{$\overline{q}$}
\FAProp(15.,10.)(20.,5.)(0.,){/Straight}{1}
\FALabel(15.,5.5)[bl]{$\tau^- $}
\FAVert(5.,10.){0}
\FAVert(15.,10.){0}
\end{feynartspicture}
\caption{$ \tau^+ $ leptono susidarymas LHC}\label{diagramos}
\end{figure}

Pagrindinis šio darbo uždavinys yra suskaičiuoti $ \tau^+ $ leptono, susidariusio iš $ W^+ $ bozono, poliarizaciją.  Didžiajame hadronų greitintuve  LHC sudurinėjami protonai, todėl $ \tau^+ $ leptonas gali susidaryti susiliejant kvarko ir antikvarko poroms: $ q\overline{q}\rightarrow W^+ \rightarrow \tau^+ \nu_{\tau} $ ir $ q\overline{q}\rightarrow Z^0,A^\gamma \rightarrow \tau^+ \tau^- $. Čia $ A^\gamma $ reiškia fotoną. Feinmano diagramos šiems procesams pavaizduotos \ref{diagramos} paveiksle.

Kaip buvo apžvelgta \ref{pdfskyrius} skyriuje, teigiamo krūvio $ \tau^+$ dalelė susidaro susiliejus $ u\overline{d},\, u\overline{s} $ kvarkams, kai sąveikos nešėjas yra $ W^+ $ bozonas, ir susiliejus $ u\overline{u},\, d\overline{d},\,s\overline{s} $ kvarkams, kai sąveikos nešėjai yra $ Z^0 $ ir $ A^\gamma $ bozonai. Matricinių elementų skaičiavimai yra pateikti priede \ref{priedas1}. Jie priklauso nuo sąveikos nešėjų ir yra labai skirtingi. Vienintelis fermionas, kurio masė įtraukiama į skaičiavimus, yra $ \tau^+ $ leptonas. Jo masė ($ 1.78$ GeV \cite{data}) yra daug didesnė už sunkiausio $ s $ kvarko masę ($ 95 $ MeV).

Diferencialinis sklaidos skerspjūvis $ \tau^+ $ dalelei išsisklaidyti LAB sistemoje kampais $ 0<\theta<\frac{\pi}{2} $ ir to skerspjūvio skaičiavimas pateiktas \ref{priedas2} ir \ref{priedas3} prieduose. Dėl to, kad susidūrimai yra simetriniai, sklaidos skerspjūvis intervale $ \frac{\pi}{2}<\theta<\pi $ yra visiškai toks pat. Sklaidos skerspjūvis, skaičiuotas su projekcijos operatoriumi, žymimas
\begin{equation}
\left(\frac{d\sigma_{AB}^n}{d\cos\!\theta}\right)
\end{equation}
(\ref{wdsigma}) ir (\ref{zdsigma}) formulės. $ AB $ reiškia kvarkų skonius, iš kurių susidarė leptonas, o $ n $ reiškia, kad dydis skaičiuotas su projekcijos operatoriumi. Vektorius $ n^\mu $ projekcijos operatoriuje pasirenkamas (\ref{vektorius}). Analogiškai, įprastas diferencialinis sklaidos skerspjūvis žymimas
\begin{equation}
\left(\frac{d\sigma_{AB}}{d\cos\!\theta}\right)
\end{equation}
be indekso $ n $.

Gautą diferencialinį sklaidos skerspjūvį padalinus iš įprasto diferencialinio sklaidos skerspjūvio (t.y. be projekcijos operatoriaus), gaunama normuota leptono vidutinės poliarizacijos projekcija į ašį $ n^\mu $, kurios erdvinė kryptis sutampa su pačios dalelės kryptimi. Tačiau eksperimentiškai neįmanoma nustatyti, kuriuo kanalu susidarė leptonas. Todėl reikia sudėti visų galimų procesų diferencialinius sklaidos skerspjūvius ir padalinti iš susumuotų įprastų diferencialinių sklaidos skerspjūvių:
\begin{align}
A^n(\cos\!\theta)=
	\frac{\displaystyle
	\left(\frac{d\sigma_{u\overline{d}}^n}{d\cos\!\theta}\right)+
	\left(\frac{d\sigma_{u\overline{s}}^n}{d\cos\!\theta}\right)+
	\left(\frac{d\sigma_{u\overline{u}}^n}{d\cos\!\theta}\right)+
	\left(\frac{d\sigma_{d\overline{d}}^n}{d\cos\!\theta}\right)+
	\left(\frac{d\sigma_{s\overline{s}}^n}{d\cos\!\theta}\right)}
	{\displaystyle
	\left(\frac{d\sigma_{u\overline{d}}}{d\cos\!\theta}\right)+
	\left(\frac{d\sigma_{u\overline{s}}}{d\cos\!\theta}\right)+
	\left(\frac{d\sigma_{u\overline{u}}}{d\cos\!\theta}\right)+
	\left(\frac{d\sigma_{d\overline{d}}}{d\cos\!\theta}\right)+
	\left(\frac{d\sigma_{s\overline{s}}}{d\cos\!\theta}\right)}
\end{align}
Šis santykis reiškia $ \tau^+ $ leptono, susidariusio Didžiajame hadronų greitintuve, vidutinės poliarizacijos vektoriaus projekciją į leptono judėjimo ašį.
 Taip pat teoriškai galima pasakyti kiekvieno proceso indėlį į bendrą normuotą projekciją, pavyzdžiui:
\begin{align}
A^n_{u\overline{d}}(\cos\!\theta)=
	\frac{\displaystyle
	\left(\frac{d\sigma_{u\overline{d}}^n}{d\cos\!\theta}\right)}
	{\displaystyle
	\left(\frac{d\sigma_{u\overline{d}}}{d\cos\!\theta}\right)+
	\left(\frac{d\sigma_{u\overline{s}}}{d\cos\!\theta}\right)+
	\left(\frac{d\sigma_{u\overline{u}}}{d\cos\!\theta}\right)+
	\left(\frac{d\sigma_{d\overline{d}}}{d\cos\!\theta}\right)+
	\left(\frac{d\sigma_{s\overline{s}}}{d\cos\!\theta}\right)}
\end{align}
Labai svarbu yra palyginti $ \tau^+ $ leptonų, susidariusių per $ W^+ $ bozoną ir $ Z^0 $, $ A^\gamma $ bozonus, poliarizacijas, nes, jei norima eksperimentiškai nustatyti leptono, susidarančio iš $ W $ bozono poliarizaciją, reikia žinoti, kokią dalį dalelių sudaro nereikalingos ar galbūt klaidingai identifikuojamos dalelės, susidariusios iš $ Z^0 $, $ A^\gamma $ bozonų.
\begin{align}
A^n_W(\cos\!\theta)=
	\frac{\displaystyle
	\left(\frac{d\sigma_{u\overline{d}}^n}{d\cos\!\theta}\right)+
	\left(\frac{d\sigma_{u\overline{s}}^n}{d\cos\!\theta}\right)}
	{\displaystyle
	\left(\frac{d\sigma_{u\overline{d}}}{d\cos\!\theta}\right)+
	\left(\frac{d\sigma_{u\overline{s}}}{d\cos\!\theta}\right)+
	\left(\frac{d\sigma_{u\overline{u}}}{d\cos\!\theta}\right)+
	\left(\frac{d\sigma_{d\overline{d}}}{d\cos\!\theta}\right)+
	\left(\frac{d\sigma_{s\overline{s}}}{d\cos\!\theta}\right)}
\end{align}
\begin{align}
A^n_{Z\gamma}(\cos\!\theta)=
	\frac{\displaystyle
	\left(\frac{d\sigma_{u\overline{u}}^n}{d\cos\!\theta}\right)+
	\left(\frac{d\sigma_{d\overline{d}}^n}{d\cos\!\theta}\right)+
	\left(\frac{d\sigma_{s\overline{s}}^n}{d\cos\!\theta}\right)}
	{\displaystyle
	\left(\frac{d\sigma_{u\overline{d}}}{d\cos\!\theta}\right)+
	\left(\frac{d\sigma_{u\overline{s}}}{d\cos\!\theta}\right)+
	\left(\frac{d\sigma_{u\overline{u}}}{d\cos\!\theta}\right)+
	\left(\frac{d\sigma_{d\overline{d}}}{d\cos\!\theta}\right)+
	\left(\frac{d\sigma_{s\overline{s}}}{d\cos\!\theta}\right)}
\end{align}
Šie dydžiai priklauso nuo kampo $ \theta $, bet taip pat galima atskirai nagrinėti atrenkant skirtingos energijos $ \tau^+ $ daleles. Didžiausią nagrinėjamo dydžio pokytį lemia mažos energijos dalelės, nes tuomet pasidaro svarbi dalelės masė. Tačiau negalima pasirinkti dalelių tik su maža energija, nes, kaip buvo minėta ankstesniame skyriuje, dalelės impulsui artėjant į 0, vektorius $ n^\mu $ tampa neapibrėžtas. Todėl reikia pasirinkti mažiausią ir didžiausią leidžiamą dalelės impulsą.

\newpage
\section{Rezultatai}

Šiame skyriuje pateikiami skaičiavimų rezultatai, kurie gauti naudojantis pasirašyta $ C\!+\!+ $ programa. Grafikai braižyti \textit{Wolfram Mathematica\textsuperscript{\textregistered}} programa. Skaičiavimams naudotų konstantų vertės yra \cite{data}:

\begin{align*}
&\alpha=\frac{1}{129},\\
&\sin^2\theta_w=0.231,\\
&M=80.385\,\mathrm{GeV},\\
&M_Z=91.188\,\mathrm{GeV},\\
&\Gamma=2.085\,\mathrm{GeV},\\
&\Gamma_Z=2.495\,\mathrm{GeV},\\
&m=1.776\,\mathrm{GeV},\\
&|V_{ud}|=0.974,\quad |V_{us}|=0.225.
\end{align*}

\begin{figure}[H]
	\centering
	\includegraphics[scale=1.5]{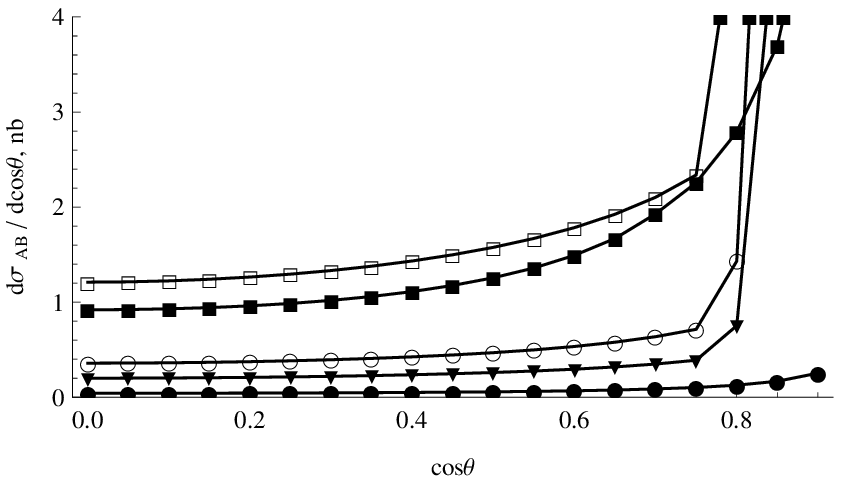}
\caption{$ \tau^+ $ leptono diferencialiniai sklaidos skerspjūviai iš skirtingų kvarkų skonių: \ding{110}$u\overline{d} $, \ding{108}$u\overline{s} $, \ding{111}$ u\overline{u} $, \ding{109}$d\overline{d} $, \ding{116}$ s\overline{s}$; įskaitytos visos $ \tau^+ $ dalelės.}\label{WZkvisas}
\end{figure}

\ref{WZkvisas} paveiksle pavaizduotas diferencialinis sklaidos skerspjūvis $ \tau^+ $ leptono susidarymui iš skirtingų kvarkų skonių įskaitant visas susidariusias $ \tau^+ $ daleles. Iš teorijos galima spėti, kad daugiausia dalelių susidaro iš $ u\overline{u} $ kvarkų, susiliejančių į fotoną, ir jie yra mažos energijos, nes fotono sąveikos stipris yra didižiausias, tačiau jo įnašas mažėja priklausomai nuo energijos kaip $ 1/\hat{s} $, kur $ \sqrt{\hat{s}} $ yra kvarkų energija jų masės centro sistemoje. Antras pagal didumą indėlis turi būti dėl $ u\overline{d} $ kvarkų, kurie susilieja turėdami energiją, artimą $ W $ bozono masei. Esant šiai rezonansinei energijai sklaidos skerspjūvis tam procesui smarkiai išauga. Kvarkų ir $ Z $ bozono sąveikos stipris yra labai mažas, todėl $ Z $ bozonas neturės didelės įtakos vidutinei leptono poliarizacijai. 
Diferencialinis sklaidos skerspjūvis labai išauga, kai kampas $ \theta $ artėja į 0, t.y. susidariusi dalelė mažai atsilenkia nuo greitintuvo ašies. Tai yra įprastas sklaidos skerspjūvio elgesys, kuris pasireiškia dėl judesio kiekio momento tvermės. Grafikuose $ \cos\!\theta $ yra nuo 0 iki 0.9, nes tai apytiksliai yra ribos, kuriose CMS detektoriuje yra išdėstyti geriausiai daleles registruojantys elementai \cite{cms}. 

\begin{figure}[H]
	\includegraphics[scale=1.5]{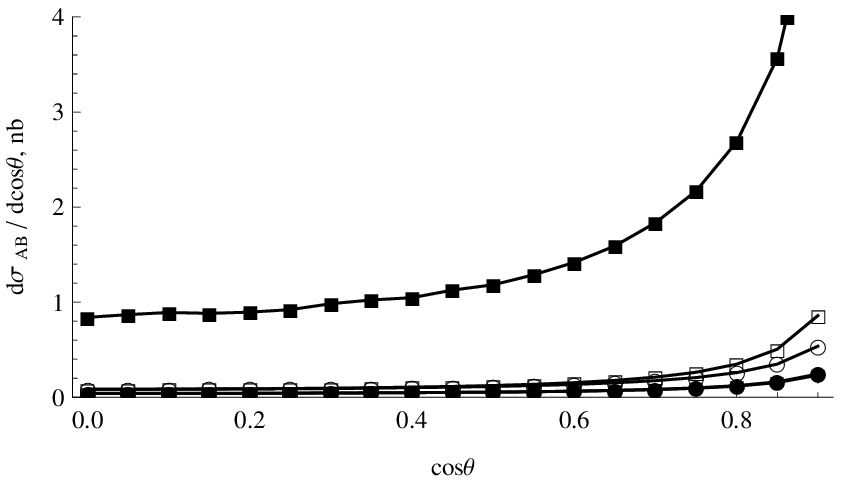}
\caption{$ \tau^+ $ leptono diferencialiniai sklaidos skerspjūviai iš skirtingų kvarkų skonių: \ding{110}$u\overline{d} $, \ding{108}$u\overline{s} $, \ding{111}$ u\overline{u} $, \ding{109}$d\overline{d} $, \ding{116}$ s\overline{s}$; įskaitytos $ \tau^+ $ dalelės su impulsu $20<|\vec{k}_2|$ GeV.}\label{WZk20}
\end{figure}

Atrinkus tik tam tikros energijos daleles ir palyginus sklaidos skerspjūvius su \ref{WZkvisas} paveiksle pavaizduotais, galima patikrinti, ar aukščiau aprašyti spėjimai yra teisingi. Pažiūrėjus į \ref{WZk20} paveikslą, kuriame įskaitytos tik $ \tau^+ $ dalelės, turinčios impulso vertę nuo 20 GeV, matoma, kad sklaidos skerspjūvis dėl $ Z^0 $ ir $ A^\gamma $ bozonų smarkiai sumažėja. Tai patvirtina, kad daugiausia šiuo kanalu susidaro mažos energijos leptonai dėl fotono.

\begin{figure}[H]
	\includegraphics[scale=1.5]{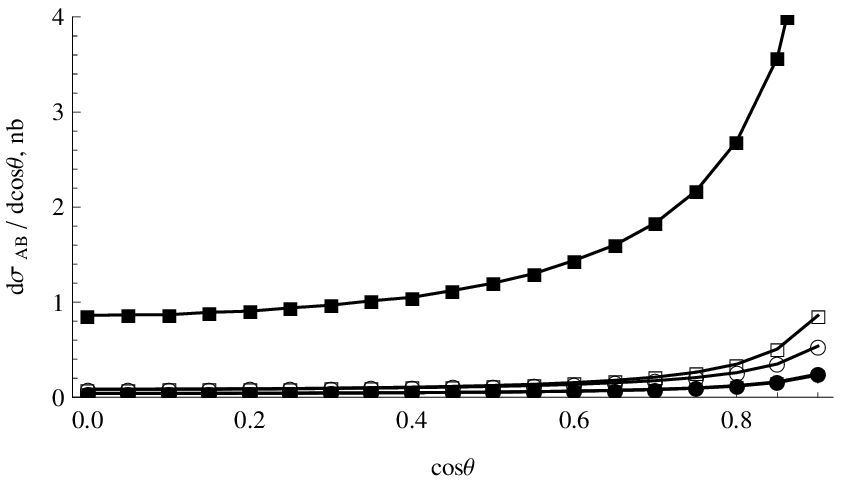}
\caption{$ \tau^+ $ leptono diferencialiniai sklaidos skerspjūviai iš skirtingų kvarkų skonių: \ding{110}$u\overline{d} $, \ding{108}$u\overline{s} $, \ding{111}$ u\overline{u} $, \ding{109}$d\overline{d} $, \ding{116}$ s\overline{s}$; įskaitytos $ \tau^+ $ dalelės su impulsu $20<|\vec{k}_2|<200 $ GeV.}\label{WZkmazas}
\end{figure}

Palyginus \ref{WZk20} paveikslą su \ref{WZkmazas} paveikslu, kuriame įskaitytos dalelės su impulsais $20<|\vec{k}_2|<200 $ GeV, galima teigti, kad beveik nesusidaro dalelių su impulsais, didesniais kaip 200 GeV.

\begin{figure}[H]
	\includegraphics[scale=1.5]{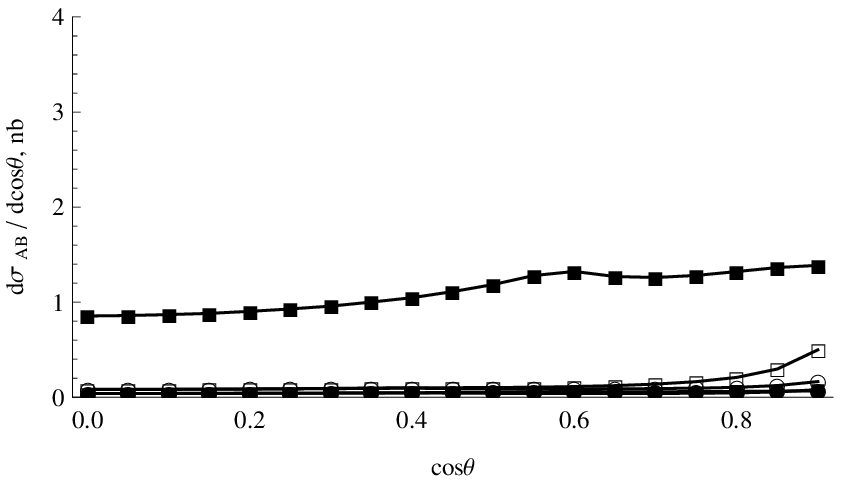}
\caption{$ \tau^+ $ leptono diferencialiniai sklaidos skerspjūviai iš skirtingų kvarkų skonių: \ding{110}$u\overline{d} $, \ding{108}$u\overline{s} $, \ding{111}$ u\overline{u} $, \ding{109}$d\overline{d} $, \ding{116}$ s\overline{s}$; įskaitytos $ \tau^+ $ dalelės su impulsu $20<|\vec{k}_2|<50 $ GeV.}\label{WZklabaimazas}
\end{figure}

\ref{WZklabaimazas} paveiksle pavaizduotas diferencialinis sklaidos skerspjūvis $ \tau^+ $ dalelėms su impulsais nuo 20 iki 50 GeV. Toks intervalas pasirinktas, kad būtų dar labiau išreikštas poliarizacijos efektas ir būtų galima matyti didelės energijos dalelių įtaką poliarizacijai.

Pasirinkus du skirtingus $ \tau^+ $ dalelių impulsų intervalus, nuo 20 iki 200 GeV ir nuo 20 iki 50 GeV, galima palyginti, kaip priklauso dalelės vidutinės poliarizacijos projekcija nuo tų dalelių impulsų ir nuo kampo $ \theta $.

\begin{figure}[H]
			\includegraphics[scale=1.5]{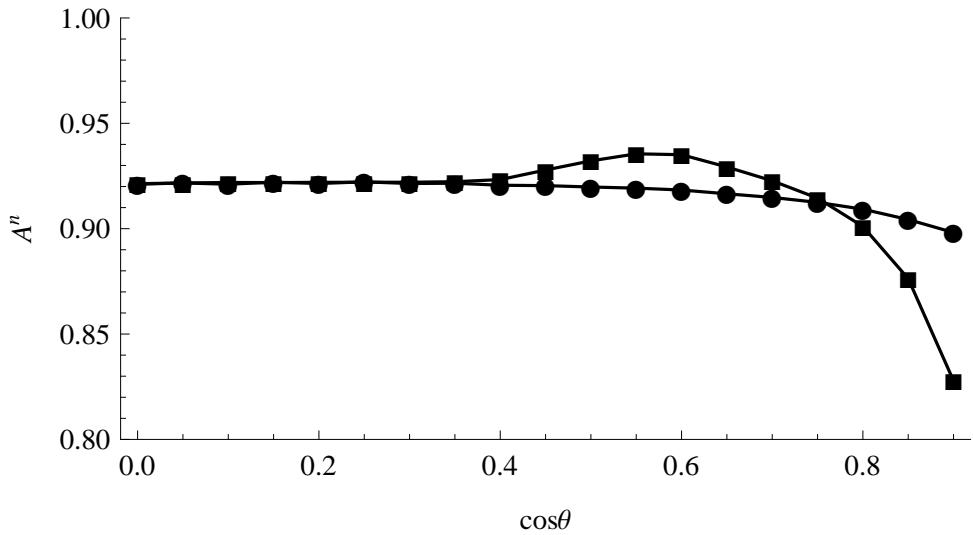}
\caption{Poliariazacijos vektoriaus projekcija $ \tau^+ $ leptono, susidariusio iš $ W^+ $ bozono: \ding{108}$20<|\vec{k}_2|<200 $ GeV, \ding{110}$20<|\vec{k}_2|<50 $ GeV}\label{ZWprob}
\end{figure}

\ref{ZWprob} paveiksle pavaizduota greitintuve susidariusios $ \tau^+ $
dalelės vidutinės poliarizacijos vektoriaus projekcija į dalelės judėjimo ašį. Pasirinkti du impulso verčių intervalai. Matoma, kad mažesnių impulsų verčių intervale vektoriaus projekcija labiau kinta. Tai paaiškinama
didesne leptono masės įtaka. Kai $ \cos\!\theta $ vertės yra 0.6 ir 0.9, matomi dideli skirtumai tarp kreivių. Tų skirtumų vertės yra
\begin{align}
\Delta A^n(0.6)=0.017\\
\Delta A^n(0.9)=0.070
\end{align}

\begin{figure}[H]
			\includegraphics[scale=1.5]{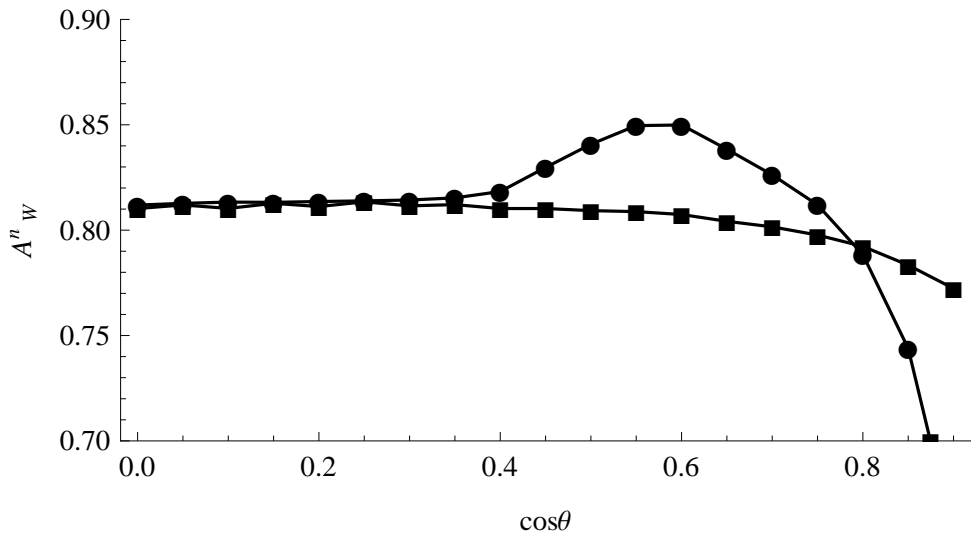}
\caption{Poliariazacijos vektoriaus projekcija $ \tau^+ $ leptono, susidariusio iš $ W^+ $ bozono: \ding{110}$20<|\vec{k}_2|<200 $ GeV, \ding{108}$20<|\vec{k}_2|<50 $ GeV}\label{Wprob}
\end{figure}

\begin{figure}[H]
			\includegraphics[scale=1.5]{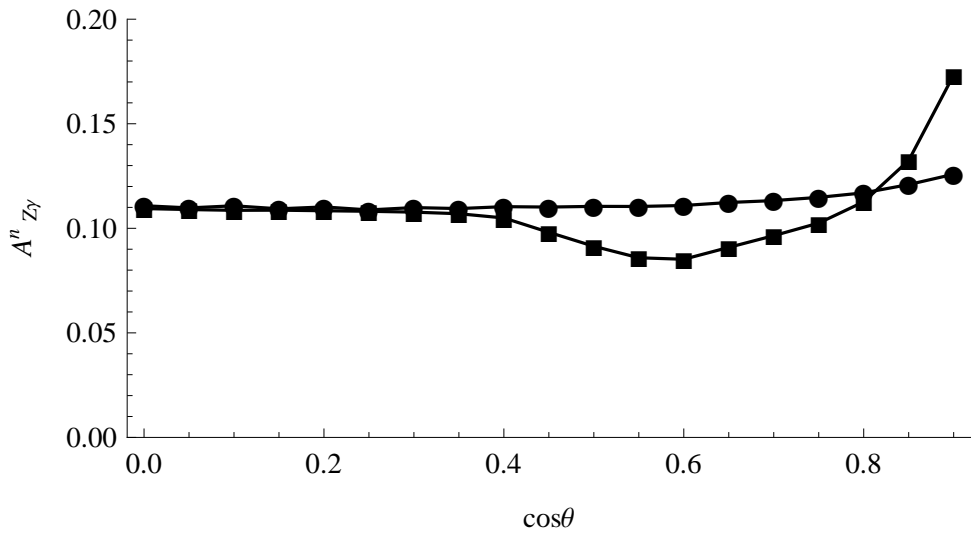}
\caption{Poliariazacijos vektoriaus projekcija $ \tau^+ $ leptono, susidariusio iš $ Z^0 $ ir $ A^\gamma $ bozonų: \ding{108}$20<|\vec{k}_2|<200 $ GeV, \ding{110}$20<|\vec{k}_2|<50 $ GeV}\label{Zprob}
\end{figure}

\ref{Wprob} ir \ref{Zprob} paveiksluose atskirai pavaizduoti atitinkamai iš $ W^+ $ ir $ Z^0,\,A^\gamma $ bozonų susidariusios dalelės vidutinės poliarizacijos vektoriaus projekcijos vertės.

\newpage
\section*{Išvados}
\addcontentsline{toc}{section}{Išvados}

Iš praeitame skyriuje aptartų rezultatų galima padaryti tokias išvadas:
\begin{itemize}
\item daugiausia pačios mažiausios energijos dalelių susidaro dėl fotono;
\item daugiausia vidutinės energijos dalelių susidaro dėl $ W^+ $ bozono;
\item į detektoriaus plotą nepataiko dalelės, kurių impulsas didesnis kaip 200 GeV;
\item leptono vidutinės poliarizacijos vektoriaus projekcija priklauso nuo dalelių impulsų verčių;
\item mažesnių impulso verčių intervale dalelių poliarizacijos vektoriaus projekcija varijuoja kampo $ \theta $ atžvilgiu smarkiau nei didesnių verčių intervale.
\item $ Z^0 $ ir $ A^\gamma $ bozonai sudaro mažąją dalį visų įvykių ir lemia mažą pokytį leptono poliarizacijos vektoriui.
\end{itemize}

\newpage
\section*{Predicting $ \tau $ lepton polarization at LHC}

\section*{Summary}
\addcontentsline{toc}{section}{Summary}

The goal of this thesis is to predict the polarization of a $ \tau $ particle produced at LHC and coming from a real $ W $ boson. This is achieved by calculating the projection of the expectation value of the polarization vector of the $ \tau $ particle. Calculations are done in the frame of the Standard Model. In this model only left-chiral currents couple to $ W $ bosons and the couplings to the $ Z $ boson are different for right-chiral and left-chiral currents. For this reason, a particle produced in $ W $ or $ Z $ decays will have a non trivial expectation value for the polarization vector. Particles coming from $ Z $ boson and photon cannot be distinguished practically from those coming from $ W $. That is why calculations for $ Z $ and $ A^\gamma $ (photon) decays into $ \tau $ particles are included. 

The initial colliding particles at LHC are protons and the most important channels for $ \tau $ production are quark fusion: $ q\overline{q}\rightarrow W^+ \rightarrow \tau^+ \nu_{\tau} $ and $ q\overline{q}\rightarrow Z^0,A^\gamma \rightarrow \tau^+ \tau^- $. Because quarks are confined in a proton, this thesis presents a thorough treatment on discribing proton's inner structure with Parton distribution functions.

The results section shows plots of differential cross sections for $ \tau $ production from different quark flavours and different energy intervals of the $ \tau $ particle. A projection of the expectation value of the polarization vector depends on the chosen energy intervals. The vector on which it is projected is chosed to be the direction of the $ \tau $ particle's motion.

\newpage
\appendix
\section*{Priedai}
\addcontentsline{toc}{section}{Priedai}
\section{Matricinių elementų skaičiavimas}\label{priedas1}

Šiame priede pateikiami matricinių elementų procesams  $ q\overline{q}\rightarrow W^+ \rightarrow \tau^+ \nu_{\tau} $ ir $ q\overline{q}\rightarrow Z^0,A^\gamma \rightarrow \tau^+ \tau^- $ skaičiavimas.

\subsection{$ W^+ $ kanalas}

Proceso $ q\overline{q}\rightarrow W^+ \rightarrow \tau^+ \nu_{\tau} $ Feinmano diagrama pavaizduota \ref{diagramos1} paveiksle.
\begin{figure}[H]
\centering
\begin{feynartspicture}(200,200)(1,1)
\FADiagram{}
\FAProp(0.,15.)(5.,10.)(0.,){/Straight}{1}
\FALabel(5.,15.5)[tr]{$q_A(p_1^\mu)$}
\FAProp(0.,5.)(5.,10.)(0.,){/Straight}{-1}
\FALabel(15.,15.5)[tl]{$\tau^+ (k_2^\mu)$}
\FAProp(5.,10.)(15.,10.)(0.,){/Sine}{0}
\FALabel(13.,10.5)[br]{$W^+(q^\mu)$}
\FAProp(15.,10.)(20.,15.)(0.,){/Straight}{-1}
\FALabel(5.,5.5)[br]{$\overline{q}_{\overline{B}}(p_2^\mu)$}
\FAProp(15.,10.)(20.,5.)(0.,){/Straight}{1}
\FALabel(15.,5.5)[bl]{$\nu_\tau(k_1^\mu) $}
\FAVert(5.,10.){0}
\FAVert(15.,10.){0}
\end{feynartspicture}
\caption{$ W^+ $ kanalas.}\label{diagramos1}
\end{figure} 

Feinmano taisyklės šiai diagramai \cite{pokorski}:
\vspace{0.5cm}

\begin{minipage}{0.3\textwidth}
	\begin{flushleft}
	kvarkams\\
		\begin{feynartspicture}(100,100)(1,1)\FADiagram{}
\FAProp(0.,18.)(10.,10.)(0.,){/Straight}{1}
\FALabel(5.,15.)[bl]{$ u_A $}
\FAProp(0.,2.)(10.,10.)(0.,){/Straight}{-1}
\FAProp(10.,10.)(20.,10.)(0.,){/Sine}{0}
\FALabel(17.,10.5)[br]{$W^+_\mu$}
\FALabel(5.,5.)[tl]{$\overline{v}_B$}
\FAVert(10.,10.){0}
	\end{feynartspicture}
	\end{flushleft}
\end{minipage}
\begin{minipage}{0.5\textwidth}
	\begin{equation}
-i\frac{e}{\sqrt{2}\sin\theta_w}\gamma^\mu V^{AB} P_L
\end{equation}
\end{minipage}

\vspace{1cm}

\begin{minipage}{0.3\textwidth}
	\begin{flushleft}
	leptonams\\
	\begin{feynartspicture}(100,100)(1,1)\FADiagram{}
\FAProp(0.,18.)(10.,10.)(0.,){/Straight}{1}
\FALabel(5.,15.)[bl]{$ \overline{u} $}
\FAProp(0.,2.)(10.,10.)(0.,){/Straight}{-1}
\FAProp(10.,10.)(20.,10.)(0.,){/Sine}{0}
\FALabel(17.,10.5)[br]{$W^+_\mu$}
\FALabel(5.,5.)[tl]{$ v $}
\FAVert(10.,10.){0}
	\end{feynartspicture}
	\end{flushleft}
\end{minipage}
\begin{minipage}{0.5\textwidth}
	\begin{equation}
-i\frac{e}{\sqrt{2}\sin\theta_w}\gamma^\mu P_L
\end{equation}
\end{minipage}
\\
čia $ e $ - elektrinis krūvis;\\
$ \theta_w $ - angl. \textit{weak mixing angle} arba Weinbergo kampas;\\
$\gamma^\mu$ - Dirako gama matricos;\\
$ A,\,B $ - kvarkų skonių indeksai;\\
$ V $ - CKM matrica (angl. \textit{Cabibbo-Kobayashi-Maskawa matrix}):\\
\begin{displaymath}
V =
\left( \begin{array}{ccc}
V_{ud} & V_{us} & V_{ub} \\
V_{cd} & V_{cs} & V_{cb} \\
V_{td} & V_{ts} & V_{tb}
\end{array} \right);
\end{displaymath}\\
$P_L$ - projekcijos operatorius, $ P_L=\frac{1}{2}(1-\gamma_5) $.\\

\begin{minipage}{0.3\textwidth}
	\begin{flushleft}
	$ W^+ $ bozono propagatorius\\
	\begin{feynartspicture}(100,100)(1,1)\FADiagram{}
\FAProp(5.,10.)(15.,10.)(0.,){/Sine}{0}
\FALabel(0.,9)[bl]{$W^+ _\mu$}
\FALabel(20.,9)[br]{$W^+ _\nu$}
\FALabel(11.,11)[br]{$ q $}
	\end{feynartspicture}
	\end{flushleft}
\end{minipage}
\begin{minipage}{0.5\textwidth}
	\begin{equation}
\frac{-i}{q^2-M^2}\left[g^{\mu\nu}-\frac{q^\mu q^\nu}{q^2-\xi M^2}(1-\xi)\right]
\end{equation}
\end{minipage}

Kadangi proceso energija gali būti tokia, kad susidaręs $ W^+ $ bozonas taps realus, reikia bozono propagatoriuje pakeisti 
\begin{equation}
\frac{1}{q^2-M^2}\longrightarrow \frac{1}{q^2-M^2+iM\Gamma}
\end{equation}
kur $ \Gamma $ yra $ W^+ $ bozono rezonanso plotis (angl. \textit{resonance width}), kuris susijęs su bozono vidutine gyvavimo trukme $ \tau $ sąryšiu $ \Gamma=1/\tau $.

Matriciniame elemente priešais $ \tau^+ $ dalelės spinorą rašomas projekcijos operatorius $ P_n=(1-\slashed{n}\gamma^5)/2 $.

Matricinis elementas gaunamas

\begin{align}
\nonumber i\mathcal{M}=\overline{v}_A(p_2) \left(-i\frac{e}{\sqrt{2}\sin\theta_w}\gamma_\mu V^{AB}P_L\right) u_B(p_1)\frac{-i}{q^2-M^2+iM\Gamma}\left[g^{\mu\nu}-\frac{q^\mu q^\nu}{q^2-\xi M^2}(1-\xi)\right]\times \\ \times\, \overline{u}_\nu(k_1) \left(-i\frac{e}{\sqrt{2}\sin\theta_w}\gamma_\nu P_L\right) P_n v_\tau(k_2)
\end{align}
Čia $ q^\mu $ yra propagatoriaus momentas, $ q^\mu=p_1^\mu+p_2^\mu $. Bozono propagatoriaus narys su kalibruotės parametru $ \xi $ lygus 0, nes kvarkų masės lygios 0 ir tuomet iš Dirako lygties $ \slashed{k}u(k)=0 $ ir $ \overline{v}(k)\slashed{k}=0 $. Sudauginus $ q^\mu \gamma_\mu $ gaunama
\begin{equation}
\overline{v}_A(p_2) \slashed{q} P_L u_B(p_1)=\overline{v}_A(p_2)  (\slashed{p}_1+\slashed{p}_2) P_L u_B(p_1)=\overline{v}_A(p_2)\slashed{p}_2 P_L u_B(p_1)- \overline{v}_A(p_2) P_L \slashed{p}_1 u_B(p_1)=0
\end{equation}

Matricinio elemento modulis pakeltas kvadratu, susumuotas ir suvidurkintas pagal fermionų poliarizacijas yra

\begin{align}\label{b}
\nonumber \frac{1}{4}\sum_\mathrm{pol.}|\mathcal{M}|^2 &=\left(\frac{e^2}{2 \sin^2\!\theta_w} |V^{AB}|\right)^2 \frac{1}{(q^2-M^2)^2+M^2 \Gamma^2}\times  \\
&\times \mathrm{Tr}\left[\slashed{p}_1 \gamma^\nu \slashed{p}_2 \gamma^\mu P_L \right]\mathrm{Tr}\left[\slashed{k}_1 \gamma_\mu P_L P_n (\slashed{k}_2-m)P_n \gamma_\nu \right]\equiv \langle|\mathcal{M}|^2\rangle
\end{align}

Pėdsakai šioje lygtyje yra
\begin{align}
\mathrm{Tr}&\left[\slashed{p}_1 \gamma^\nu \slashed{p}_2 \gamma^\mu P_L\right]=2(p_1^\mu p_2^\nu+p_2^\mu p_1^\nu-g^{\mu\nu}(p_1\cdot p_2)+ i\epsilon^{\sigma\nu\lambda\mu}p_{1 \sigma}p_{2\lambda})
\end{align}

Čia praleidžiamas daugiklis $ 1/4 $ iš dviejų projekcijos operatorių. Jis bus įrašytas pačioje skaičiavimo pabaigoje.
\begin{align}
\nonumber \mathrm{Tr}&\left[\slashed{k}_1 \gamma_\mu P_L P_n (\slashed{k}_2-m)P_n \gamma_\nu \right]=\\
\nonumber &= \mathrm{Tr}\bigg[\gamma_\nu \slashed{k}_1 \gamma_\mu P_L(\slashed{k}_2-m-(\slashed{k}_2-m)\slashed{n}\gamma^5-\slashed{n}\gamma^5(\slashed{k}_2- m)+\slashed{n}\gamma^5(\slashed{k}_2- m)\slashed{n}\gamma^5)\bigg]=\\
\nonumber &=\mathrm{Tr}\bigg[\gamma_\nu \slashed{k}_1 \gamma_\mu P_L(\slashed{k}_2-(-m)\slashed{n}\gamma^5-\slashed{n}\gamma^5(- m)+\slashed{n}\gamma^5(\slashed{k}_2)\slashed{n}\gamma^5)\bigg]=\\
\nonumber &=\mathrm{Tr}\bigg[\gamma_\nu \slashed{k}_1 \gamma_\mu P_L(\slashed{k}_2+2m\slashed{n}\gamma^5 +\slashed{n}\slashed{k}_2\slashed{n})\bigg]=\\ 
\nonumber &= \mathrm{Tr}\bigg[\gamma_\nu \slashed{k}_1 \gamma_\mu P_L(\slashed{k}_2+2m\slashed{n}\gamma^5 +2(n\cdot k_2)\slashed{n}-\slashed{k}_2 n^2)\bigg]=\\
\nonumber &=(1-n^2)\mathrm{Tr}\bigg[\slashed{k}_1 \gamma_\mu \slashed{k}_2\gamma_\nu P_L\bigg]+2(m +(n\cdot k_2))\mathrm{Tr}\bigg[\slashed{k}_1 \gamma_\mu \slashed{n}\gamma_\nu P_L\bigg]=\\
\nonumber &=2(1-n^2)\bigg(k_{1\mu}k_{2\nu}+k_{2\mu}k_{1\nu}-g_{\mu\nu}(k_1\cdot k_2)+i\epsilon_{\alpha\mu\beta\nu}k_1^\alpha k_2^\beta\bigg)+\\
&\quad +4(m +(n\cdot k_2))\bigg(k_{1\mu}n_\nu+n_\mu k_{1\nu}-g_{\mu\nu}(k_1\cdot n)+i\epsilon_{\alpha\mu\beta\nu}k_1^\alpha n^\beta\bigg)
\end{align}

Sudauginus pėdsakus gaunama
\begin{align}\label{sand}
&16(1-n^2)\bigg((k_1\cdot p_2)(k_2\cdot p_1)\bigg)+32\big(m +(n\cdot k_2)\big)\bigg((k_1\cdot p_2)(n\cdot p_1)\bigg)
\end{align}

Invariantiniai Mandelstamo kintamieji čia apibrėžiami taip:
\begin{align}
&s=(p_1^\mu+p_2^\mu)^2\\
&t=(p_1^\mu-k_2^\mu)^2=(p_2^\mu-k_1^\mu)^2
\end{align}
ir matricinis elementas užrašomas (įskaičius daugiklį 1/4):
\begin{equation}\label{matel1}
\langle|\mathcal{M}|^2\rangle_I^n=\frac{4\pi^2\alpha^2 |V^{AB}|^2}{\sin^4\!\theta_w}\frac{1}{(s-M^2)^2+M^2 \Gamma^2} \bigg[\frac{(1-n^2)}{4}\Big(t(t-m^2)\Big)
+\big(m +(n\cdot k_2)\big)\Big(-t(n\cdot p_1)\Big)\bigg]
\end{equation}
Matricinis elementas pažymėtas su indeksu $ I $, kad būtų atskirtas nuo žemiau aprašytų elementų, ir turi viršutinį indeksą, kuris reiškia, kad šis elementas apskaičiuotas su projekcijos operatoriumi $ P_n $.

Kadangi LAB sistemoje yra fiksuojama $ \tau^+ $ dalelės kryptis greitintuvo ašies atžvilgiu ir skaičiuojant (\ref{matel1}) matricinį elementą buvo pasirinkta, kad kvarkas juda teigiama $ \hat{z} $ ašies kryptimi, tai reikia dar įskaičiuoti atvejį, kad kvarkas juda neigiama $ \hat{z} $ ašies kryptimi, o $ \tau^+ $ dalelė ir toliau sudarys kampą $ \theta $ su teigiama $ \hat{z} $ ašimi (\ref{schema} pav. schema). Matriciniame elemente tai atitiks pakeitimą $ p_1^\mu \leftrightarrow p_2^\mu $. Atlikus pakeitimą (\ref{sand}) lygtyje, antrasis matricinis elementas užrašomas su indeksu $ II $
\begin{equation}
\langle|\mathcal{M}|^2\rangle_{II}^n=\frac{4\pi^2\alpha^2 |V^{AB}|^2}{\sin^4\!\theta_w}\frac{1}{(s-M^2)^2+M^2 \Gamma^2} \bigg[\frac{(1-n^2)}{4}\Big(u(u-m^2)\Big)
+\big(m +(n\cdot k_2)\big)\Big(-u(n\cdot p_2)\Big)\bigg]
\end{equation}
kur $ u $ yra kitas Mandelstamo kintamasis
\begin{equation}
u=(p_1^\mu-k_1^\mu)^2=(p_2^\mu-k_2^\mu)^2
\end{equation}

\begin{figure}[t]
\setlength{\unitlength}{1mm}
\begin{picture}(140, 50)
\put(15,10){\vector(1,0){30}}
\put(60,15){\vector(2,1){30}}
\put(80,31){$\tau^+(\vec{k}_2) $}
\put(70,13){$\theta$}
\put(105,10){\vector(-1,0){30}}
\put(30,12){$\vec{p}_1 $}
\put(90,12){$\vec{p}_2 $}
\put(119, 10){\vector(1,0){15}}
\put(135,10){$\hat{z} $}
\end{picture}
\caption{Kinematinė schema}\label{schema}
\end{figure}

Įprastiniai matriciniai elementai (be projekcijos operatoriaus) gaunami vektorių $ n^\mu $ prilyginus 0:
\begin{equation}
\langle|\mathcal{M}|^2\rangle_I=\frac{4\pi^2\alpha^2 |V^{AB}|^2}{\sin^4\!\theta_w}\frac{1}{(s-M^2)^2+M^2 \Gamma^2} \bigg[t(t-m^2)\bigg]
\end{equation}
\begin{equation}
\langle|\mathcal{M}|^2\rangle_{II}=\frac{4\pi^2\alpha^2 |V^{AB}|^2}{\sin^4\!\theta_w}\frac{1}{(s-M^2)^2+M^2 \Gamma^2} \bigg[u(u-m^2)\bigg]
\end{equation}

Vektorius $ n^\mu $ apibrėžtas kovariantiškai:
\begin{equation}
n^\mu=\frac{1}{|\vec{k}|}\left(-m T^\mu + \frac{E}{m}k^\mu\right)
\end{equation}
Sandaugos su šiuo vektoriumi yra:
\begin{equation}
(p_1\cdot n)=\frac{m^2-t}{2m}\left(\frac{|\vec{k}_2|-E_2\cos\!\theta}{E_2-|\vec{k}_2|\cos\!\theta}\right)
\end{equation}
\begin{equation}
(p_2\cdot n)=\frac{s+t}{2m}\left(\frac{|\vec{k}_2|+E_2\cos\!\theta}{E_2+|\vec{k}_2|\cos\!\theta}\right)
\end{equation}

\subsection{$ Z^0$ ir $ A^\gamma $ kanalas}
\begin{figure}[H]
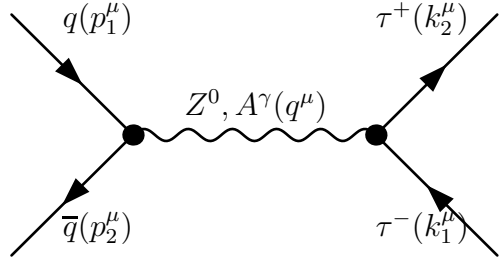

\centering
\begin{feynartspicture}(200,200)(1,1)
\FADiagram{}
\FAProp(0.,15.)(5.,10.)(0.,){/Straight}{1}
\FALabel(5.,15.5)[tr]{$q(p_1^\mu)$}
\FAProp(0.,5.)(5.,10.)(0.,){/Straight}{-1}
\FALabel(15.,15.5)[tl]{$\tau^+(k_2^\mu)$}
\FAProp(5.,10.)(15.,10.)(0.,){/Sine}{0}
\FALabel(13.,10.5)[br]{$Z^0, A^\gamma (q^\mu)$}
\FAProp(15.,10.)(20.,15.)(0.,){/Straight}{1}
\FALabel(5.,5.5)[br]{$\overline{q}(p_2^\mu)$}
\FAProp(15.,10.)(20.,5.)(0.,){/Straight}{-1}
\FALabel(15.,5.5)[bl]{$\tau^- (k_1^\mu)$}
\FAVert(5.,10.){0}
\FAVert(15.,10.){0}
\end{feynartspicture}
\caption{$ Z^0$ ir $ A^\gamma $ kanalas.}\label{diagrama2}
\end{figure}

Proceso $ q\overline{q}\rightarrow Z^0,A^\gamma \rightarrow \tau^+ \tau^- $ Feinmano diagrama pavaizduota \ref{diagrama2} paveiksle.

Feinmano taisyklės šioms diagramoms \cite{pokorski}
\vspace{0.5cm}

\begin{minipage}{0.3\textwidth}
	\begin{flushleft}
	kvarkams\\
\begin{feynartspicture}(100,100)(1,1)\FADiagram{}
\FAProp(0.,18.)(10.,10.)(0.,){/Straight}{1}
\FALabel(5.,15.)[bl]{$ u $}
\FAProp(0.,2.)(10.,10.)(0.,){/Straight}{-1}
\FAProp(10.,10.)(20.,10.)(0.,){/Sine}{0}
\FALabel(17.,10.5)[br]{$Z^0_\mu$}
\FALabel(5.,5.)[tl]{$\overline{v}$}
\FAVert(10.,10.){0}
	\end{feynartspicture}
	\end{flushleft}
\end{minipage}
\begin{minipage}{0.5\textwidth}
	\begin{equation}
(\pm)i\frac{e}{2\sin\theta_w \cos\theta_w}\gamma_\mu (c_1P_L-c_2P_R)\delta^{AB}
\end{equation}
\end{minipage}

\begin{minipage}{0.3\textwidth}
	\begin{flushleft}
	\begin{feynartspicture}(100,100)(1,1)\FADiagram{}
\FAProp(0.,18.)(10.,10.)(0.,){/Straight}{1}
\FALabel(5.,15.)[bl]{$ u $}
\FAProp(0.,2.)(10.,10.)(0.,){/Straight}{-1}
\FAProp(10.,10.)(20.,10.)(0.,){/Sine}{0}
\FALabel(17.,10.5)[br]{$A^\gamma_\mu$}
\FALabel(5.,5.)[tl]{$\overline{v}$}
\FAVert(10.,10.){0}
	\end{feynartspicture}
	\end{flushleft}
\end{minipage}
\begin{minipage}{0.5\textwidth}
	\begin{equation}
-iQe\gamma_\mu \delta^{AB}
\end{equation}
\end{minipage}

\begin{minipage}{0.3\textwidth}
	\begin{flushleft}
	leptonams\\
	\begin{feynartspicture}(100,100)(1,1)\FADiagram{}
\FAProp(0.,18.)(10.,10.)(0.,){/Straight}{1}
\FALabel(5.,15.)[bl]{$ \overline{u} $}
\FAProp(0.,2.)(10.,10.)(0.,){/Straight}{-1}
\FAProp(10.,10.)(20.,10.)(0.,){/Sine}{0}
\FALabel(17.,10.5)[br]{$Z^0_\mu$}
\FALabel(5.,5.)[tl]{$ v $}
\FAVert(10.,10.){0}
	\end{feynartspicture}
	\end{flushleft}
\end{minipage}
\begin{minipage}{0.5\textwidth}
	\begin{equation}
+i\frac{e}{2\sin\theta_w \cos\theta_w}\gamma_\mu(c_3 P_L-c_4 P_R)
\end{equation}
\end{minipage}

\begin{minipage}{0.3\textwidth}
	\begin{flushleft}
	\begin{feynartspicture}(100,100)(1,1)\FADiagram{}
\FAProp(0.,18.)(10.,10.)(0.,){/Straight}{1}
\FALabel(5.,15.)[bl]{$ \overline{u} $}
\FAProp(0.,2.)(10.,10.)(0.,){/Straight}{-1}
\FAProp(10.,10.)(20.,10.)(0.,){/Sine}{0}
\FALabel(17.,10.5)[br]{$A^\gamma_\mu$}
\FALabel(5.,5.)[tl]{$ v $}
\FAVert(10.,10.){0}
	\end{feynartspicture}
	\end{flushleft}
\end{minipage}
\begin{minipage}{0.5\textwidth}
	\begin{equation}
+ie\gamma_\mu
\end{equation}
\end{minipage}

Koeficientai šiose taisyklėse yra

\begin{minipage}{0.45\textwidth}\begin{flushleft}
\begin{align*}
&d,\,\, s\,\, \mathrm{kvarkams}:\\
&c_1=1-\frac{2}{3}\sin^2\theta_w\\
&c_2=\frac{2}{3}\sin^2\theta_w\\
&c_3=1-2\sin^2\theta_w\\
&c_4=2\sin^2\theta_w\\
&Q=-\frac{1}{3}\\
&(\pm i)\rightarrow +i
\end{align*}
\end{flushleft}\end{minipage}
\begin{minipage}{0.5\textwidth}\begin{flushright}
\begin{align*}
&u\,\, \mathrm{kvarkams}:\\
&c_1=1-\frac{4}{3}\sin^2\theta_w\\
&c_2=\frac{4}{3}\sin^2\theta_w\\
&c_3=1-2\sin^2\theta_w\\
&c_4=2\sin^2\theta_w\\
&Q=\frac{2}{3}\\
&(\pm i)\rightarrow -i
\end{align*}
\end{flushright}\end{minipage}

\vspace{0.5cm}
\begin{minipage}{0.3\textwidth}
	\begin{flushleft}
	$ Z^0_\mu $ bozono propagatorius\\
\begin{feynartspicture}(100,100)(1,1)\FADiagram{}
\FAProp(5.,10.)(15.,10.)(0.,){/Sine}{0}
\FALabel(0.,9)[bl]{$Z^0 _\mu$}
\FALabel(20.,9)[br]{$Z^0 _\nu$}
\FALabel(11.,11)[br]{$ q $}
	\end{feynartspicture}
	\end{flushleft}
\end{minipage}
\begin{minipage}{0.5\textwidth}
	\begin{equation}
\frac{-i}{q^2-M_Z^2}\left[g^{\mu\nu}-(1-\xi)\frac{q^\mu q^\nu}{q^2-\xi M_Z^2}\right]
\end{equation}
\end{minipage}

\begin{minipage}{0.3\textwidth}
	\begin{flushleft}
	$ A^\gamma_\mu $ bozono propagatorius\\
	\begin{feynartspicture}(100,100)(1,1)\FADiagram{}
\FAProp(5.,10.)(15.,10.)(0.,){/Sine}{0}
\FALabel(0.,9)[bl]{$A^\gamma _\mu$}
\FALabel(20.,9)[br]{$A^\gamma _\nu$}
\FALabel(11.,11)[br]{$ q $}
	\end{feynartspicture}
	\end{flushleft}
\end{minipage}
\begin{minipage}{0.5\textwidth}
	\begin{equation}
\frac{-i}{q^2}\left[g^{\mu\nu}-(1-\xi)\frac{q^\mu q^\nu}{q^2}\right]
\end{equation}
\end{minipage}

Čia taip pat reikia $ Z^0 $ bozono propagatoriuje pakeisti 
\begin{equation}
\frac{1}{q^2-M_Z^2}\longrightarrow \frac{1}{q^2-M_Z^2+iM_Z\Gamma_Z}
\end{equation}
kur $ \Gamma_Z $ yra $ Z^0 $ bozono rezonanso plotis.

Matricinis elementas $ \tau^+ \tau^- $ susidarymui dėl $ A^\gamma $ ir $ Z^0 $ bozonų susiliejant kvarkams:
\begin{align}
\nonumber i\mathcal{M}=& (\pm i) \frac{e^2}{4\sin^2\theta_w \cos^2\theta_w}\,\frac{1}{q^2-M_Z^2+iM_Z\Gamma_Z}\Big(g^{\mu\nu}-(1-\xi)\frac{q^\mu q^\nu}{q^2-\xi M_Z^2}\Big)\times\\
\nonumber &\times \Big[\overline{v}(p_2)\gamma_\mu (c_1P_L-c_2P_R)u(p_1)\Big]\Big[\overline{u}(k_1)\gamma_\nu (c_3P_L-c_4P_R)P_n v(k_2)\Big]-\\
&-iQ e^2 \frac{1}{q^2}\Big(g^{\mu\nu}-(1-\xi)\frac{q^\mu q^\nu}{q^2}\Big) \Big[\overline{v}(p_2)\gamma_\mu u(p_1)\Big]\Big[\overline{u}(k_1)\gamma_\nu P_n v(k_2)\Big]
\end{align}
Prieš $ \tau^+ $ dalelės spinorą įrašytas projekcijos iperatorius $ P_n $. Ženklas prie interferencijos nario visais atvejais ($ u $, $ d $ ir $ s $ kvarkams) vienodas. Nariai su kalibruotės parametru $ \xi $ lygūs 0. 

Matricinis elementas pakėlus kvadratu, susumavus pagal visų dalelių poliarizacijas ir suvidurkinus yra
\begin{align}
\nonumber \frac{1}{4}\sum_{\mathrm{pol.}}|\mathcal{M}|^2=&\quad\frac{1}{4}\left(\frac{e^2}{4\sin^2\!\theta_w \cos^2\!\theta_w}\right)^2\frac{1}{(q^2-M_Z^2)^2+M_Z^2\Gamma_Z^2} \times\\
\nonumber &\quad\times\mathrm{Tr}\big[\slashed{p}_2 \gamma^\mu(c_1P_L-c_2P_R)\slashed{p}_1 \gamma^\nu(c_1P_L-c_2P_R) \big]\times\\
\nonumber & \quad\times\mathrm{Tr}\big[(\slashed{k}_1+m) \gamma_\mu(c_3P_L-c_4P_R)P_n(\slashed{k}_2-m)P_n \gamma_\nu(c_3 P_L-c_4P_R)\big]+\\
\nonumber &+\frac{1}{4}\left(\frac{Qe^2}{q^2}\right)^2 \mathrm{Tr}\big[\slashed{p}_2 \gamma^\mu\slashed{p}_1 \gamma^\nu \big] \mathrm{Tr}\big[(\slashed{k}_1+m) \gamma_\mu P_n (\slashed{k}_2-m)P_n \gamma_\nu\big]+\\
\nonumber &+\frac{1}{4}\left(\frac{e^2}{4\sin^2\!\theta_w \cos^2\!\theta_w}|Q|e^2\frac{1}{q^2}\,\frac{1}{(q^2-M_Z^2)^2+M_Z^2\Gamma_Z	^2}\right)\mathrm{Tr}\big[\slashed{p}_2 \gamma^\mu \slashed{p}_1 \gamma^\nu (c_1P_L-c_2P_R)\big] \\
\nonumber &\quad\Bigg\lbrace (q^2-M_Z^2)\Big(2\mathrm{Tr}\big[\slashed{k}_1\gamma_\mu P_n(\slashed{k}_2-m)P_n \gamma_\nu(c_3P_L-c_4P_R)\big]+\\
\nonumber &\quad +m(c_3-c_4)\mathrm{Tr}\big[\gamma_\mu P_n(\slashed{k}_2-m)P_n \gamma_\nu \big]\Big)+\\
&\quad + iM_Z\Gamma_Z m (c_3+c_4) \mathrm{Tr}\big[\gamma_\mu\gamma^5 P_n(\slashed{k}_2-m)P_n \gamma_\nu \big]\Bigg\rbrace
\end{align} 

Atskirai suskaičiuojamas kiekvienas narys. Kol kas bus praleidžiamas daugiklis $ 1/4 $ iš dviejų projekcijos operatorių. Jis bus įrašytas pačioje skaičiavimo pabaigoje.

\subsubsection{$ A^\gamma$ bozonas}

$ A^\gamma $ bozono naryje leptonų pėdsakas:
\begin{align}
\nonumber &\mathrm{Tr}\big[(\slashed{k}_1+m) \gamma_\mu P_n (\slashed{k}_2-m)P_n \gamma_\nu\big]=\\
\nonumber &=\mathrm{Tr}\big[(\slashed{k}_1+m) \gamma_\mu(\slashed{k}_2-m-\slashed{n}\gamma^5(\slashed{k}_2-m)-(\slashed{k}_2-m)\slashed{n} \gamma^5+\slashed{n}\gamma^5(\slashed{k}_2-m)\slashed{n} \gamma^5) \gamma_\nu \big]=\\
&=\mathrm{Tr}\big[\slashed{k}_1 \gamma_\mu(\slashed{k}_2+2m\slashed{n}\gamma^5 +2(n\cdot k_2)\slashed{n}-\slashed{k}_2n^2) \gamma_\nu\big]+m\mathrm{Tr}\big[\gamma_\mu(-m-\slashed{n}\gamma^5\slashed{k}_2-\slashed{k}_2\slashed{n} \gamma^5+m n^2) \gamma_\nu\big]
\end{align}
Iš karto praleidžiu narius, kurie bus antisimetriniai tenzoriai, nes juos sudauginus su simetriniu kvarkų tenzoriumi rezultatas lygus 0. Lieka
\begin{align}
\nonumber (1-n^2)&\mathrm{Tr}\big[\slashed{k}_1 \gamma_\mu \slashed{k}_2 \gamma_\nu\big]+2(n\cdot k_2)\mathrm{Tr}\big[\slashed{k}_1 \gamma_\mu \slashed{n}\gamma_\nu\big]-m^2(1-n^2)\mathrm{Tr}\big[\gamma_\mu \gamma_\nu\big]=\\
\nonumber &=4(1-n^2)\big(k_{1\mu}k_{2\nu}+k_{2\mu}k_{1\nu}-g_{\mu\nu}(k_1\cdot k_2)\big)+\\
&\quad+8(n\cdot k_2)\big(k_{1\mu}n_\nu+n_\mu k_{1\nu}-g_{\mu\nu}(k_1\cdot n)\big) -4m^2(1-n^2)g_{\mu\nu}
\end{align}

Visas $ A^\gamma $ narys:
\begin{align}
\nonumber \frac{1}{4}\left(\frac{Qe^2}{q^2}\right)^2 &\mathrm{Tr}\big[\slashed{p}_2 \gamma^\mu\slashed{p}_1 \gamma^\nu \big] \mathrm{Tr}\big[(\slashed{k}_1+m) \gamma_\mu P_n (\slashed{k}_2-m)P_n \gamma_\nu\big]=\\
\nonumber =&\frac{1}{4}\left(\frac{Qe^2}{q^2}\right)^2 4\big[p_1^\mu p_2^\nu+p_2^\mu p_1^\nu-g^{\mu\nu}(p_1\cdot p_2)\big]\times\\
\nonumber &\times \Big[ 4(1-n^2)\big(k_{1\mu}k_{2\nu}+k_{2\mu}k_{1\nu}-g_{\mu\nu}(k_1\cdot k_2)\big)+\\
\nonumber &\quad+8(n\cdot k_2)\big(k_{1\mu}n_\nu+n_\mu k_{1\nu}-g_{\mu\nu}(k_1\cdot n)\big) -4m^2(1-n^2)g_{\mu\nu}\Big]=\\
\nonumber =&\left(\frac{Qe^2}{q^2}\right)^2 \Big[ 8(1-n^2)\big((p_1\cdot k_1)(p_2\cdot k_2)+(p_1\cdot k_2)(p_2\cdot k_1)\big)+\\
\nonumber &\qquad\qquad+16(n\cdot k_2)\big((p_1\cdot k_1)(p_2\cdot n)+(p_1\cdot n)(p_2\cdot k_1)\big)+8m^2(1-n^2)(p_1\cdot p_2)\Big]=\\
\nonumber =&\left(\frac{Qe^2}{s}\right)^2 \Big[ 2(1-n^2)\big((s+t-m^2)^2+(m^2-t)^2\big)+\\
&\qquad\qquad+8(n\cdot k_2)\big((s+t-m^2)(p_2\cdot n)+(p_1\cdot n)(m^2-t)\big)+4m^2(1-n^2)s\Big]
\end{align}

\subsubsection{$ Z^0 $ bozonas}

Visas $ Z^0 $ narys 
\begin{align}
\nonumber \frac{1}{4}\left(\frac{e^2}{4\sin^2\!\theta_w \cos^2\!\theta_w}\right)^2&\frac{1}{(q^2-M_Z^2)^2+M_Z^2\Gamma_Z^2} \times\\
\nonumber &\quad\times\mathrm{Tr}\big[\slashed{p}_2 \gamma^\mu \slashed{p}_1 \gamma^\nu(c_1^2 P_L+c_2^2 P_R) \big]\times\\
\nonumber & \quad\times\Big(\mathrm{Tr}\big[\slashed{k}_1 \gamma_\mu P_n(\slashed{k}_2-m)P_n \gamma_\nu(c_3^2 P_L+ c_4^2 P_R)\big] +\\
&\qquad +m\mathrm{Tr}\big[ \gamma_\mu P_n(\slashed{k}_2-m)P_n \gamma_\nu(-c_3 c_4)(P_L+P_R)\big]\Big)
\end{align}

Pėdsakai čia yra:
\begin{align}
\nonumber &\mathrm{Tr}\big[\slashed{p}_2 \gamma^\mu \slashed{p}_1 \gamma^\nu(c_1^2 P_L+c_2^2 P_R) \big]=\\
&=2(c_1^2+c_2^2)\big[p_1^\mu p_2^\nu+p_2^\mu p_1^\nu-g^{\mu\nu}(p_1\cdot p_2)\big]+2i(c_1^2-c_2^2)\epsilon^{\alpha\mu\beta\nu}p_{2\alpha} p_{1\beta}
\end{align}
\begin{align}
\nonumber &\mathrm{Tr}\big[\slashed{k}_1 \gamma_\mu P_n(\slashed{k}_2-m)P_n \gamma_\nu(c_3^2 P_L+ c_4^2 P_R)\big]=\\
\nonumber &=\mathrm{Tr}\big[\slashed{k}_1 \gamma_\mu (\slashed{k}_2-m-\slashed{n}\gamma^5(\slashed{k}_2-m)-(\slashed{k}_2-m)\slashed{n} \gamma^5+\slashed{n}\gamma^5(\slashed{k}_2-m)\slashed{n} \gamma^5) \gamma_\nu(c_3^2 P_L+ c_4^2 P_R)\big]=\\
\nonumber &=\mathrm{Tr}\big[\slashed{k}_1 \gamma_\mu(\slashed{k}_2+2m\slashed{n}\gamma^5+2(n\cdot k_2)\slashed{n}-\slashed{k}_2 n^2 )\gamma_\nu(c_3^2 P_L+c_4^2P_R)\big]=\\
\nonumber &=(1-n^2)\mathrm{Tr}\big[\slashed{k}_1 \gamma_\mu \slashed{k}_2 \gamma_\nu(c_3^2 P_L+c_4^2P_R)\big]+\mathrm{Tr}\big[\slashed{k}_1 \gamma_\mu(2m\slashed{n}\gamma^5+2(n\cdot k_2)\slashed{n})\gamma_\nu(c_3^2 P_L+c_4^2P_R)\big]=\\
\nonumber &=2(1-n^2)\Big[\big(c_3^2+c_4^2\big)\big(k_{1\mu}k_{2\nu}+k_{2\mu}k_{1\nu}-g_{\mu\nu}(k_1\cdot k_2)\big)+\big(c_3^2-c_4^2\big)i\epsilon_{\alpha\mu\beta\nu}k_1^\alpha k_2^\beta \Big]+\\
\nonumber &\quad+4m\Big[\big(c_3^2-c_4^2\big)\big(k_{1\mu}n_\nu+n_\mu k_{1\nu}-g_{\mu\nu}(k_1\cdot n)\big)+\big(c_3^2+c_4^2\big)i\epsilon_{\alpha\mu\beta\nu}k_1^\alpha n^\beta \Big]+\\
&\quad+4(n\cdot k_2)\Big[\big(c_3^2+c_4^2\big)\big(k_{1\mu}n_\nu+n_\mu k_{1\nu}-g_{\mu\nu}(k_1\cdot n)\big)+\big(c_3^2-c_4^2\big)i\epsilon_{\alpha\mu\beta\nu}k_1^\alpha n^\beta \Big]
\end{align}

\begin{align}
\nonumber &m\mathrm{Tr}\big[ \gamma_\mu P_n(\slashed{k}_2-m)P_n \gamma_\nu(-c_3 c_4)(P_L+P_R)\big]=\\
\nonumber &=(-c_3 c_4 )m\mathrm{Tr}\big[\gamma_\mu (\slashed{k}_2-m-\slashed{n}\gamma^5(\slashed{k}_2-m)-(\slashed{k}_2-m)\slashed{n} \gamma^5+\slashed{n}\gamma^5(\slashed{k}_2-m)\slashed{n} \gamma^5) \gamma_\nu\big]=\\
\nonumber &=(-c_3 c_4 )m\mathrm{Tr}\big[\gamma_\mu(-m+\slashed{n}\slashed{k}_2\gamma^5-\slashed{k}_2\slashed{n} \gamma^5+mn^2)\gamma_\nu\big]=\\
&=4 c_3 c_4 m^2(1-n^2)g_{\mu\nu}-8c_3 c_4 mi\epsilon_{\alpha\mu\beta\nu}k_2^\alpha n^\beta
\end{align}

Įrašius šiuos pėdsakus:
\begin{align}
\nonumber \frac{1}{4}&\left(\frac{e^2}{4\sin^2\!\theta_w \cos^2\!\theta_w}\right)^2\frac{1}{(q^2-M_Z^2)^2+M_Z^2\Gamma_Z^2} \times\\
\nonumber &\times\Big(2(c_1^2+c_2^2)\Big[p_1^\mu p_2^\nu+p_2^\mu p_1^\nu-g^{\mu\nu}(p_1\cdot p_2)\Big]+2i(c_1^2-c_2^2)\epsilon^{\alpha\mu\beta\nu}p_{2\alpha} p_{1\beta}\Big)\times\\
\nonumber & \times\Big(2(1-n^2) \Big[\big(c_3^2+c_4^2\big)\big(k_{1\mu}k_{2\nu}+k_{2\mu}k_{1\nu}-g_{\mu\nu}(k_1\cdot k_2)\big)+\big(c_3^2-c_4^2\big)i\epsilon_{\alpha\mu\beta\nu}k_1^\alpha k_2^\beta \Big]+\\
\nonumber &\quad+4m\Big[\big(c_3^2-c_4^2\big)\big(k_{1\mu}n_\nu+n_\mu k_{1\nu}-g_{\mu\nu}(k_1\cdot n)\big)+\big(c_3^2+c_4^2\big)i\epsilon_{\alpha\mu\beta\nu}k_1^\alpha n^\beta \Big]+\\
\nonumber &\quad+4(n\cdot k_2)\Big[\big(c_3^2+c_4^2\big)\big(k_{1\mu}n_\nu+n_\mu k_{1\nu}-g_{\mu\nu}(k_1\cdot n)\big)+\big(c_3^2-c_4^2\big)i\epsilon_{\alpha\mu\beta\nu}k_1^\alpha n^\beta \Big]+\\
&\quad+4 c_3 c_4 m^2(1-n^2)g_{\mu\nu}-8c_3 c_4 mi\epsilon_{\alpha\mu\beta\nu}k_2^\alpha n^\beta\Big)=\\
\nonumber =&\frac{1}{4}\left(\frac{e^2}{4\sin^2\!\theta_w \cos^2\!\theta_w}\right)^2\frac{1}{(q^2-M_Z^2)^2+M_Z^2\Gamma_Z^2} \times\\
\nonumber &\times \Bigg(8(c_1^2+c_2^2)(c_3^2+c_4^2)(1-n^2)\Big((p_1\cdot k_1)(p_2\cdot k_2)+(p_1\cdot k_2)(p_2\cdot k_1)\Big)+\\
\nonumber &\quad +16(c_1^2+c_2^2)(c_3^2-c_4^2)m\Big((p_1\cdot k_1)(p_2\cdot n)+(p_1\cdot n)(p_2\cdot k_1)\Big)+\\
\nonumber &\quad +16(c_1^2+c_2^2)(c_3^2+c_4^2)(n\cdot k_2)\Big((p_1\cdot k_1)(p_2\cdot n)+(p_1\cdot n)(p_2\cdot k_1)\Big)+\\
\nonumber &\quad-16(c_1^2+c_2^2)c_3 c_4 m^2(1-n^2)(p_1\cdot p_2)+\\
\nonumber &\quad+8(c_1^2-c_2^2)(c_3^2-c_4^2)(1-n^2)\Big(-(p_1\cdot k_1)(p_2\cdot k_2)+(p_1\cdot k_2)(p_2\cdot k_1)\Big)+\\
\nonumber &\quad+16(c_1^2-c_2^2)(c_3^2+c_4^2)m\Big(-(p_1\cdot k_1)(p_2\cdot n)+(p_1\cdot n)(p_2\cdot k_1)\Big)+\\
\nonumber &\quad+16(c_1^2-c_2^2)(c_3^2-c_4^2)(n\cdot k_2)\Big(-(p_1\cdot k_1)(p_2\cdot n)+(p_1\cdot n)(p_2\cdot k_1)\Big)+\\
&\quad+32(c_1^2-c_2^2)c_3 c_4 m\Big(-(p_1\cdot n)(p_2\cdot k_2)+(p_1\cdot k_2)(p_2\cdot n)\Big)\Bigg)=\\
\nonumber &=\left(\frac{e^2}{4\sin^2\!\theta_w \cos^2\!\theta_w}\right)^2\frac{1}{(s-M_Z^2)^2+M_Z^2\Gamma_Z^2} \times\\
\nonumber &\times \Bigg(\frac{1}{2}(c_1^2+c_2^2)(c_3^2+c_4^2)(1-n^2)\Big((s+t-m^2)^2+(m^2-t)^2\Big)+\\
\nonumber &\quad +2(c_1^2+c_2^2)(c_3^2-c_4^2)m\Big((s+t-m^2)(p_2\cdot n)+(p_1\cdot n)(m^2-t)\Big)+\\
\nonumber &\quad +2(c_1^2+c_2^2)(c_3^2+c_4^2)(n\cdot k_2)\Big((s+t-m^2)(p_2\cdot n)+(p_1\cdot n)(m^2-t)\Big)+\\
\nonumber &\quad-2(c_1^2+c_2^2)c_3 c_4 m^2(1-n^2)s+\\
\nonumber &\quad+\frac{1}{2}(c_1^2-c_2^2)(c_3^2-c_4^2)(1-n^2)\Big(-(s+t-m^2)^2+(m^2-t)^2\Big)+\\
\nonumber &\quad+2(c_1^2-c_2^2)(c_3^2+c_4^2)m\Big(-(s+t-m^2)(p_2\cdot n)+(p_1\cdot n)(m^2-t)\Big)+\\
\nonumber &\quad+2(c_1^2-c_2^2)(c_3^2-c_4^2)(n\cdot k_2)\Big(-(s+t-m^2)(p_2\cdot n)+(p_1\cdot n)(m^2-t)\Big)+\\
&\quad+4(c_1^2-c_2^2)c_3 c_4 m\Big(-(p_1\cdot n)(s+t-m^2)+(m^2-t)(p_2\cdot n)\Big)\Bigg)
\end{align}

\subsubsection{Interferencija}

Visas interferencijos narys (pėdsakai analogiški $ Z^0 $ bozono atvejui):
\begin{align}
\nonumber &\frac{1}{4}\left(\frac{e^2}{4\sin^2\!\theta_w \cos^2\!\theta_w}|Q|e^2\frac{1}{q^2}\,\frac{1}{(q^2-M_Z^2)^2+M_Z^2\Gamma_Z	^2}\right)\mathrm{Tr}\big[\slashed{p}_2 \gamma^\mu \slashed{p}_1 \gamma^\nu (c_1P_L-c_2P_R)\big] \times\\
\nonumber &\times\Bigg\lbrace (q^2-M_Z^2)\Big(2\mathrm{Tr}\big[\slashed{k}_1\gamma_\mu P_n(\slashed{k}_2-m)P_n \gamma_\nu(c_3P_L-c_4P_R)\big]+\\
\nonumber &\quad +m(c_3-c_4)\mathrm{Tr}\big[\gamma_\mu P_n(\slashed{k}_2-m)P_n \gamma_\nu \big]\Big)+\\
&\quad + iM_Z\Gamma_Z m (c_3+c_4) \mathrm{Tr}\big[\gamma_\mu\gamma^5 P_n(\slashed{k}_2-m)P_n \gamma_\nu \big]\Bigg\rbrace=\\
\nonumber =&\frac{1}{4}\left(\frac{e^2}{4\sin^2\!\theta_w \cos^2\!\theta_w}|Q|e^2\frac{1}{q^2}\,\frac{1}{(q^2-M_Z^2)^2+M_Z^2\Gamma_Z^2}\right)\times\\
\nonumber &\times\bigg(2(c_1-c_2)\big(p_1^\mu p_2^\nu+p_2^\mu p_1^\nu-g^{\mu\nu}(p_1\cdot p_2)\big)+2i(c_1+c_2)\epsilon^{\alpha\mu\beta\nu}p_{2\alpha} p_{1\beta} \bigg)\times\\
\nonumber &\times\Bigg\lbrace (q^2-M_Z^2)\Big(4(1-n^2)\Big[\big(c_3-c_4\big)\big(k_{1\mu}k_{2\nu}+k_{2\mu}k_{1\nu}-g_{\mu\nu}(k_1\cdot k_2)\big)+\big(c_3+c_4\big)i\epsilon_{\alpha\mu\beta\nu}k_1^\alpha k_2^\beta \Big]+\\
\nonumber &\quad+8m\Big[\big(c_3+c_4\big)\big(k_{1\mu}n_\nu+n_\mu k_{1\nu}-g_{\mu\nu}(k_1\cdot n)\big)+\big(c_3-c_4\big)i\epsilon_{\alpha\mu\beta\nu}k_1^\alpha n^\beta \Big]+\\
\nonumber &\quad+8(n\cdot k_2)\Big[\big(c_3-c_4\big)\big(k_{1\mu}n_\nu+n_\mu k_{1\nu}-g_{\mu\nu}(k_1\cdot n)\big)+\big(c_3+c_4\big)i\epsilon_{\alpha\mu\beta\nu}k_1^\alpha n^\beta \Big]-\\
\nonumber &\quad-4m^2(c_3-c_4)(1-n^2)g_{\mu\nu}+8m(c_3-c_4)i\epsilon_{\alpha\mu\beta\nu}k_2^\alpha n^\beta \Big)\\
&\quad + iM_Z\Gamma_Z m (c_3+c_4)8(n_\mu k_{2\nu}-k_{2\mu}n_\nu) \Bigg\rbrace=\\
\nonumber =&\left(\frac{e^2}{4\sin^2\!\theta_w \cos^2\!\theta_w}|Q|e^2\frac{1}{q^2}\,\frac{1}{(q^2-M_Z^2)^2+M_Z^2\Gamma_Z^2}\right)\times\\
\nonumber &\times\Bigg\lbrace (q^2-M_Z^2)\Bigg(4(c_1-c_2)(c_3-c_4)(1-n^2)\Big((p_1\cdot k_1)(p_2\cdot k_2)+(p_1\cdot k_2)(p_2\cdot k_1)\Big)+\\
\nonumber &\quad +8(c_1-c_2)(c_3+c_4)m\Big((p_1\cdot k_1)(p_2\cdot n)+(p_1\cdot n)(p_2\cdot k_1)\Big)+\\
\nonumber &\quad +8(c_1-c_2)(c_3-c_4)(n\cdot k_2)\Big((p_1\cdot k_1)(p_2\cdot n)+(p_1\cdot n)(p_2\cdot k_1)\Big)+\\
\nonumber &\quad+4(c_1-c_2)(c_3- c_4) m^2(1-n^2)(p_1\cdot p_2)+\\
\nonumber &\quad+4(c_1+c_2)(c_3+c_4)(1-n^2)\Big(-(p_1\cdot k_1)(p_2\cdot k_2)+(p_1\cdot k_2)(p_2\cdot k_1)\Big)+\\
\nonumber &\quad+8(c_1+c_2)(c_3-c_4)m\Big(-(p_1\cdot k_1)(p_2\cdot n)+(p_1\cdot n)(p_2\cdot k_1)\Big)+\\
\nonumber &\quad+8(c_1+c_2)(c_3+c_4)(n\cdot k_2)\Big(-(p_1\cdot k_1)(p_2\cdot n)+(p_1\cdot n)(p_2\cdot k_1)\Big)+\\
\nonumber &\quad+8(c_1+c_2)(c_3- c_4) m\Big(-(p_1\cdot k_2)(p_2\cdot n)+(p_1\cdot n)(p_2\cdot k_2)\Big)\Bigg)+\\
&\quad+4M_Z \Gamma_Z (c_1+c_2)(c_3+c_4)m \epsilon^{\alpha\mu\beta\nu}p_{2\alpha}p_{1\beta} (k_{2\mu}n_\nu-n_\mu k_{2\nu})\Bigg\rbrace=
\end{align}
\begin{align}
\nonumber =&\left(\frac{e^2}{4\sin^2\!\theta_w \cos^2\!\theta_w}|Q|e^2\frac{1}{s}\,\frac{1}{(s-M_Z^2)^2+M_Z^2\Gamma_Z^2}\right)\times\\
\nonumber &\times\Bigg\lbrace (s-M_Z^2)\Bigg((c_1-c_2)(c_3-c_4)(1-n^2)\Big((s+t-m^2)^2+(m^2-t)^2\Big)+\\
\nonumber &\quad +4(c_1-c_2)(c_3+c_4)m\Big((s+t-m^2)(p_2\cdot n)+(p_1\cdot n)(m^2-t)\Big)+\\
\nonumber &\quad +4(c_1-c_2)(c_3-c_4)(n\cdot k_2)\Big((s+t-m^2)(p_2\cdot n)+(p_1\cdot n)(m^2-t)\Big)+\\
\nonumber &\quad+2(c_1-c_2)(c_3- c_4) m^2(1-n^2)s+\\
\nonumber &\quad+(c_1+c_2)(c_3+c_4)(1-n^2)\Big(-(s+t-m^2)^2+(m^2-t)^2\Big)+\\
\nonumber &\quad+4(c_1+c_2)(c_3-c_4)m\Big(-(s+t-m^2)(p_2\cdot n)+(p_1\cdot n)(m^2-t)\Big)+\\
\nonumber &\quad+4(c_1+c_2)(c_3+c_4)(n\cdot k_2)\Big(-(s+t-m^2)(p_2\cdot n)+(p_1\cdot n)(m^2-t)\Big)+\\
\nonumber &\quad+4(c_1+c_2)(c_3- c_4) m\Big(-(m^2-t)(p_2\cdot n)+(p_1\cdot n)(s+t-m^2)\Big)\Bigg)+\\
&\quad+4M_Z \Gamma_Z (c_1+c_2)(c_3+c_4)m \epsilon^{\alpha\mu\beta\nu}p_{2\alpha}p_{1\beta} (k_{2\mu}n_\nu-n_\mu k_{2\nu})\Bigg\rbrace
\end{align}

\subsubsection{$ Z^0$ ir $ A^\gamma $ matricinis elementas}

Visas $ Z^0$ ir $ A^\gamma $ kanalo matricinis elementas yra (įskaičius daugiklį 1/4):
\begin{align}
\nonumber \langle|\mathcal{M}|^2\rangle_I^n =&\left(\frac{Qe^2}{s}\right)^2 \Big[ \frac{(1-n^2)}{2}\big((s+t-m^2)^2+(m^2-t)^2+2m^2s\big)+\\
\nonumber &\qquad\qquad+2(n\cdot k_2)\big((s+t-m^2)(p_2\cdot n)+(p_1\cdot n)(m^2-t)\big)\Big]+\\
\nonumber &+\left(\frac{e^2}{4\sin^2\!\theta_w \cos^2\!\theta_w}\right)^2\frac{1}{(s-M_Z^2)^2+M_Z^2\Gamma_Z^2} \times\\
\nonumber &\times \Bigg(\frac{1}{8}(c_1^2+c_2^2)(c_3^2+c_4^2)(1-n^2)\Big((s+t-m^2)^2+(m^2-t)^2\Big)+\\
\nonumber &\quad +\frac{1}{2}(c_1^2+c_2^2)(c_3^2-c_4^2)m\Big((s+t-m^2)(p_2\cdot n)+(p_1\cdot n)(m^2-t)\Big)+\\
\nonumber &\quad +\frac{1}{2}(c_1^2+c_2^2)(c_3^2+c_4^2)(n\cdot k_2)\Big((s+t-m^2)(p_2\cdot n)+(p_1\cdot n)(m^2-t)\Big)-\\
\nonumber &\quad-\frac{1}{2}(c_1^2+c_2^2)c_3 c_4 m^2(1-n^2)s+\\
\nonumber &\quad+\frac{1}{8}(c_1^2-c_2^2)(c_3^2-c_4^2)(1-n^2)\Big(-(s+t-m^2)^2+(m^2-t)^2\Big)+\\
\nonumber &\quad+\frac{1}{2}(c_1^2-c_2^2)(c_3^2+c_4^2)m\Big(-(s+t-m^2)(p_2\cdot n)+(p_1\cdot n)(m^2-t)\Big)+\\
\nonumber &\quad+\frac{1}{2}(c_1^2-c_2^2)(c_3^2-c_4^2)(n\cdot k_2)\Big(-(s+t-m^2)(p_2\cdot n)+(p_1\cdot n)(m^2-t)\Big)+\\
\nonumber &\quad+(c_1^2-c_2^2)c_3 c_4 m\Big(-(p_1\cdot n)(s+t-m^2)+(m^2-t)(p_2\cdot n)\Big)\Bigg)+\\
\nonumber &+\left(\frac{e^2}{4\sin^2\!\theta_w \cos^2\!\theta_w}|Q|e^2\frac{1}{s}\,\frac{1}{(s-M_Z^2)^2+M_Z^2\Gamma_Z^2}\right)\times\\
\nonumber &\times\Bigg\lbrace (s-M_Z^2)\Bigg(\frac{1}{4}(c_1-c_2)(c_3-c_4)(1-n^2)\Big((s+t-m^2)^2+(m^2-t)^2\Big)+\\
\nonumber &\quad +(c_1-c_2)(c_3+c_4)m\Big((s+t-m^2)(p_2\cdot n)+(p_1\cdot n)(m^2-t)\Big)+\\
\nonumber &\quad +(c_1-c_2)(c_3-c_4)(n\cdot k_2)\Big((s+t-m^2)(p_2\cdot n)+(p_1\cdot n)(m^2-t)\Big)+\\
\nonumber &\quad+\frac{1}{2}(c_1-c_2)(c_3- c_4) m^2(1-n^2)s+\\
\nonumber &\quad+\frac{1}{4}(c_1+c_2)(c_3+c_4)(1-n^2)\Big(-(s+t-m^2)^2+(m^2-t)^2\Big)+\\
\nonumber &\quad+(c_1+c_2)(c_3-c_4)m\Big(-(s+t-m^2)(p_2\cdot n)+(p_1\cdot n)(m^2-t)\Big)+\\
\nonumber &\quad+(c_1+c_2)(c_3+c_4)(n\cdot k_2)\Big(-(s+t-m^2)(p_2\cdot n)+(p_1\cdot n)(m^2-t)\Big)+\\
\nonumber &\quad+(c_1+c_2)(c_3- c_4) m\Big(-(m^2-t)(p_2\cdot n)+(p_1\cdot n)(s+t-m^2)\Big)\Bigg)+\\
&\quad+M_Z \Gamma_Z (c_1+c_2)(c_3+c_4)m \epsilon^{\alpha\mu\beta\nu}p_{2\alpha}p_{1\beta} (k_{2\mu}n_\nu-n_\mu k_{2\nu})\Bigg\rbrace
\end{align}

Sukeitus kvarko ir antikvarko momentus, matricinis elementas gaunamas:

\begin{align}
\nonumber \langle|\mathcal{M}|^2\rangle_{II}^n =&\left(\frac{Qe^2}{s}\right)^2 \Big[ \frac{(1-n^2)}{2}\big((s+t-m^2)^2+(m^2-t)^2+2m^2s\big)+\\
\nonumber &\qquad\qquad+2(n\cdot k_2)\big((s+t-m^2)(p_2\cdot n)+(p_1\cdot n)(m^2-t)\big)\Big]+\\
\nonumber &+\left(\frac{e^2}{4\sin^2\!\theta_w \cos^2\!\theta_w}\right)^2\frac{1}{(s-M_Z^2)^2+M_Z^2\Gamma_Z^2} \times\\
\nonumber &\times \Bigg(\frac{1}{8}(c_1^2+c_2^2)(c_3^2+c_4^2)(1-n^2)\Big((s+t-m^2)^2+(m^2-t)^2\Big)+\\
\nonumber &\quad +\frac{1}{2}(c_1^2+c_2^2)(c_3^2-c_4^2)m\Big((s+t-m^2)(p_2\cdot n)+(p_1\cdot n)(m^2-t)\Big)+\\
\nonumber &\quad +\frac{1}{2}(c_1^2+c_2^2)(c_3^2+c_4^2)(n\cdot k_2)\Big((s+t-m^2)(p_2\cdot n)+(p_1\cdot n)(m^2-t)\Big)-\\
\nonumber &\quad-\frac{1}{2}(c_1^2+c_2^2)c_3 c_4 m^2(1-n^2)s+\\
\nonumber &\quad+\frac{1}{8}(c_1^2-c_2^2)(c_3^2-c_4^2)(1-n^2)\Big((s+t-m^2)^2-(m^2-t)^2\Big)+\\
\nonumber &\quad+\frac{1}{2}(c_1^2-c_2^2)(c_3^2+c_4^2)m\Big((s+t-m^2)(p_2\cdot n)-(p_1\cdot n)(m^2-t)\Big)+\\
\nonumber &\quad+\frac{1}{2}(c_1^2-c_2^2)(c_3^2-c_4^2)(n\cdot k_2)\Big((s+t-m^2)(p_2\cdot n)-(p_1\cdot n)(m^2-t)\Big)+\\
\nonumber &\quad+(c_1^2-c_2^2)c_3 c_4 m\Big((p_1\cdot n)(s+t-m^2)-(m^2-t)(p_2\cdot n)\Big)\Bigg)+\\
\nonumber &+\left(\frac{e^2}{4\sin^2\!\theta_w \cos^2\!\theta_w}|Q|e^2\frac{1}{s}\,\frac{1}{(s-M_Z^2)^2+M_Z^2\Gamma_Z^2}\right)\times\\
\nonumber &\times\Bigg\lbrace (s-M_Z^2)\Bigg(\frac{1}{4}(c_1-c_2)(c_3-c_4)(1-n^2)\Big((s+t-m^2)^2+(m^2-t)^2\Big)+\\
\nonumber &\quad +(c_1-c_2)(c_3+c_4)m\Big((s+t-m^2)(p_2\cdot n)+(p_1\cdot n)(m^2-t)\Big)+\\
\nonumber &\quad +(c_1-c_2)(c_3-c_4)(n\cdot k_2)\Big((s+t-m^2)(p_2\cdot n)+(p_1\cdot n)(m^2-t)\Big)+\\
\nonumber &\quad+\frac{1}{2}(c_1-c_2)(c_3- c_4) m^2(1-n^2)s+\\
\nonumber &\quad+\frac{1}{4}(c_1+c_2)(c_3+c_4)(1-n^2)\Big((s+t-m^2)^2-(m^2-t)^2\Big)+\\
\nonumber &\quad+(c_1+c_2)(c_3-c_4)m\Big((s+t-m^2)(p_2\cdot n)-(p_1\cdot n)(m^2-t)\Big)+\\
\nonumber &\quad+(c_1+c_2)(c_3+c_4)(n\cdot k_2)\Big((s+t-m^2)(p_2\cdot n)-(p_1\cdot n)(m^2-t)\Big)+\\
\nonumber &\quad+(c_1+c_2)(c_3- c_4) m\Big((m^2-t)(p_2\cdot n)-(p_1\cdot n)(s+t-m^2)\Big)\Bigg)+\\
&\quad+M_Z \Gamma_Z (c_1+c_2)(c_3+c_4)m \epsilon^{\alpha\mu\beta\nu}p_{1\alpha}p_{2\beta} (k_{2\mu}n_\nu-n_\mu k_{2\nu})\Bigg\rbrace
\end{align}

Įprastiniai $ Z^0$ ir $ A^\gamma $ kanalo matriciniai elementai yra:
\begin{align}
\nonumber \langle|\mathcal{M}|^2\rangle_I =&\left(\frac{Qe^2}{s}\right)^2 \Big[ 2\big((s+t-m^2)^2+(m^2-t)^2+2m^2s\big)\Big]+\\
\nonumber &+\left(\frac{e^2}{4\sin^2\!\theta_w \cos^2\!\theta_w}\right)^2\frac{1}{(s-M_Z^2)^2+M_Z^2\Gamma_Z^2} \times\\
\nonumber &\times \Bigg(\frac{1}{2}(c_1^2+c_2^2)(c_3^2+c_4^2)\Big((s+t-m^2)^2+(m^2-t)^2\Big)-\\
\nonumber &\quad-2(c_1^2+c_2^2)c_3 c_4 m^2s+\\
\nonumber &\quad+\frac{1}{2}(c_1^2-c_2^2)(c_3^2-c_4^2)\Big(-(s+t-m^2)^2+(m^2-t)^2\Big)+\\
\nonumber &+\left(\frac{e^2}{4\sin^2\!\theta_w \cos^2\!\theta_w}|Q|e^2\frac{1}{s}\,\frac{1}{(s-M_Z^2)^2+M_Z^2\Gamma_Z^2}\right)\times\\
\nonumber &\times\Bigg\lbrace (s-M_Z^2)\Bigg((c_1-c_2)(c_3-c_4)\Big((s+t-m^2)^2+(m^2-t)^2\Big)+\\
\nonumber &\quad+2(c_1-c_2)(c_3- c_4) m^2s+\\
\nonumber &\quad+(c_1+c_2)(c_3+c_4)\Big(-(s+t-m^2)^2+(m^2-t)^2\Big)\Bigg\rbrace
\end{align}

\begin{align}
\nonumber \langle|\mathcal{M}|^2\rangle_{II} =&\left(\frac{Qe^2}{s}\right)^2 \Big[ 2\big((s+t-m^2)^2+(m^2-t)^2+2m^2s\big)\Big]+\\
\nonumber &+\left(\frac{e^2}{4\sin^2\!\theta_w \cos^2\!\theta_w}\right)^2\frac{1}{(s-M_Z^2)^2+M_Z^2\Gamma_Z^2} \times\\
\nonumber &\times \Bigg(\frac{1}{2}(c_1^2+c_2^2)(c_3^2+c_4^2)\Big((s+t-m^2)^2+(m^2-t)^2\Big)-\\
\nonumber &\quad-2(c_1^2+c_2^2)c_3 c_4 m^2s+\\
\nonumber &\quad+\frac{1}{2}(c_1^2-c_2^2)(c_3^2-c_4^2)\Big((s+t-m^2)^2-(m^2-t)^2\Big)+\\
\nonumber &+\left(\frac{e^2}{4\sin^2\!\theta_w \cos^2\!\theta_w}|Q|e^2\frac{1}{s}\,\frac{1}{(s-M_Z^2)^2+M_Z^2\Gamma_Z^2}\right)\times\\
\nonumber &\times\Bigg\lbrace (s-M_Z^2)\Bigg((c_1-c_2)(c_3-c_4)\Big((s+t-m^2)^2+(m^2-t)^2\Big)+\\
\nonumber &\quad+2(c_1-c_2)(c_3- c_4) m^2s+\\
\nonumber &\quad+(c_1+c_2)(c_3+c_4)\Big((s+t-m^2)^2-(m^2-t)^2\Big)\Bigg\rbrace
\end{align}

Vektorius $ n^\mu $ apibrėžtas kovariantiškai:
\begin{equation}
n^\mu=\frac{1}{|\vec{k}|}\left(-m T^\mu + \frac{E}{m}k^\mu\right)
\end{equation}
Sandaugos su šiuo vektoriumi yra:
\begin{equation}
(p_1\cdot n)=\frac{m^2-t}{2m}\left(\frac{|\vec{k}_2|-E_2\cos\!\theta}{E_2-|\vec{k}_2|\cos\!\theta}\right)
\end{equation}
\begin{equation}
(p_2\cdot n)=\frac{s+t-m^2}{2m}\left(\frac{|\vec{k}_2|+E_2\cos\!\theta}{E_2+|\vec{k}_2|\cos\!\theta}\right)
\end{equation}

\newpage 
\section{Diferencialinis sklaidos skerspjūvis}\label{priedas2}

Diferencialinis sklaidos skerspjūvis skaičiuojamas masės centro atskaitos sistemoje (angl. \textit{center-of-mass frame}, CM) ir išreiškiamas Mandelstamo kintamaisiais, kad paskui būtų lengva jį įvertinti LAB sistemoje. Atskirai skaičiuojami du atvejai: kai galinėje būsenoje yra viena masyvi $ \tau^+ $ dalelė ir kai yra dvi masyvios dalelės.

\subsection{Viena masyvi dalelė}

Diferencialinio sklaidos skerspjūvio formulė yra 
\begin{align}
d\sigma&=\frac{1}{3}\,\frac{\langle|\mathcal{M}|^2\rangle}{4(p_1\cdot p_2)}\,\frac{d^3\vec{k}_1}{(2\pi)^3 2E_1}\,\frac{d^3\vec{k}_2}{(2\pi)^3 2E_2}(2\pi)^4 \delta^4(p_1^\mu+p_2^\mu-k_1^\mu-k_2^\mu),
\end{align}
o daugiklis $ 1/3 $ atsiranda dėl vidurkinimo pagal kvarkų spalvas: yra 9 galimybės kvarkui ir antikvarkui turėti spalvos ir antispalvos poras, iš kurių 3 yra palankios - kai spalvos yra vienos rūšies.

Masės centro sistemoje pilutinis impulsas lygus 0, todėl
\begin{equation}
\vec{p}_1+\vec{p}_2=0,
\end{equation}
o pilnutinė energija yra $ \sqrt{s} $.
\begin{align}\label{pradzia}
\nonumber d\sigma&=\frac{\langle|\mathcal{M}|^2\rangle}{6s(2\pi)^2 4E_1 E_2}d^3\vec{k}_1 d^3\vec{k}_2 \,\,\delta^3(-\vec{k}_1-\vec{k}_2)\delta(\sqrt{s}-E_1-E_2)\\
&=\frac{\langle|\mathcal{M}|^2\rangle}{24s(2\pi)^2}\frac{d^3\vec{k}_2}{E_1 E_2}\delta(\sqrt{s}-E_1-E_2)
\end{align}
Iš pradžių integruojama pagal neutrino impulsą, nes jis vis tiek nėra stebimas. Kadangi neutrinas neturi masės, tai iš $ \delta $ funkcijos gaunamas	 neutrino impulsas $ \vec{k}_1=-\vec{k}_2 $ ir jo energija $ E_1=|\vec{k}_2| $. Integralas pagal $ \tau^+ $ trimatį momentą perrašomas į sferinę koordinačių sistemą ir integravimas pagal momento modulį pakeičiamas į integravimą pagal energiją:
\begin{align}
&d^3\vec{k}_2=|\vec{k}_2|^2 d|\vec{k}_2|d\Omega;\\
&d|\vec{k}_2|=\frac{d\sqrt{E_2^2-m^2}}{dE_2}dE_2=\frac{E_2}{|\vec{k}_2|}dE_2;\\
&\delta(\sqrt{s}-E_1-E_2)=\left(1+\frac{dE_1}{dE_2}\right)^{-1}\delta\left(E_2-\frac{s+m^2}{2\sqrt{s}}\right)=\frac{E_1}{\sqrt{s}}\delta\!\left(E_2-\frac{s+m^2}{2\sqrt{s}}\right)
\end{align}
Įrašius šiuos pakeitimus, diferencialinis sklaidos skerspjūvis tampa
\begin{align}
\nonumber d\sigma&=\frac{\langle|\mathcal{M}|^2\rangle}{24s(2\pi)^2}\frac{|\vec{k}_2| }{\sqrt{s}}dE_2 d\Omega \,\,\delta\!\left(E_2-\frac{s+m^2}{2\sqrt{s}}\right)\\
\nonumber &=\frac{\langle|\mathcal{M}|^2\rangle}{24s(2\pi)^2\sqrt{s}}\sqrt{\frac{(s+m^2)^2}{4s}-m^2} d\Omega\\
&=\frac{\langle|\mathcal{M}|^2\rangle}{48s^2(2\pi)^2}(s-m^2) d\Omega
\end{align}
Kadangi matricinis elementas nepriklauso nuo azimutinio kampo $ \varphi $, sklaidos skerspjūvį galima suintegruoti pagal $ \varphi $, o integralą pagal $ \cos\!\theta $ pakeisti į integralą pagal Mandelstamo kintamąjį $ t $. Masės centro sistemoje šis kintamasis yra
\begin{align}
&t=-\frac{1}{2}(s-m^2)(1-\cos\theta)
\end{align}
ir jo diferencialas
\begin{align}
&d\cos\theta=\frac{2}{s-m^2}dt
\end{align}
Taigi, invariantinis diferencialinis sklaidos skerspjūvis, išreikštas per Mandelstamo kintamuosius, yra
\begin{align}\label{dsigma1}
\frac{d\sigma}{dt}&=\frac{1}{48s^2 \pi}\langle|\mathcal{M}|^2\rangle
\end{align}
Šį diferencialinį sklaidos skerspjūvį pažymėsiu priklausomai nuo to, kuris matricinis elementas naudojamas:
\begin{equation}
\left(\frac{d\sigma}{dt}\right)_I=\frac{1}{48s^2 \pi}\langle|\mathcal{M}|^2\rangle_I,\quad \left(\frac{d\sigma}{dt}\right)_{II}=\frac{1}{48s^2 \pi}\langle|\mathcal{M}|^2\rangle_{II}
\end{equation}

\subsection{Dvi masyvios dalelės}

Diferencialinio sklaidos skerspjūvio skaičiavimą galima pradėti nuo (\ref{pradzia}) formulės:
\begin{align}
\nonumber d\sigma&=\frac{\langle|\mathcal{M}|^2\rangle}{24s(2\pi)^2}\frac{d^3\vec{k}_2}{E_1 E_2}\delta(\sqrt{s}-E_1-E_2)
\end{align}
Tačiau dabar $ k_1^\mu $ yra $ \tau^- $ dalelės momentas. Jos impulsas $ \vec{k}_1=-\vec{k}_2 $, bet energija dabar yra $ E_1=\sqrt{|\vec{k}_2|^2+m^2}=E_2 $. Delta funkcija perrašoma:
\begin{align}
\delta(\sqrt{s}-2E_2)=\frac{1}{2}\delta\!\left(E_2-\frac{\sqrt{s}}{2}\right)
\end{align}
Įrašius šiuos pakeitimus, diferencialinis sklaidos skerspjūvis tampa
\begin{align}
\nonumber d\sigma&=\frac{\langle|\mathcal{M}|^2\rangle}{24s(2\pi)^2}\frac{|\vec{k}_2| }{\sqrt{s}}dE_2 d\Omega \,\,\delta\!\left(E_2-\frac{\sqrt{s}}{2}\right)\\
&=\frac{\langle|\mathcal{M}|^2\rangle}{48s(2\pi)^2}\frac{\sqrt{s-4m^2}}{\sqrt{s}} d\Omega
\end{align}
Mandelstamo kintamasis $ t $ lygus
\begin{align}
&t=-m^2-\frac{s}{2}+\frac{\sqrt{s}}{2}\sqrt{s-4m^2}\cos\!\theta
\end{align}
ir jo diferencialas
\begin{align}
&d\cos\theta=\frac{2}{\sqrt{s}\sqrt{s-4m^2}}dt
\end{align}
Invariantinis diferencialinis sklaidos skerspjūvis, išreikštas per Mandelstamo kintamuosius, yra
\begin{align}\label{dsigma2}
\frac{d\sigma}{dt}&=\frac{1}{48s^2 \pi}\langle|\mathcal{M}|^2\rangle
\end{align}
Toks pat, kaip ir vienos masyvios dalelės atveju. Čia skiriasi tik integravimo ribos pagal $ t $.

\newpage
\section{Dalelių kinematika LAB sistemoje}\label{priedas3}

Norint skaičiuoti diferencialinį sklaidos skerspjūvį, kuris priklausytų nuo poliarinio kampo $ \theta $ LAB sistemoje, reikia kintamąjį $ t $ išreikšti per $ \cos\!\theta $ LAB sistemoje, ir tuomet integruoti pagal protonų momentų dalis $ x_1 $ ir $ x_2 $ šioje sistemoje. Skaičiavimas vėl skiriasi priklausomai nuo to, ar galinėje būsenoje yra viena, ar dvi dalelės.

\subsection{Viena masyvi dalelė}

Pradinių kvarkų momentai yra tokie:
\begin{equation}
p_1^\mu=x_1P\left(\begin{array}{c}1\\0\\0\\1 \end{array}\right),\quad p_2^\mu=x_2P\left(\begin{array}{c}1\\0\\0\\-1 \end{array}\right)
\end{equation}
kur $ P $ yra protono energija LAB sistemoje.
Mandelstamo kintamasis $ t $ išreiškiamas iš dviejų lygčių, pašalinant $ \tau^+ $ dalelės trimatį impulsą $ \vec{k}_2 $:
\begin{align}
\label{mandel1}&t=(p_1^\mu-k_2^\mu)^2=m^2-2x_1P\left(\sqrt{|\vec{k}_2|^2+m^2}-|\vec{k}_2|\cos\theta\right)\\
\label{mandel2}&u=(p_2^\mu-k_2^\mu)^2=m^2-2x_2P\left(\sqrt{|\vec{k}_2|^2+m^2}+|\vec{k}_2|\cos\theta\right)=m^2-t-s
\end{align}
Paskutinėje lygtyje panaudotas Mandelstamo kintamųjų sąryšis
\begin{equation}
s+t+u=\sum_i m_i
\end{equation}
ir Mandelstamo kintamasis $ s $ LAB sistemoje yra
\begin{equation}
s=(p_1^\mu+p_2^\mu)^2=4P^2x_1x_2
\end{equation}
Gaunami du sprendiniai:
\begin{align}\label{tpm}
\nonumber t_{\pm}&=\frac{1}{(x_1+x_2)^2-(x_1-x_2)^2\cos^2 \theta}\times\\
\nonumber &\times\Big(m^2 x_2(x_1+x_2+\cos^2 \theta(x_1-x_2))-4P^2x_1^2 x_2(x_1+x_2+\cos^2 \theta(x_2-x_1))\pm \\
&\pm 2\cos\theta x_1 x_2\sqrt{m^4+16P^4 x_1^2 x_2^2+4P^2m^2(-x_1^2-x_2^2+(x_1-x_2)^2\cos^2\theta)}\Big)
\end{align}
Iš (\ref{tpm}) lygties randama
\begin{align}
\nonumber \frac{dt_\pm}{d\cos\theta}&=\frac{2x_2}{(x_1+x_2)^2-(x_1-x_2)^2 \cos^2\!\theta}\Bigg(\cos\theta m^2 (x_1-x_2)+4\cos\theta P^2 x_1^2 (x_1-x_2)\pm\\
\nonumber &\qquad \pm\frac{4\cos^2\!\theta m^2 P^2 x_1(x_1-x_2)^2}{\sqrt{m^4+16P^4 x_1^2 x_2^2+4P^2m^2(-x_1^2-x_2^2+(x_1-x_2)^2\cos^2\!\theta)}}\pm\\
\nonumber &\qquad\pm x_1\sqrt{m^4+16P^4 x_1^2 x_2^2+4P^2m^2(-x_1^2-x_2^2+(x_1-x_2)^2\cos^2\!\theta)}\Bigg)+\\
\nonumber & +\frac{2x_2\cos\theta(x_1-x_2)^2}{((x_1+x_2)^2-(x_1-x_2)^2 \cos^2\!\theta)^2}\Bigg(m^2(x_1+x_2+\cos^2\!\theta(x_1-x_2))-\\
\nonumber &\qquad-4P^2x_1^2(x_1+x_2+\cos^2\!\theta(x_2-x_1))\pm\\
&\qquad\pm 2\cos\theta x_1\sqrt{m^4+16P^4 x_1^2 x_2^2+4P^2m^2(-x_1^2-x_2^2+(x_1-x_2)^2\cos^2\!\theta}\Bigg)
\end{align}

Matriciniuose elementuose su projekcijos operatoriais yra $ \tau^+ $ dalelės impulso modulis, todėl reikia gauti ir jo išraišką. Iš (\ref{mandel1}) ir (\ref{mandel2}) lygčių, eliminavus $ \cos\!\theta $, gaunama
\begin{equation}
|\vec{k}_2|=\frac{1}{4Px_1 x_2}\sqrt{t^2(x_1-x_2)^2+2t(m^2+4P^2x_1^2)(x_1-x_2)x_2+(m^2-4P^2x_1^2)^2x_2^2}
\end{equation}

Dydis $ t $ priklauso nuo pradinių kvarkų momentų dalių $ x_1 $ ir $ x_2 $, kurios nusako masės centro judėjimą LAB sistemoje. Jis gali turėti vieną arba dvi vertes priklausomai nuo tų momentų dalių \cite{landau}. Kintamieji $ x_1 $ ir $ x_2 $ gali turėti tik tokias vertes, kad užtektų energijos susidaryti $ \tau^+ $ dalelei. Masės centro sistemoje turi galioti tokia nelygybė: $ \sqrt{s}\geq m $. Ši nelygybė yra skaliarinė, todėl turi galioti ir laboratorinėje sistemoje:
\begin{equation}
4P^2x_1x_2\geq m^2
\end{equation} 
Gaunamos tokios kintamųjų integravimo ribos
\begin{equation}
\int_{m^2/4P^2}^1 dx_1 \int_{m^2/4P^2x_1}^1 dx_2
\end{equation}

\begin{figure}[H]
\centering
		\includegraphics[scale=1]{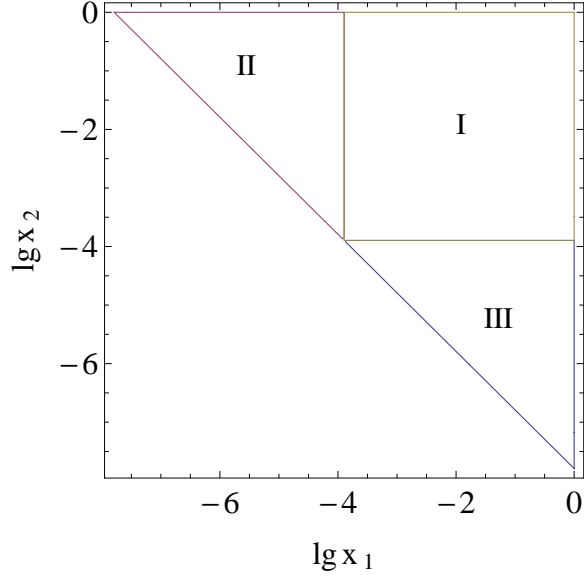}
\caption{Integravimo sritis pagal $ x_1 $ ir $ x_2 $ $ W^+ $ bozono atveju. $ \tau $ leptono masė $ m=1.776 $ GeV; protono energija $ P=7 $ TeV.}\label{regionW}
\end{figure}

Kintamųjų $ x_1 $ ir $ x_2 $ integravimo sritys skyla į dvi: pirma sritis yra ta, kur $ t $ įgyja vieną leistiną vertę, antra - kur dvi vertes. Tos sritys randamos iš kvadratinės šaknies (\ref{tpm}) lygtyje. Kintamasis $ t $ turės vieną reikšmę ir kampo $ \theta $ kosinusas galės įgyti visas reikšmes, kai integravimo sritis yra
\begin{equation}
\int_{m/2P}^1 dx_1 \int_{m/2P}^1 dx_2
\end{equation}
ir (\ref{tpm}) lygtyje imamas + ženklas. Ši sritis \ref{regionW} paveiksle pavadinta $ I $.

Kintamasis $ t $ turės dvi reikšmes ir kampo $ \theta $ kosinusas turės mažiausią galimą reikšmę, kai integravimo sritis yra
\begin{align}
\int_{m^2/4P^2}^{m/2P}dx_1 \int_{m^2/4P^2x_1}^1 dx_2 +\int_{m^2/4P^2}^{m/2P}dx_2 \int_{m^2/4P^2x_2}^1 dx_1
\end{align}
ir minimali kosinuso vertė yra 
\begin{equation}
\cos\!\theta_{min}=\frac{\sqrt{-(4P^2x_1^2-m^2)(4P^2x_2^2-m^2)}}{2Pm\sqrt{(x_1-x_2)^2}}
\end{equation}
Integravimo sritis \ref{regionW} paveiksle pavadinta $ II $ ir $ III $.

Sklaidos skerspjūvio formulėje (\ref{dsigma1}) diferencialas $ dt $ perrašomas
\begin{equation}
dt=\left(\frac{dt}{d\cos\!\theta}\right)d\cos\!\theta
\end{equation}
Srityje, kurioje galima viena $ t $ reikšmė, šis diferencialo pakeitimas yra toks paprastas. Tačiau srityse, kuriose galimos dvi $ t $ reikšmės, yra ir du skirtingi diferencialai. Kai $ x_1>x_2 $, sistemos masės centras juda teigiama $ \hat{z} $ ašies kryptimi. Kadangi yra maksimalus dalelės išlėkimo kampas, tai dalelė negali išlėkti kampu, didesniu už $ \frac{\pi}{2} $. Ir atvirkščiai, kai $ x_1<x_2 $ sistemos masės centras juda neigiama $ \hat{z} $ ašies kryptimi ir dalelė negali išlėkti kampu, mažesniu už $ \frac{\pi}{2} $ (čia kalbama tik apie sritį, kurioje galimos dvi $ t $ vertės). Šiame darbe pasirinktos kampo $ \theta $ vertės tarp 0 ir $ \frac{\pi}{2} $. Todėl pasirenkama integravimo sritis, kur $ x_1>x_2 $, \ref{regionW} paveiksle pažymėta $ III $. Diferencialas šioje srityje yra
\begin{equation}
dt=\left(\frac{dt_+}{d\cos\!\theta}\right)d\cos\!\theta-\left(\frac{dt_-}{d\cos\!\theta}\right)d\cos\!\theta
\end{equation}

Integralas pagal $ \cos\!\theta $ perrašomas taip (sritis, kur $ x_1>x_2 $):
\begin{equation}
\int_{m^2/4P^2}^{m/2P}dx_2 \int_{m^2/4P^2x_2}^1 dx_1 \left[\left(\frac{d\sigma}{dt} \right)_{t_+}\left(\frac{dt_+}{d\cos\!\theta}\right)- \left(\frac{d\sigma}{dt} \right)_{t_-}\left(\frac{dt_-}{d\cos\!\theta}\right)\right]
\end{equation}
Čia $ \left(\frac{d\sigma}{dt} \right) $ yra sklaidos skerspjūvis (\ref{dsigma1}), priklausantis nuo $ t_+ $ arba $ t_- $.

Visas diferencialinis sklaidos skerspjūvis, įskaitant skirtingus kvarkų skonius, $ \tau^+ $ dalelei išsisklaidyti LAB sistemoje kampais $ 0<\theta<\frac{\pi}{2} $ yra
\begin{align}\label{wdsigma}
\nonumber \left(\frac{d\sigma}{d\cos\!\theta}\right)_{LAB}=&\sum_{AB=u\overline{d}, u\overline{s}}\int_\frac{m}{2P}^1 dx_1 \int_\frac{m}{2P}^1 dx_2 \times \\
\nonumber &\times \Bigg[ \Big(\frac{d\sigma(t_+)}{dt}\Big)_I \Big(\frac{dt_+}{d\cos\!\theta}\Big) f_A(x_1)f_B(x_2)+\Big(\frac{d\sigma(t_+)}{dt}\Big)_{II} \Big(\frac{dt_+}{d\cos\!\theta}\Big) f_A(x_2)f_B(x_1) \Bigg]+\\
\nonumber &+\sum_{AB=u\overline{d}, u\overline{s}}\int_\frac{m^2}{4P^2}^\frac{m}{2P} dx_2 \int_\frac{m^2}{4P^2 x_2}^1 dx_1 \times \\
\nonumber &\times \Bigg[ \Big(\frac{d\sigma(t_+)}{dt}\Big)_I \Big(\frac{dt_+}{d\cos\!\theta}\Big) f_A(x_1)f_B(x_2)+\Big(\frac{d\sigma(t_+)}{dt}\Big)_{II} \Big(\frac{dt_+}{d\cos\!\theta}\Big) f_A(x_2)f_B(x_1)-\\
&\quad -\Big(\frac{d\sigma(t_-)}{dt}\Big)_I \Big(\frac{dt_-}{d\cos\!\theta}\Big) f_A(x_1)f_B(x_2)-\Big(\frac{d\sigma(t_-)}{dt}\Big)_{II} \Big(\frac{dt_-}{d\cos\!\theta}\Big) f_A(x_2)f_B(x_1)  \Bigg]
\end{align}

\subsection{Dvi masyvios dalelės}

Dalelių kinematika šiuo atveju skiriasi dėl to, kad galinėje būsenoje yra dvi masyvios $ \tau $ dalelės. Mandelstamo kintamasis $ t $ išreiškiamas iš dviejų lygčių, pašalinant $ \tau^+ $ dalelės trimatį impulsą $ \vec{k}_2 $:
\begin{align}
\label{mandel1Z}&t=m^2-2x_1P\left(\sqrt{|\vec{k}_2|^2+m^2}-|\vec{k}_2|\cos\theta\right)\\
\label{mandel2Z}&u=m^2-2x_2P\left(\sqrt{|\vec{k}_2|^2+m^2}+|\vec{k}_2|\cos\theta\right)=2m^2-t-s
\end{align}
Gaunami du sprendiniai:
\begin{align}\label{tpmZ}
\nonumber t_\pm=&\frac{1}{(x_1+x_2)^2-(x_1-x_2)^2\cos^2\!\theta}\times\\
\nonumber &\times\Big(m^2((x_1+x_2)^2-(x_1-x_2)^2\cos^2\!\theta)-4 P^2 x_1^2 x_2(x_1+x_2+\cos^2\!\theta(x_2-x_1))\pm\\
&\pm 4\cos\!\theta P x_1 x_2\sqrt{4P^2x_1^2 x_2^2-m^2((x_1+x_2)^2-(x_1-x_2)^2\cos^2\!\theta)}\Big)
\end{align}
Iš (\ref{tpmZ}) lygties randama
\begin{align}
\nonumber dt_\pm=&\frac{1}{(x_1+x_2)^2-(x_1-x_2)^2\cos^2\!\theta}\Big(-8P^2 x_1^2 x_2 \cos\!\theta (x_2-x_1)\pm\\
\nonumber &\quad \pm\frac{4P x_1x_2 \cos^2\!\theta m^2 (x_1-x_2)^2}{\sqrt{4P^2x_1^2 x_2^2-m^2((x_1+x_2)^2-(x_1-x_2)^2\cos^2\!\theta)}}\pm\\
\nonumber &\quad \pm 4Px_1 x_2 \sqrt{4P^2x_1^2 x_2^2-m^2((x_1+x_2)^2-(x_1-x_2)^2\cos^2\!\theta)}\Big)+\\
\nonumber &+\frac{2\cos\!\theta (x_1-x_2)^2}{((x_1+x_2)^2-(x_1-x_2)^2\cos^2\!\theta)^2}\Big(-4P^2 x_1^2 x_2(x_1+x_2+\cos^2\!\theta(x_2-x_1))\pm\\
&\quad \pm 4Px_1 x_2 \cos\!\theta \sqrt{4P^2x_1^2 x_2^2-m^2((x_1+x_2)^2-(x_1-x_2)^2\cos^2\!\theta)}\Big)
\end{align}
Iš (\ref{mandel1Z}) ir (\ref{mandel1Z}) lygčių, eliminavus $ \cos\!\theta $, gaunama
\begin{equation}
|\vec{k}_2|=\frac{1}{4Px_1 x_2}\sqrt{(s x_1+t(x_1-x_2))^2+m^4(x_1-x_2)^2-2m^2(sx_1(x_1+x_2)+t(x_1-x_2)^2)}
\end{equation}

Dydis $ t $ priklauso nuo pradinių kvarkų momentų dalių $ x_1 $ ir $ x_2 $, kurios nusako masės centro judėjimą LAB sistemoje. Jis gali turėti vieną arba dvi vertes priklausomai nuo tų momentų dalių \cite{landau}. Kintamieji $ x_1 $ ir $ x_2 $ gali turėti tik tokias vertes, kad užtektų energijos susidaryti dviem $ \tau^+ \tau^- $ dalelėm. Masės centro sistemoje turi galioti tokia nelygybė: $ \sqrt{s}\geq 2m $. Ši nelygybė yra skaliarinė, todėl turi galioti ir laboratorinėje sistemoje:
\begin{equation}
4P^2x_1x_2\geq 4m^2
\end{equation} 
Gaunamos tokios kintamųjų integravimo ribos
\begin{equation}
\int_{m^2/P^2}^1 dx_1 \int_{m^2/P^2x_1}^1 dx_2
\end{equation}

\begin{figure}[H]
	\centering
		\includegraphics[scale=1]{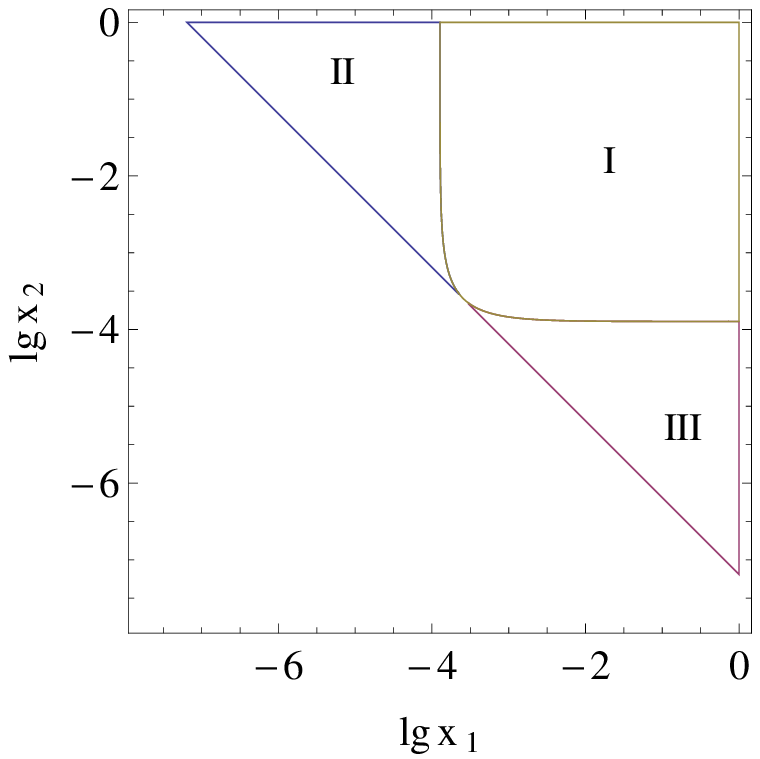}
\caption{Integravimo sritis pagal $ x_1 $ ir $ x_2 $ $ Z^0 $ ir $ A^\gamma $ bozonų atveju. $ \tau $ leptono masė $ m=1.776 $ GeV; protono energija $ P=7 $ TeV.}\label{regionZ}
\end{figure}

Kintamųjų $ x_1 $ ir $ x_2 $ integravimo sritys skyla į dvi: pirma sritis yra ta, kur $ t $ įgyja vieną leistiną vertę, antra - kur dvi vertes. Tos sritys randamos iš kvadratinės šaknies (\ref{tpmZ}) lygtyje. Kintamasis $ t $ turės vieną reikšmę ir kampo $ \theta $ kosinusas galės įgyti visas reikšmes, kai integravimo sritis yra
\begin{equation}
\int_\frac{m}{2P-m}^1 dx_1 \int_\frac{m}{2P-m/x_1}^1 dx_2
\end{equation}
ir (\ref{tpm}) lygtyje imamas + ženklas. Ši sritis \ref{regionZ} paveiksle pavadinta $ I $.

Kintamasis $ t $ turės dvi reikšmes ir kampo $ \theta $ kosinusas turės mažiausią galimą reikšmę, kai integravimo sritis yra
\begin{equation}
\int_{\frac{m}{P}}^{1}dx_1 \int_{\frac{m^2}{P^2 x_1}}^{\frac{m}{2P-m/x_1}} dx_2+\int_{\frac{m}{P}}^{1}dx_2 \int_{\frac{m^2}{P^2 x_2}}^{\frac{m}{2P-m/x_2}} dx_1
\end{equation}
ir minimali kosinuso vertė yra 
\begin{equation}
\cos\!\theta_{min}=\frac{\sqrt{m^2(x_1+x_2)^2-4P^2 x_1^2 x_2^2}}{m\sqrt{(x_1-x_2)^2}},
\end{equation}
Integravimo sritis \ref{regionZ} paveiksle pavadinta $ II $ ir $ III $.

Sklaidos skerspjūvio formulėje (\ref{dsigma2}) diferencialas $ dt $ perrašomas
\begin{equation}
dt=\left(\frac{dt}{d\cos\!\theta}\right)d\cos\!\theta
\end{equation}
Srityje, kurioje galima viena $ t $ reikšmė, šis diferencialo pakeitimas yra toks paprastas. Srityje, kurioje galimos dvi $ t $ reikšmės, pasirenkama integravimo sritis, kur $ x_1>x_2 $, \ref{regionZ} paveiksle pažymėta $ III $. Diferencialas šioje srityje yra
\begin{equation}
dt=\left(\frac{dt_+}{d\cos\!\theta}\right)d\cos\!\theta-\left(\frac{dt_-}{d\cos\!\theta}\right)d\cos\!\theta
\end{equation}

Integralas pagal $ \cos\!\theta $ perrašomas taip (sritis, kur $ x_1>x_2 $):
\begin{equation}
\int_{\frac{m}{P}}^{1}dx_1 \int_{\frac{m^2}{P^2 x_1}}^{\frac{m}{2P-m/x_1}} dx_2 \left[\left(\frac{d\sigma}{dt} \right)_{t_+}\left(\frac{dt_+}{d\cos\!\theta}\right)- \left(\frac{d\sigma}{dt} \right)_{t_-}\left(\frac{dt_-}{d\cos\!\theta}\right)\right]
\end{equation}
Čia $ \left(\frac{d\sigma}{dt} \right) $ yra sklaidos skerspjūvis (\ref{dsigma2}), priklausantis nuo $ t_+ $ arba $ t_- $.

Visas diferencialinis sklaidos skerspjūvis, įskaitant skirtingus kvarkų skonius, $ \tau^+ $ dalelei išsisklaidyti LAB sistemoje kampais $ 0<\theta<\frac{\pi}{2} $ yra
\begin{align}\label{zdsigma}
\nonumber \left(\frac{d\sigma}{d\cos\!\theta}\right)_{LAB}=&\sum_{AB=u\overline{u}, d\overline{d},s\overline{s}}\int_\frac{m}{2P-m}^1 dx_1 \int_\frac{m}{2P-m/x_1}^1 dx_2 \times \\
\nonumber &\times \Bigg[ \Big(\frac{d\sigma(t_+)}{dt}\Big)_I \Big(\frac{dt_+}{d\cos\!\theta}\Big) f_A(x_1)f_B(x_2)+\Big(\frac{d\sigma(t_+)}{dt}\Big)_{II} \Big(\frac{dt_+}{d\cos\!\theta}\Big) f_A(x_2)f_B(x_1) \Bigg]+\\
\nonumber &+\sum_{AB=u\overline{u}, d\overline{d},s\overline{s}}\int_{\frac{m}{P}}^{1}dx_1 \int_{\frac{m^2}{P^2 x_1}}^{\frac{m}{2P-m/x_1}} dx_2 \times \\
\nonumber &\times \Bigg[ \Big(\frac{d\sigma(t_+)}{dt}\Big)_I \Big(\frac{dt_+}{d\cos\!\theta}\Big) f_A(x_1)f_B(x_2)+\Big(\frac{d\sigma(t_+)}{dt}\Big)_{II} \Big(\frac{dt_+}{d\cos\!\theta}\Big) f_A(x_2)f_B(x_1)-\\
&\quad -\Big(\frac{d\sigma(t_-)}{dt}\Big)_I \Big(\frac{dt_-}{d\cos\!\theta}\Big) f_A(x_1)f_B(x_2)-\Big(\frac{d\sigma(t_-)}{dt}\Big)_{II} \Big(\frac{dt_-}{d\cos\!\theta}\Big) f_A(x_2)f_B(x_1)  \Bigg]
\end{align}

\end{document}